\documentclass[aps,prc,amsfonts,preprintnumbers,superscriptaddress,showpacs,nofootinbib,10pt]{revtex4-1}

\usepackage{bm} 
\usepackage{braket} 
\usepackage{amsmath} 

\usepackage{graphicx,color}

\newcommand{\GC}{G_\text{C}}
\newcommand{\GQ}{G_\text{Q}}
\newcommand{\GM}{G_\text{M}}

\newcommand{\GMNucl}{G_\text{M}}
\newcommand{\GE}{G_\text{E}}

\newcommand{\mN}{m_N}
\newcommand{\md}{m_d}
\newcommand{\mpi}{M_\pi}

\newcommand{\fpi}{F_\pi}
\newcommand{\gA}{g_A}
\newcommand{\nn}{\nonumber \\ }

\newcommand{\be}{\begin{equation}}
\newcommand{\ee}{\end{equation}}
\newcommand{\bea}{\begin{eqnarray}}
\newcommand{\eea}{\end{eqnarray}}

\begin{document}

\title{High-accuracy calculation of the deuteron charge and quadrupole form factors
in chiral effective field theory}



\author{A.~A.~Filin}
\email{arseniy.filin@rub.de}
\affiliation{Ruhr-Universit\"at Bochum, Fakult\"at f\"ur Physik und Astronomie,
Institut f\"ur Theoretische Physik II,
  D-44780 Bochum, Germany}
\author{D.~M\"oller}
\email{daniel.moeller-x8g@rub.de}
\affiliation{Ruhr-Universit\"at Bochum, Fakult\"at f\"ur Physik und Astronomie,
Institut f\"ur Theoretische Physik II,
  D-44780 Bochum, Germany}
\author{V.~Baru}
\email{vadimb@tp2.rub.de}
\affiliation{Helmholtz-Institut f\"ur Strahlen- und Kernphysik and Bethe Center for Theoretical Physics, Universit\"at Bonn, D-53115 Bonn, Germany}
\affiliation{Institute for Theoretical and Experimental Physics NRC ``Kurchatov Institute'', Moscow 117218, Russia  }
\affiliation{P.N. Lebedev Physical Institute of the Russian Academy of Sciences, 119991, Leninskiy Prospect 53, Moscow, Russia}
\author{E.~Epelbaum}
\email[]{evgeny.epelbaum@rub.de}
\affiliation{Ruhr-Universit\"at Bochum, Fakult\"at f\"ur Physik und Astronomie,
Institut f\"ur Theoretische Physik II,
  D-44780 Bochum, Germany}
\author{H.~Krebs}
\email[]{hermann.krebs@rub.de}
\affiliation{Ruhr-Universit\"at Bochum, Fakult\"at f\"ur Physik und Astronomie,
Institut f\"ur Theoretische Physik II,
  D-44780 Bochum, Germany}
\author{P.~Reinert}
\email[]{patrick.reinert@rub.de}
\affiliation{Ruhr-Universit\"at Bochum, Fakult\"at f\"ur Physik und Astronomie,
Institut f\"ur Theoretische Physik II,
  D-44780 Bochum, Germany}

\date{\today}

\begin{abstract}
We present a comprehensive analysis of the deuteron charge and
quadrupole form factors based on the latest two-nucleon potentials
and charge density operators derived in chiral effective
field theory. The single- and two-nucleon contributions to the charge density
are expressed in terms of the proton and neutron form factors, for
which the most up-to-date empirical parametrizations are employed.
By adjusting the fifth-order short-range terms in the two-nucleon
charge density operator to reproduce the world data on the momentum-transfer
dependence of the deuteron charge and quadrupole form factors, we
predict the values of the structure radius and the quadrupole
moment of the deuteron:
$r_{\rm str}=1.9729\substack{+0.0015\\ -0.0012}\ \text{fm},\
Q_d=0.2854\substack{+0.0038\\ -0.0017}\ \text{fm}^2. $
A comprehensive and systematic analysis of various sources of
uncertainty in our predictions is performed.
Following the strategy advocated in our recent publication
Phys.~Rev.~Lett.~\textbf{124}, 082501 (2020), we employ
the extracted structure radius together with the accurate atomic data for
the deuteron-proton mean-square charge radii difference to update the
determination of the neutron charge radius, for which we find:
$r_n^2=-0.105\substack{+0.005\\ -0.006} \, \text{fm}^2$.
Given the observed rapid convergence of the deuteron form factors
in the momentum-transfer range of $Q \simeq 1-2.5$~fm$^{-1}$,
we argue that this intermediate-energy domain is particularly
sensitive to the details of the nucleon form factors and can be used
to test different parametrizations.
\end{abstract}


 \maketitle

\section{Introduction}\label{sec:intro}

Chiral effective field theory (EFT) is becoming a precision tool for
analyzing low-energy few-nucleon reactions and nuclear
structure~\cite{Epelbaum:2008ga,Epelbaum:2012vx,Epelbaum:2019kcf,Machleidt:2011zz}.
The chiral expansion of the nucleon-nucleon (NN) force has been recently
pushed to fifth order (N$^4$LO)~\cite{Entem:2014msa} and even beyond~\cite{Entem:2015xwa}. The
last-generation chiral EFT NN potentials of Ref.~\cite{Reinert:2017usi} provide an excellent description
of the neutron-proton and proton-proton scattering data, which, at the
highest considered order, is even better than the one achieved using
so-called high-precision phenomenological potentials such as the CD Bonn~\cite{Machleidt:2000ge},
Nijm I, II and Reid93~\cite{Stoks:1994wp} and AV18~\cite{Wiringa:1994wb} models. The
essential feature of these chiral NN forces is the usage of a
semi-local regulator~\cite{Epelbaum:2014efa,Epelbaum:2014sza}, see
also Refs.~\cite{Gezerlis:2013ipa,Piarulli:2014bda}, which allows one to significantly reduce the amount
of finite-cutoff artifacts in the long-range part of the interaction.
For an alternative regularization approach using a non-local cutoff
see Ref.~\cite{Entem:2017gor}. The chiral NN potentials of
Ref.~\cite{Reinert:2017usi} also  provide a clear evidence of the two-pion
exchange, which is determined in a parameter-free way by the chiral
symmetry of QCD along with the empirical information on pion-nucleon
scattering from the recent analysis in the framework of the Roy-Steiner equations~\cite{Hoferichter:2015tha,Hoferichter:2015hva}.
In the most recent work of Ref.~\cite{Reinert:2020mcu},  the potential of Ref.~\cite{Reinert:2017usi}  was updated to include
also the charge-independence-breaking and charge-symmetry-breaking NN interactions up through N$^4$LO.

In parallel with these
studies, a simple and universal algorithm for quantifying
truncation errors in chiral EFT without reliance on cutoff variation
was formulated in Ref.~\cite{Epelbaum:2014efa} and validated in
Ref.~\cite{Epelbaum:2014sza}.  This approach has been successfully
applied to a variety of low-energy hadronic observables, see
e.g.~Refs.~\cite{Binder:2015mbz,Binder:2018pgl,Epelbaum:2018ogq,Skibinski:2016dve,Yao:2016vbz,Siemens:2017opr,Lynn:2019vwp,NevoDinur:2018hdo,Blin:2018pmj,Lonardoni:2018nob}. In
Refs.~\cite{Furnstahl:2015rha,Melendez:2017phj,Wesolowski:2018lzj,Epelbaum:2019zqc}, it was re-interpreted and
further scrutinized within a Bayesian approach.

These developments provide a solid basis for applications beyond the
two-nucleon system and offer highly nontrivial possibilities to test
chiral EFT by pushing the expansion to high orders. In this
paper, we focus on the charge and quadrupole elastic form factors (FFs) of
the deuteron.

The electromagnetic FFs of the deuteron
certainly belong to the most extensively studied observables in
nuclear physics, see Refs.~\cite{Garcon:2001sz,Gilman:2001yh,Marcucci:2015rca} for review articles.
A large variety of theoretical approaches ranging from
non-relativistic quantum mechanics to fully covariant models have been
applied to this problem since the 1960s, see
Ref.~\cite{Phillips:2003pa} for an overview. The electromagnetic
structure of the deuteron has also been investigated in the framework of
pionless~\cite{Chen:1999tn} and chiral~\cite{Phillips:1999am,Walzl:2001vb,Phillips:2003jz,Phillips:2006im,Valderrama:2007ja,Piarulli:2012bn,Epelbaum:2013naa}
EFT.

In spite of the extensive existing theoretical work, there is a strong
motivation to take a fresh look at the
deuteron FFs in the framework of chiral EFT.  First of all, the
calculation of the  deuteron charge FF
with unprecedented accuracy, by employing consistent NN interactions and charge density operators up to the fifth order in the chiral expansion,
provides direct access to the structure radius of the deuteron and through that to  the neutron charge radius, as elaborated  in Ref.~\cite{Filin:2019eoe}.
Similarly, the quantitative description of the quadrupole FF,
supplemented with the comprehensive error analysis,
opens the possibility to extract  the quadrupole moment of
the deuteron that is known very accurately  and thus probes our understanding of the nuclear forces and currents.
In this context, it is worth mentioning the tendency of modern nuclear interactions derived in chiral EFT to significantly
underpredict the radii of medium-mass and heavy nuclei, see e.g.~\cite{Cipollone:2014hfa}.
The existing calculations for $A \ge 16$ systems do, however, not take into account contributions to the three-nucleon force beyond
third order of the chiral expansion (N$^2$LO), exchange currents and relativistic corrections
and also suffer from uncertainties intrinsic to truncations of the many-body Hilbert space.
It is, therefore,  of great importance to test the role of these effects in \emph{consistent}
calculations of electromagnetic few-nucleon
processes at high orders in chiral EFT along with a careful error analysis.
Focusing on the few-nucleon sector has an advantage of avoiding potential uncertainties associated
with many-body methods. In particular, no additional softening of the
interactions by using e.g.~Similarity Renormalization Group
transformation~\cite{Bogner:2007rx} is necessary for the  light nuclei
like $^2$H, $^3$H, $^3$He and $^4$He.
It is also interesting and important to test the performance and applicability range of the newest high-precision chiral NN
potentials of Refs.~\cite{Reinert:2020mcu,Reinert:2017usi} and the charge density operators
by studying the momentum-transfer ($Q$) dependence of the deuteron FFs and their
convergence with respect to the chiral  expansion.
This provides a rather non-trivial test of the applicability range of
chiral EFT since the deuteron FFs decrease by several orders of magnitude with increasing
values of $Q$. Therefore, a correction to the charge
operator that is small at $Q^2 = 0$ may, potentially, have a large
impact  at higher-$Q^2$ values.

In this paper, we perform a detailed analysis of the deuteron charge and quadrupole FFs in chiral
EFT. We include all contributions to the charge-density operator at fourth
order (N$^3$LO) relative to the leading single-nucleon operator and
take into account the  short-range operators at N$^4$LO.
The strength of the N$^4$LO short-range operators is adjusted to obtain the best fits to the
experimental data for the deuteron charge and quadrupole FFs.
We demonstrate that both the single- and two-nucleon charge density operators
can be expressed in terms of
the nucleon FFs and  exploit this fact in the  calculation of the deuteron FFs. This allows us to avoid reliance
on the strict chiral expansion for the nucleon FFs by employing
the corresponding empirical parametrizations.
Since  the errors related to the truncation of the chiral expansion
are still very small in the momentum range of  $Q \simeq 1-2.5$~fm$^{-1}$,
this  intermediate energy domain  appears to be particularly  sensitive to the
nucleon FFs and thus can be used to test the consistency of the
employed up-to-date nucleon FFs with the deuteron FFs.

Once the two NN contact terms in the charge density operator are
determined from a fit to the  world data on  the deuteron FFs,
we arrive at a parameter-free
prediction for the quantities at $Q=0$, namely the structure radius
and the quadrupole moment of the deuteron.
It is worth mentioning that the nucleon FFs do not contribute to the extracted deuteron observables at $Q=0$.
We perform various consistency checks of our theoretical approach
and demonstrate that   (i)  our results show only a  mild residual cutoff
dependence;  (ii) the results for the deuteron FFs, the structure
radius and the quadrupole moment are   basically insensitive
to the choice of off-shell parameters entering the NN potentials and the charge density operator. However, this is only true
as long as the NN potentials and the charge density  are calculated
consistently, which implies  that the nucleon FFs must be included
both in  the one and  two-body charge density operators, as advocated below.
Finally, we perform a detailed error analysis of the obtained results
by addressing various sources of uncertainties.

In Ref.~\cite{Filin:2019eoe}, we already employed this approach to
extract the structure radius from the charge deuteron FF.  Here, we
provide additional details of the calculation and update the analysis of
Ref.~\cite{Filin:2019eoe} in the following aspects:  (a) we employ the
latest version of the NN potential from Ref.~\cite{Reinert:2020mcu}
that includes the relevant isospin breaking corrections,  (ii)  we carry out a combined analysis of
both the charge and quadrupole deuteron FFs,
(iii) relying on our  Bayesian estimate of the truncation error from the
chiral expansion, in the fits to the FF data we extend the momentum range to $Q\sim 6 $~fm$^{-1}$
as compared to $Q\sim 4 $~fm$^{-1}$ used in Ref.~\cite{Filin:2019eoe}.

Our paper is organized as follows. In Section~\ref{sec:formalism}, we
discuss a general formalism to calculate the form factors of the
deuteron. Sections~\ref{sec:charge} and~\ref{sec:regul} are devoted to the chiral expansion
and regularization of the charge density operator. In Section~\ref{sec:charge}
we also give a short overview of the nucleon FFs used as input in our calculations.
Section~\ref{sec:relcorr} deals with the treatment of the relativistic
corrections. Next, the notation for various contributions to the form factors,
their chiral order and relations to the structure radius and the quadrupole moment
are specified in Section~\ref{sec:analyticresults}.
Our results for the momentum-transfer dependence of the charge and
quadrupole FFs are presented in Section~\ref{sec:results}.
After fixing the short-range charge density operator from the  best fit  to the experimental data we  extract the values of the deuteron structure
radius,  the neutron charge radius and the deuteron quadrupole moment  and analyze various sources of uncertainties. Also, we discuss the convergence of the chiral
expansion for both the deuteron FFs and the extracted quantities at $Q=0$. The main results of our
study are summarized in Section~\ref{sec:summary}, where we also
discuss their impact on the determination of the neutron charge radius using high-accuracy atomic data on the deuteron-proton charge radius difference.

\section{Formalism}\label{sec:formalism}

\subsection{Elastic electron-deuteron scattering}

\begin{figure}
\includegraphics[width=0.5\textwidth]{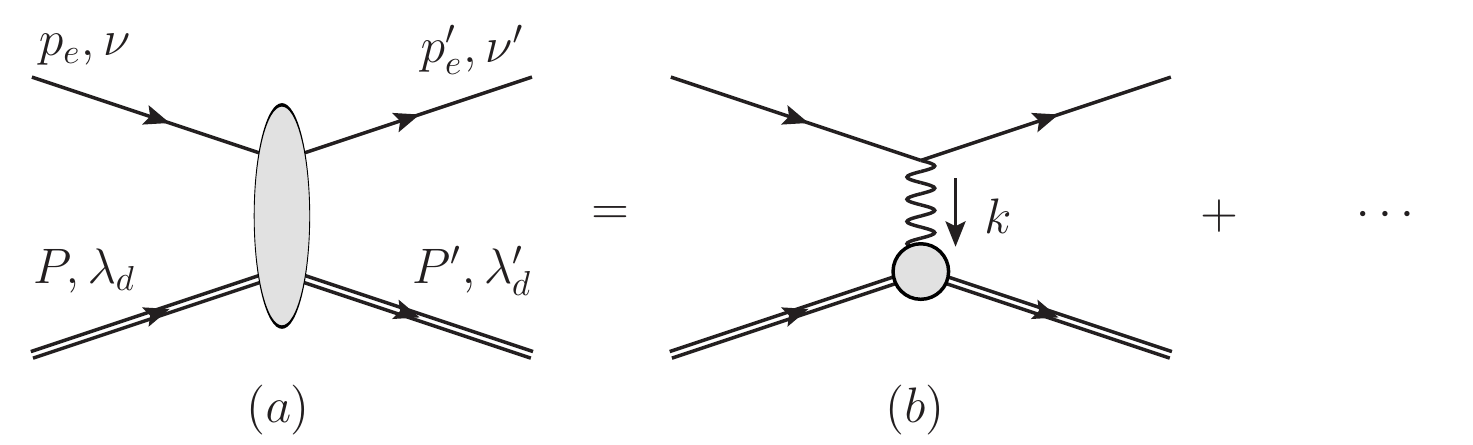}
\caption{\label{fig:edscattering}
Diagrams representing elastic electron-deuteron scattering.
Diagram (a) shows a general contribution to the  elastic
electron-deuteron scattering process and the corresponding
kinematics. Diagram (b) visualizes the one-photon-exchange
contribution, while the ellipses refer to  multi-photon-exchange
processes suppressed by powers of the fine structure constant.
Single, double and wiggly lines correspond to electrons, deuterons, and photons respectively.
}
\end{figure}

The kinematics of elastic electron-deuteron scattering is visualized
in Fig.~\ref{fig:edscattering} (a) and can be defined as
\begin{eqnarray}
  d(P, \lambda_d) + e^{-}(p_e, \nu) \; \to \;  d(P', \lambda_{d}') + e^{-}(p_{e}', \nu'),
\end{eqnarray}
where variables in brackets denote the momentum and spin projection of the corresponding particle.
Throughout this work, we focus on the one-photon-exchange mechanism,
see Fig.~\ref{fig:edscattering} (b), which provides a direct relation
between the electron-deuteron scattering observables and
the deuteron form factors.
Each additional photon exchange is suppressed by one power of the
fine-structure constant.
Thus, in line with the conclusions of  Ref.~\cite{Dong:2009zzc},
these corrections will be neglected below --- see Sec.~\ref{Sec:twophoton}
for a more detailed discussion.
The one-photon-exchange amplitude of elastic electron-deuteron
scattering, see Fig.~\ref{fig:edscattering} (b),
can be factorized into leptonic and hadronic parts~\cite{Arnold:1979cg}:
\begin{eqnarray}
  \mathcal{M} = e \bar{u}(p_{e}', \nu') \gamma_\mu u(p_e, \nu) \frac{1}{k^2}
  \Braket{P', \lambda_d'| J^\mu| P, \lambda_d},
\end{eqnarray}
where $e$ is the magnitude of the electron charge,
$u$ and $\bar{u}$ are the spinors of the initial and final electrons normalized as $\bar{u}(p, \nu)u(p, \nu)=2 m_e$
with $m_e$ being the electron mass, $\gamma_\mu$ are the Dirac
matrices and $k=P'-P$ is the four-momentum of the exchanged photon.
For convenience, we define a  quantity $Q^2$, which is positive in the space-like region,  and the
corresponding dimensionless variable $\eta$ via
\begin{eqnarray}
  Q^2 := -k_\mu k^\mu = -k^2  = - {(P'-P)}^2 \ge 0,
  \qquad
  \eta := \frac{Q^2}{4 \md^2},
\end{eqnarray}
where $\md = 1.875612 942 57(57)$ GeV stands for the deuteron mass~\cite{Tanabashi:2018oca}. Using
Lorentz invariance, time-reversal invariance as well as
parity and current conservation,  the most general form of the matrix element
of the deuteron electromagnetic current  $\Braket{P', \lambda_d'| J^\mu| P, \lambda_d}$
can be expressed as~\cite{Garcon:2001sz,Arnold:1980zj}
\begin{eqnarray}
  \label{eq:mostgeneralgammadeutint}
  \Braket{P', \lambda_d'| J^\mu| P, \lambda_d} &=&
  - e \,
    G_1(Q^2) \left( \xi^*(P', \lambda_d') \cdot \xi(P, \lambda_d) \right)  {(P'+P)}^\mu
    \nonumber
    \\
    &&- e\, G_2(Q^2) \left( \xi^\mu(P, \lambda_d) \left( \xi^*(P', \lambda_d') \cdot k \right)
                      - \xi^{*\mu}(P', \lambda_d') \left( \xi(P, \lambda_d) \cdot k \right)  \right)
    \nonumber
    \\
    &&+ e\, G_3(Q^2) \frac{\left( \xi(P, \lambda_d) \cdot k \right) \left( \xi^*(P', \lambda_d') \cdot k \right)  {(P'+P)}^\mu}{2 \md^2}
  ,
\end{eqnarray}
where dimensionless, real, Lorentz-scalar functions $G_1(Q^2)$, $G_2(Q^2)$, and $G_3(Q^2)$
parametrize the photon-deuteron interaction, and
  the deuteron polarization four-vectors, $\xi(P, \lambda_d)$ and $\xi(P', \lambda_d')$,  satisfy the following constraints
\begin{eqnarray}
  \xi(P, \lambda_d) \cdot P = 0, \qquad \xi(P', \lambda_d') \cdot P' = 0.
  \label{eq:polvecconstr}
\end{eqnarray}

\subsection{The electromagnetic form factors of the deuteron}

In practice, instead of the scalar functions $G_i(Q^2)$ from
Eq.~\eqref{eq:mostgeneralgammadeutint},
one usually introduces the deuteron charge, magnetic and quadrupole form
factors $\GC(Q^2) $, $\GM(Q^2)$ and $\GQ(Q^2)$, respectively, which are related to $G_i(Q^2)$ via
 the following equations:
 \begin{eqnarray}
   \label{FFdef}
  \GC(Q^2) &=& G_1(Q^2) + \frac{2}{3} \eta \, \GQ(Q^2),
  \nn
  \GM(Q^2) &=& G_2(Q^2),
  \nn
  \GQ(Q^2) &=& G_1(Q^2) - G_2(Q^2) + \left( 1 + \eta \right)  G_3(Q^2).
\end{eqnarray}
At $Q^2=0$, these form factors are normalized according to~\cite{Garcon:2001sz}
\begin{eqnarray}
  \GC(0) = 1,
  \qquad
  \GM(0) = \frac{\md}{m_p} \mu_d
          \simeq 1.714,
  \qquad
  \GQ(0) = \md^2 Q_d
          \simeq 25.83,
  \label{eq:GCatzero}
\end{eqnarray}
where $\GC(0) = 1$ corresponds to the electric charge conservation,
$Q_d = (0.2859 \pm 0.0003) \, \text{fm}^2$~\cite{Ericson:1982ei,Bishop:1979zz}
is the deuteron quadrupole moment,
$\mu_d = 0.8574382311(48)$~\cite{Mohr:2015ccw} is the deuteron magnetic moment in the units of nuclear magnetons,
and $m_p$ stands for the proton mass.
The derivative of
$\GC(Q^2)$  with respect to $Q^2$ taken at  $Q^2=0$ is related to the deuteron charge radius, as discussed in
Section~\ref{sec:analyticresults}.

\subsection{From observables to form factors}

Using the one-photon exchange approximation,
the unpolarized elastic  electron-deuteron differential cross section in the laboratory frame reads
\begin{eqnarray}\label{Eq:XS}
  \frac{d \sigma}{d \Omega}(Q^2, \theta)
  =
  {\left. \frac{d \sigma}{d \Omega} \right|}_\text{NS}
  \left[ A(Q^2) + B(Q^2) \tan^2 (\theta/2) \right],
\end{eqnarray}
where a no-structure pointlike cross section, $ {\left. \frac{d \sigma}{d \Omega} \right|}_\text{NS}$,
is defined as the product of the Mott differential cross section, $\sigma_{\rm Mott}$, multiplied with the recoil factor
\begin{eqnarray}\nonumber
  {\left. \frac{d \sigma}{d \Omega} \right|}_\text{NS}
  =  \sigma_{\rm Mott}
  \frac{1}{\left( 1+ \frac{2E}{\md} \sin^2 (\theta/2) \right) }, \hspace{1cm}
\sigma_{\rm Mott}
  = {\left( \frac{\alpha}{2E} \right)}^2
  \frac{\cos^2(\theta/2)}{\sin^4(\theta/2)}.
\end{eqnarray}
Here $E$  is the energy of the incoming electron, $\theta$ is the scattering
angle of the electron in the laboratory frame and
$\alpha$ is the fine-structure constant.
The elastic structure functions $A$ and $B$ are related to the
deuteron form factors given in Eq.~\eqref{FFdef} via
\begin{eqnarray}
  A(Q^2) &=& \GC^2(Q^2) + \frac{2}{3} \eta \GM^2(Q^2) +  \frac{8}{9} \eta^2 \GQ^2(Q^2),
  \nn
  B(Q^2) &=& \frac{4}{3} \eta \left(  1 +  \eta \right) \GM^2(Q^2).
\end{eqnarray}
While the unpolarized electron-deuteron  scattering cross section in Eq.~\eqref{Eq:XS} provides   access to the  magnetic FF via its relation to the  structure function $B(Q^2)$,
it does not allow one to extract the charge and quadrupole FFs  individually as they contribute to $A(Q^2)$ in a linear combination.
A complementary information on  these form factors  can be extracted from polarization data. In particular, the experimentally measurable
tensor analyzing power $T_{20}(Q^2, \theta)$ gives additional relation:
 \begin{eqnarray}
-\sqrt{2} \left[ A(Q^2) + B(Q^2) \tan^2 (\theta/2) \right] T_{20}(Q^2, \theta)&=&\nn
 \frac{8}{3} \eta \GC(Q^2) \GQ(Q^2)  +  \frac{8}{9} \eta^2 \GQ^2(Q^2) &+& \frac{1}{3} \eta \left(1+2 (1+\eta)\tan^2 (\theta/2) \right) \GM^2(Q^2).
\end{eqnarray}
Therefore, all three deuteron FFs can be extracted individually from a combined analysis of the structure functions $A(Q^2)$ and $B(Q^2)$ together with the polarization observable $T_{20}$.

\subsection{Experimental data base}\label{sec:exper}

In Ref.~\cite{Abbott:2000ak},  a  rigorous extraction of the charge, quadrupole and magnetic deuteron form factors from
the available world data for elastic electron-deuteron scattering was
performed  in the 4-momentum transfer range of $Q = 0-7$~fm$^{-1}$.  This analysis also includes polarization data of Ref.~\cite{Abbott:2000fg} from JLab. 
In addition, there is one more recent  measurement
of tensor polarization observables in elastic electron-deuteron scattering from Novosibirsk~\cite{Nikolenko:2003zq}.
Therefore, in what follows, we employ the world data for the deuteron form factors
extracted in Refs.~\cite{Abbott:2000ak,Nikolenko:2003zq}
as experimental input except for the data point for $\GQ$ at
$Q=2.788$ fm$^{-1}$ given in Table 1  of Ref.~\cite{Abbott:2000ak},
for which we believe the uncertainty have been misprinted. Indeed,
unlike the  data point at $Q=2.788$ fm$^{-1}$ shown in Fig. 1 in Ref.~\cite{Abbott:2000ak}
(see the square with the  strongly asymmetric uncertainty),
 the uncertainty quoted in Table 1 is symmetric and an order of
 magnitude  smaller than the one shown in the plot.  The error for
 $\GQ$ at this energy is also significantly smaller than those for the other energies within the same experiment.

In a recent review article~\cite{Marcucci:2015rca}, a parametrization of the world data on the deuteron form factors was
provided that has much smaller uncertainties than in the previous extractions.
While we do not use this parametrization in our fits, we will use it for the sake of comparison.

\subsection{A comment on the two-photon exchange corrections}\label{Sec:twophoton}

Unlike the extensive investigations of the two-photon exchange (TPE)
contributions to electron-proton scattering, there are very few works
focusing on the study of the TPE corrections for the deuteron electromagnetic FFs.
Specifically, in Ref.~\cite{Dong:2009zzc} a gauge invariant set of
diagrams for the TPE corrections to electron-deuteron scattering was
identified and estimated under certain assumptions for
the photon momentum in the loops.  As a result, the effect of  the TPE
on the charge and quadrupole form factors was found to be very small (less than $1\%$).
Meanwhile, in their previous investigation~\cite{Dong:2009gp}, the
authors found an order of magnitude larger effect  from TPE on the
deuteron FFs when only one subset of diagrams
was included. A significant suppression of  the TPE corrections in
Ref.~\cite{Dong:2009zzc} is therefore presumably related to the
restoration of  gauge invariance once the complete set of diagrams is
included.
The enhanced role of TPE effects was also claimed in
Ref.~\cite{Kobushkin:2009pc}, which might again be related
to the incomplete set of diagrams considered in that work.
In the current study we, therefore, rely on the conclusions of
Ref.~\cite{Dong:2009zzc} and neglect the TPE
contributions. It would be interesting to have a fresh look at this in
future studies.

\subsection{Deuteron form factors in the Breit frame}

Deuteron form factors are Lorentz-scalars and can be calculated in any frame,
but for practical calculations it is convenient to choose the Breit frame.
In the Breit frame, the kinematic variables take the simple form
\begin{eqnarray}\label{eq:kinem_breit}
  k = (0, \bm{k}),
  \qquad
  P = \left( P_0, -\frac{\bm{k}}{2} \right) ,
  \qquad
  P' = \left( P_0, +\frac{\bm{k}}{2} \right) ,
  \qquad
  P_0 = \sqrt{\md^2 + \frac{\bm{k}^2}{4}} = \md \sqrt{1 + \eta},
  \qquad
  \bm{k}^2 = Q^2,
\end{eqnarray}
where the direction of the photon momentum $\bm{k}$ is chosen along
the positive $z$ axis.
The polarization vectors of the incoming and outgoing deuterons
in the Breit frame
can be derived by boosting the corresponding rest-frame polarization vectors.
For the \emph{incoming} deuteron, one obtains
\begin{eqnarray}
  \label{eq:DeutPolVecInitial}
  \xi^\mu(P, \pm 1) = \left( 0, \frac{\mp 1}{\sqrt{2}}, \frac{-i}{\sqrt{2}}, 0 \right),
  \qquad
  \xi^\mu(P, 0) = \left(-\sqrt{\eta},0,0,\sqrt{1+\eta}\right),
\end{eqnarray}
where the second argument of $\xi^\mu$ denotes the spin projection of the
deuteron onto the $z$-axis.
Similarly, the polarization vector of the \emph{outgoing} deuteron in the Breit frame reads
\begin{eqnarray}
  \label{eq:DeutPolVecFinal}
  \xi^{ * \mu}(P', \pm 1) = \left( 0, \frac{\mp 1}{\sqrt{2}}, \frac{+i}{\sqrt{2}}, 0 \right),
  \qquad
  \xi^{ * \mu}(P', 0) = \left(
  \sqrt{\eta},0,0,\sqrt{1+\eta}
  \right) ,
\end{eqnarray}
where the sign of the zeroth component of  the polarization vector is
opposite from that of the incoming deuteron.
As expected, these definitions of $\xi$ explicitly satisfy the constraints in Eq.~\eqref{eq:polvecconstr}.

To calculate the deuteron FFs, we express them in terms of the matrix elements
$\Braket{P', \lambda_d'| J^\mu| P, \lambda_d}$ defined in
Eq.~\eqref{eq:mostgeneralgammadeutint}.
First, we simplify Eq.~\eqref{eq:mostgeneralgammadeutint}
using the relations
\begin{eqnarray}
  \label{eq:simplifyxi}
  \xi^*(P', \lambda_d') \cdot \xi(P, \lambda_d)  &=& (-1) (\delta_{ \lambda_d',  \lambda_d}
  + 2 \eta \,  \delta_{ \lambda_d', 0} \delta_{ \lambda_d, 0})\,,
  \nn
  \xi(P, \lambda_d) \cdot k  &=& (-2 \md) \sqrt{\eta} \sqrt{1+\eta}\, \delta_{ \lambda_d, 0}\,,
  \nn
  \xi^*(P', \lambda_d') \cdot k  &=& (-2 \md) \sqrt{\eta} \sqrt{1+\eta}\, \delta_{ \lambda_d', 0}\,,
\end{eqnarray}
which can be derived using the explicit form of the deuteron
polarization
vectors
in the Breit frame given in Eqs.~\eqref{eq:DeutPolVecInitial} and~\eqref{eq:DeutPolVecFinal}.
Simplifying the zeroth and three-vector components in Eq.~\eqref{eq:mostgeneralgammadeutint}
one obtains
\begin{eqnarray}
  \Braket{P', \lambda_d'| J^0| P, \lambda_d}
  &=& 2 P_0 \left\{
  G_1(Q^2) \delta_{\lambda_d, \lambda_d'}
  + 2 \eta \delta_{\lambda_d, 0} \delta_{\lambda_d', 0}
  \left(
  G_1(Q^2) - G_2(Q^2) + (1+\eta) G_3(Q^2)
  \right)
  \right\} \,,
  \nonumber
  \\
  \Braket{P', \lambda_d'| J^i| P, \lambda_d}
  &=& 2 P_0 \sqrt{\eta} \, G_2(Q^2)
  \left(
  \xi^i (P, \lambda_d) \delta_{\lambda_d', 0} -\xi^{* i } (P', \lambda_d') \delta_{\lambda_d, 0}
   \right)\,.
\end{eqnarray}
Using Eqs.~(\ref{FFdef}), (\ref{eq:DeutPolVecInitial}) and
(\ref{eq:DeutPolVecFinal}), we finally obtain
\begin{eqnarray}
  \GC(Q^2) &=& \frac{1}{3 e } \frac{1}{2P_0} \left( \braket{P',1|J_B^0|P,1} + \braket{P',0|J_B^0|P,0} + \braket{P', {-1}|J_B^0|P,{-1}} \right),
  \label{eq:defgc}
  \\
  \GQ(Q^2) &=& \frac{1}{2  e  \eta} \frac{1}{2P_0}\left( \braket{P', 0|J_B^0|P,0} - \braket{P',1|J_B^0|P,1}\right),
  \label{eq:defgq}
  \\
  \GM(Q^2) &=& \frac{1}{\sqrt{\eta} e } \frac{1}{2P_0} \Braket{P',1|\frac{J_B^{x}+i J_B^{y}}{\sqrt{2}}|P,0},
  \label{eq:defgm}
\end{eqnarray}
where $J_B^\mu  = (J_B^0,J_B^x,J_B^y,J_B^z)$ are contravariant
components of the four-vector current in the Breit frame.

\subsection{Matrix elements of the electromagnetic current }

\begin{figure}
\includegraphics[width=0.85\textwidth]{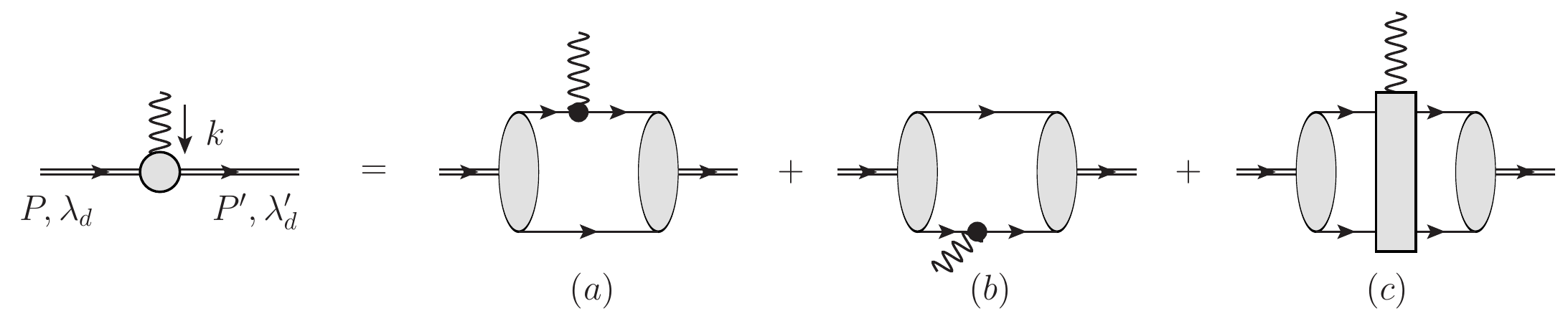}
\caption{\label{fig:convolution}
The matrix element $\Braket{P', \lambda_d'| J^\mu| P, \lambda_d}$
written as a sum of single-nucleon contributions (a) and (b)
and the two-nucleon contribution (c).
Single, double and wiggly lines refer to nucleons, deuteron particles
and photons, respectively. Black dots and the gray rectangle denote
the full photon-nucleon interaction vertex and the two-nucleon current
operator.
}
\end{figure}

In the Breit frame, the deuteron
form factors are expressed in terms of the matrix elements of the electromagnetic current  convolved with the deuteron wave functions,
$\Braket{P', \lambda_d'| J^\mu| P, \lambda_d}$,
according to Eqs.~\eqref{eq:defgc}-\eqref{eq:defgm}. The matrix elements read
\begin{eqnarray}
 \frac{1}{2P_0}\Braket{P', \lambda_d' | J_B^\mu| P, \lambda_d }  =
 \! \! \int \frac{d^3 l_1}{{(2\pi)}^3}  \frac{d^3 l_2}{{(2\pi)}^3}
  \, \psi^\dagger_{\lambda_d'}\left(\bm{l}_2 +\frac{\bm{k}}{4}, \bm{v}_B\right )
  \,
   J^\mu_B \,
  \psi_{\lambda_d}\left(\bm{l}_1 -\frac{\bm{k}}{4}, -\bm{v}_B\right ),
\label{Eq:ME}
\end{eqnarray}
where $J^\mu_B$ is the four-vector current calculated in the Breit frame,
$\psi_{\lambda}$ is the deuteron wave function with the polarization $\lambda$
and the  deuteron in the final (initial) state moves with the
velocity $\bm{v}_B $ ($-\bm{v}_B $) with  $\bm{v}_B = \bm{k}/(2\sqrt{\bm{k}^2/4+\md^2}) =   \hat{\bm{k}} \sqrt{\eta/(1+\eta)}$
and the momenta are defined in Eq.~\eqref{eq:kinem_breit}.
This matrix element is visualized in Fig.~\ref{fig:convolution}, where diagrams (a) and (b)
involve the single-nucleon electromagnetic current
while diagram (c)  corresponds to the matrix element of the two-nucleon current.

In this paper, we calculate the deuteron FFs in the framework of chiral EFT utilizing an expansion
around the non-relativistic limit\footnote{See Refs.~\cite{Arnold:1979cg,Marcucci:2015rca,Gross:2019thk} for related
studies using manifestly covariant approaches.}
and taking into account relativistic corrections as required by power counting.
Specifically, we start with the expressions for the single- and two-nucleon   charge density operators, whose chiral expansion will be summarized in
the next section.
Using the deuteron wave functions at the corresponding order in the
chiral expansion and employing consistently regularized expressions for the charge
density operators in the partial wave basis, we calculate numerically
the corresponding convolution integrals.

\section{Chiral expansion of the charge density operator}\label{sec:charge}

The nuclear electromagnetic charge and current operators have been
recently worked out to N$^3$LO in chiral EFT
by our group using the method of unitary transformation
~\cite{Kolling:2009iq,Kolling:2011mt,Krebs:2019aka} and
by the JLab-Pisa group employing time-ordered perturbation theory
~\cite{Pastore:2008ui,Pastore:2009is,Pastore:2011ip}, see also
Ref.~\cite{Park:1995pn} for a pioneering study along this line.
Following our works on the derivation of the electromagnetic currents
~\cite{Kolling:2009iq,Kolling:2011mt,Krebs:2019aka} and nuclear forces
~\cite{Epelbaum:2014sza,Reinert:2017usi,Epelbaum:2014efa,Bernard:2007sp,Bernard:2011zr,Krebs:2012yv,Krebs:2013kha,Epelbaum:2014sea}, in this study
we employ the Weinberg power counting for the operators constructed in
chiral  EFT. The hierarchy of the operators is  based on  the
expansion parameter $q \in \{ p/ \Lambda_b, \; M_\pi /\Lambda_b \} $ with  $p$ being a typical soft scale and
$\Lambda_b^2 \sim m_N M_{\pi}$ (with $M_{\pi}$ for the pion mass) referring to
 the breakdown scale of the chiral expansion.
This implies that the contributions to the charge and current operators
appear at orders $q^{-3}$ (LO),  $q^{-1}$ (NLO),  $q^{0}$ (N$^2$LO),
$q^{1}$ (N$^3$LO) and $q^{2}$ (N$^4$LO). Notice that the JLab-Pisa
group employed the counting scheme with $\mN
\sim \Lambda_b$ used in the single-nucleon sector, so that their NLO
corrections appear already at order $q^{-2}$.   We further emphasize
that
the expressions for the two-nucleon charge and current densities in
Refs.~\cite{Kolling:2009iq,Kolling:2011mt,Krebs:2019aka} and
~\cite{Pastore:2008ui,Pastore:2009is,Pastore:2011ip} do not completely
agree with each other. The differences are, however, irrelevant for
the calculation of the deuteron charge and quadrupole form factors.
For a comprehensive review of the electroweak currents and a detailed
comparison between the two sets of calculations see
Ref.~\cite{Krebs:2020pii}.

\subsection{Single-nucleon contributions to the charge density operator}

At the chiral order we are working, the single-nucleon
contributions to the charge density operator in the kinematics $N(p) + \gamma(k) \to N(p^\prime)$
take a well-known form (see Refs.~\cite{Friar:1997js,Krebs:2019aka} and references therein)
\begin{eqnarray}
  \label{eq:rho1n}
  \rho_\text{1N} = e \left( 1 - \frac{{\bm{k}}^2}{8 \mN^2} \right) \GE({\bm{k}}^2)
  + i e \frac{2 \GMNucl({\bm{k}}^2) - \GE({\bm{k}}^2)}{4 \mN^2} (\bm{\sigma} \cdot \bm{k} \times \bm{p}).
\end{eqnarray}
Here, $\GE({\bm{k}}^2)$ and $\GMNucl({\bm{k}}^2)$ are the electric and
magnetic form factors of the nucleon respectively,
and $e$ is the absolute value of the electron charge.
The single-nucleon form factors can be written
in terms of the isospin projectors and the corresponding form factors
of the proton and neutron
\begin{eqnarray}
  \GE({\bm{k}}^2) = \GE^p({\bm{k}}^2) \frac{1 + \tau_3}{2} + \GE^n({\bm{k}}^2) \frac{1 - \tau_3}{2}, \nn
  \GMNucl({\bm{k}}^2) = \GMNucl^p({\bm{k}}^2) \frac{1 + \tau_3}{2} + \GMNucl^n({\bm{k}}^2) \frac{1 - \tau_3}{2}.
\end{eqnarray}
For convenience, we also introduce the isoscalar nucleon form factors which are relevant for  electron-deuteron scattering
\begin{eqnarray}
  \GE^S({\bm{k}}^2) := \GE^p({\bm{k}}^2) + \GE^n({\bm{k}}^2),
  \qquad
  \GMNucl^S({\bm{k}}^2) := \GMNucl^p({\bm{k}}^2) + \GMNucl^n({\bm{k}}^2).
  \label{eq:isoscalarFFs}
\end{eqnarray}
In order to facilitate the comparison with phenomenological studies,
it is also convenient to decompose the single nucleon charge density
from Eq.~\eqref{eq:rho1n} into
\begin{eqnarray}
	\label{eq:rho_decomp}
 	\rho_\text{1N} = \rho^{\rm Main}_\text{1N} +  \rho_\text{1N}^\text{DF}+ \rho_\text{1N}^\text{SO},
\end{eqnarray}
 with
\begin{eqnarray}
	  \label{eq:rho1Nmain}
  \rho_\text{1N}^\text{Main} = e \GE({\bm{k}}^2), \hspace{1.cm}
    \rho_\text{1N}^\text{DF} = e \left(  - \frac{{\bm{k}}^2}{8 \mN^2} \right) \GE({\bm{k}}^2), \hspace{1.cm}
    \rho_\text{1N}^\text{SO} =
   i e \frac{2 \GMNucl({\bm{k}}^2) - \GE({\bm{k}}^2)}{4 \mN^2} \bm{\sigma} \cdot \bm{k} \times \bm{p},
\end{eqnarray}
where, apart from the main contribution, DF and SO stand for the Darwin-Foldy and spin-orbit contributions, respectively.
Terms involving order-$\mathcal{O}(\mN^{-4})$ corrections to the
charge density are beyond the accuracy of our study.

The chiral expansion of the electromagnetic FFs of the nucleon is well known to
converge slowly as they turn out to be dominated by contributions
of vector mesons~\cite{Kubis:2000zd,Schindler:2005ke}, which are not included as explicit degrees of
freedom in chiral EFT. To minimize the impact of the slow convergence
of the EFT expansion of the nucleon FFs on two-nucleon observables,
the following two approaches can be employed:
\begin{itemize}
	\item
		Instead of looking at the individual FFs of the deuteron  $\GC$ and $\GQ$, one
		calculates the ratio $\GC/\GQ$ as done e.g.~in Refs.~\cite{Phillips:2003jz,Phillips:2006im}.
		This is advantageous if one can neglect the contribution of the magnetic form factor $\GMNucl({\bm{k}}^2)$ in Eq.~\eqref{eq:rho1n}.
		However, in addition to this, one also needs  to assume either that
		the contributions from two-nucleon charge densities can be neglected
		altogether or that they scale with $\GE({\bm{k}}^2)$ in the same way as the one-body densities.
		Then, the quantity $\GE({\bm{k}}^2)$  drops out in the ratio $\GC/\GQ$.
		In this study, we show  that two-nucleon charge density operators should indeed
		be proportional to $\GE({\bm{k}}^2)$, see Sec.~\ref{Sec:twoN} for details.
		We also note that due to the numerical smallness of the SO contribution, which is the only term proportional to $\GMNucl({\bm{k}}^2)$,
		considering this ratio may, in practice, indeed provide quite accurate results.
		On the other hand, formally, this approximation is not valid at the accuracy level of our analysis.

  \item
    	Instead of relying on the strict chiral expansion of the nucleon FFs
    	one can employ empirical parametrizations extracted from experimental data,
    	as done e.g.~in Ref.~\cite{Valderrama:2007ja}.
\end{itemize}
In this work, we utilize the second approach and use up-to-date parametrizations extracted
from experimental data as will be described in the next section.
The uncertainty of our results associated with the
single-nucleon FFs will be addressed in Section~\ref{subsec:Error1NFF}.

\subsection{Input for nucleon form factors}\label{Sec:NuclFF}

\begin{figure}[ht]
\includegraphics[width=0.85\textwidth]{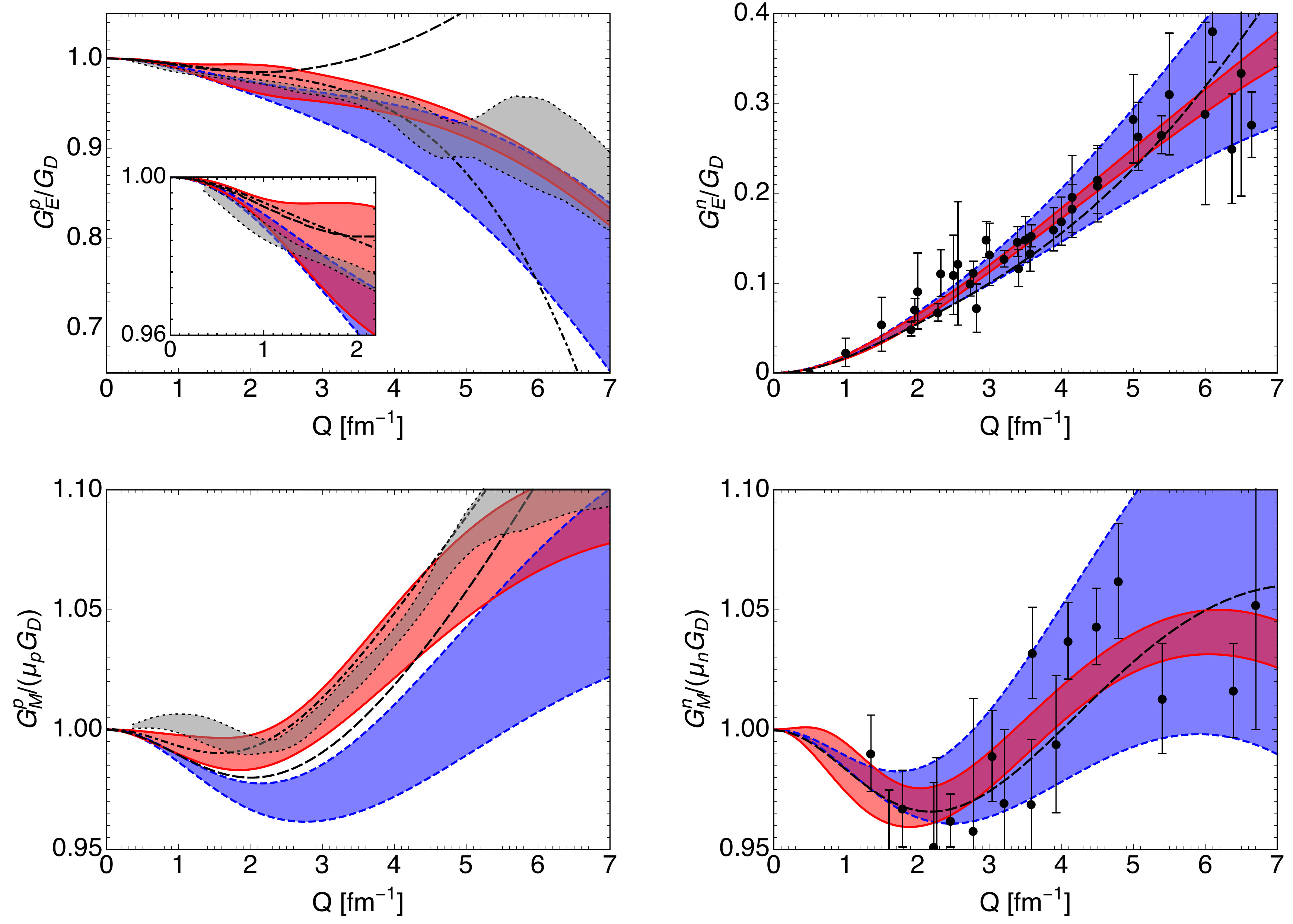}
\centering
  \caption{\label{fig:1NFF} (Color online) The proton (left panel) and neutron (right panel) form factors
normalized to the dipole form factor $G_\text{D}(Q^2) = {(1+Q^2/\Lambda_D^2)}^{-2}$ with $\Lambda_D^2 = 0.71 \text{ GeV}^2$.
The gray bands between the dotted lines correspond to the proton form factors extracted in Ref.~\cite{Bernauer:2013tpr} from a combined fit to  all data including polarized ratio measurements. The uncertainty corresponds to the
combined statistical and systematic uncertainties taken in quadrature, among which the  sensitivity to the functional form of the spline used in the fits is the largest. The red bands between the solid lines represent the results of the global analysis
of the world proton observables  and the neutron FFs  from  Ref.~\cite{Ye:smallrp}, see also  Ref.~\cite{Ye:2017gyb} for the published version and  text for the details.
The blue bands between the dashed lines  show the results of the SC approach of Ref.~\cite{Belushkin:2006qa}
from a simultaneous dispersive analysis
of    all four FFs (data are also shown as dots --- see Refs.~\cite{Punjabi:2015bba,Ye:2017gyb,Belushkin:2006qa} for more details) in both the space-like and time-like regions.
The dashed lines show an update of the analysis of Ref.~\cite{Belushkin:2006qa} which is based on the
fit to the MAMI data for electron-proton scattering and simultaneously to the world data for the neutron form factors~\cite{Lorenz:2012tm}.
The dot-dashed lines represent the results for the proton FFs extracted using  the dispersive approach  of Ref.~\cite{Lorenz:2014yda} from a global  analysis of the world data for electron-proton scattering. No errors for the nucleon FFs were given
 in Refs.~\cite{Lorenz:2012tm} and~\cite{Lorenz:2014yda}.
}
\end{figure}

\begin{figure}
\includegraphics[width=0.75\textwidth]{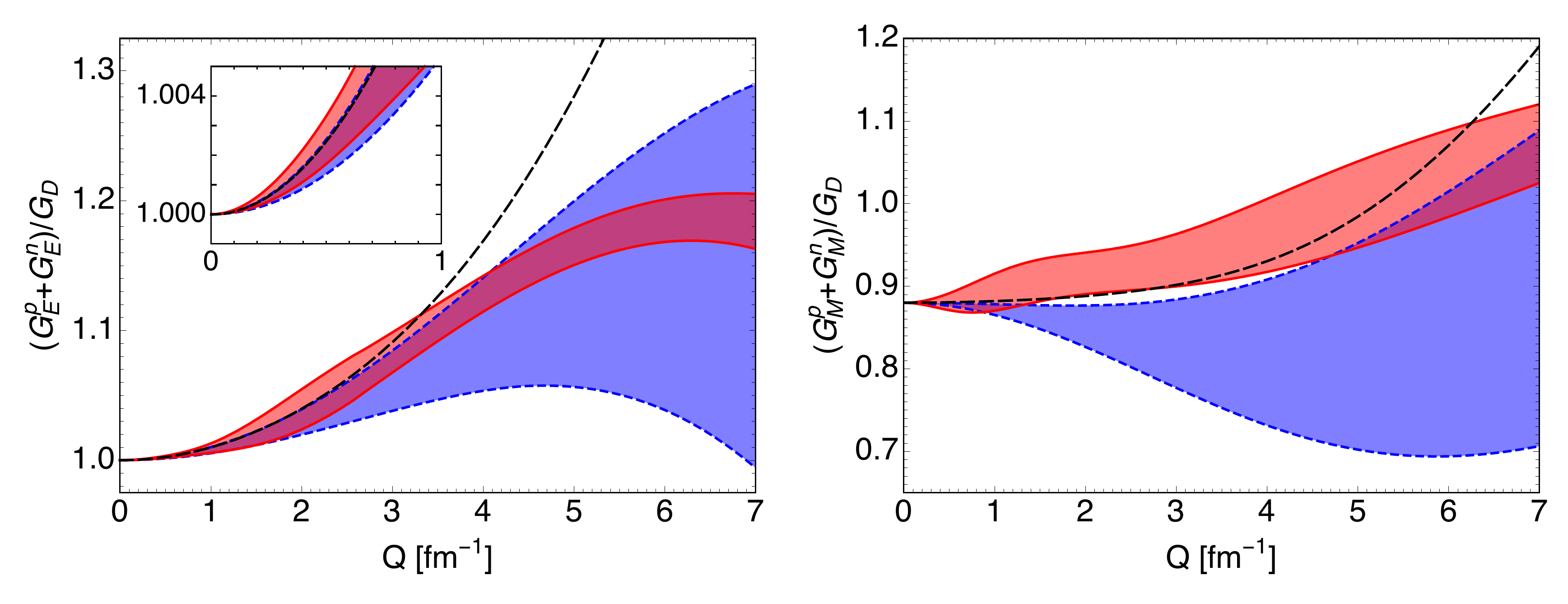}
\caption{\label{fig:1NFFscal} (Color online) Isoscalar nucleon electric (left panel) and magnetic (right panel)
  form factors normalized to the dipole form factor $G_\text{D}(Q^2)$.
For remaining notation see Fig.~\ref{fig:1NFF}.
}
\end{figure}

The electromagnetic form factors of the proton and neutron probe the
charge and magnetization distributions of the nucleons  via the
interaction of electromagnetic currents
and  have been investigated experimentally for more than 70 years  using electron scattering --- see e.g. Refs.~\cite{Punjabi:2015bba,Pacetti:2015iqa, Drechsel:2007sq,Perdrisat:2006hj, Arrington:2006zm} for selected review articles.

The most recent extraction of the proton form factors was carried out in Refs.~\cite{Ye:2017gyb,Ye:smallrp},
where a  global analysis of all existing data was done including the corrections for different normalization of various data,
and effects from TPE.
The results of Ref.~\cite{Ye:smallrp} are shown in Fig.~\ref{fig:1NFF} (left panel) by red bands confined by solid lines.
These fits were constrained at low $Q^2$ by the latest CODATA-2018 values for the proton charge radii%
\footnote{
	The difference between the nucleon form factor parametrizations presented
	in the original work of Ref.~\cite{Ye:2017gyb} and its update Ref.~\cite{Ye:smallrp}
	lies in the value for the proton charge radius used as input.
	Ref.~\cite{Ye:smallrp} employs the more recent (CODATA-2018) value consistent
	with the measurements from muonic hydrogen Lamb shift~\cite{Pohl:2010zza}
	as well as with the latest atomic hydrogen measurements of the Rydberg constant~\cite{Beyer:2017gug}
	and the Lamb shift~\cite{Bezginov:2019mdi}, while  Ref.~\cite{Ye:2017gyb} relies on the
	larger value for the proton charge radii taken from CODATA-2014~\cite{Mohr:2015ccw}.
	The effect of the proton charge radius on the shape of the proton FFs is relevant only at very low $Q$ (lower than $1$~fm$^{-1}$).
	At larger $Q$, the shape of the proton form factor is strongly constrained by other experimental data.
}~\cite{CODATA2018} and by the magnetic radii from the Particle Data Group (PDG)~\cite{Tanabashi:2018oca} while  at high $Q^2$  a  power-law falloff was enforced.
Another global analysis of the proton data was carried out by the A1 collaboration in Refs.~\cite{Bernauer:2010wm,Bernauer:2013tpr},
where specific functional form for the form factors was assumed to fit the world data and no constraints
on the proton radii were imposed.
Apart from some differences%
\footnote{
	The difference  in $\GMNucl^p$  might be at least partly  related to the
	fact that the world average value for the magnetic radii of the proton~\cite{Tanabashi:2018oca} used as input
	in Ref.~\cite{Ye:2017gyb}  has some tension with the value extracted by the A1 collaboration in
	Ref.~\cite{Bernauer:2010wm}.
	}
in $\GMNucl^p$ at low $Q^2$ and very large differences in the estimated uncertainties,
the extracted electric and magnetic form factors of the proton in
Refs.~\cite{Ye:2017gyb,Ye:smallrp} and~\cite{Bernauer:2013tpr}  are
essentially consistent with each other, cf.~red and gray bands in
Fig.~\ref{fig:1NFF}.

A determination of the neutron form factors is much more complicated than for the proton,
since there are no free-neutron targets and it is, therefore,
necessary to analyze experimental data on nuclear targets like $^2$H or $^3$He.
A reliable extraction of the neutron form factors  from such data requires a detailed
 understanding of the nuclear corrections (involving
nuclear wave functions,  final state interaction,  meson exchange currents etc.).
The results of the most up-to-date parametrization of  the neutron
FFs carried out in Ref.~\cite{Ye:smallrp}  are presented in
Fig.~\ref{fig:1NFF}  (see red  bands between solid lines in the right
panel).

Already in Refs.~\cite{Hohler:1976ax, Mergell:1995bf},
it was pointed out  that   analyticity and unitarity
put strong constraints on the nucleon FFs.
Using the spectral-function-based  dispersive approach,
the nucleon FFs  were obtained in Ref.~\cite{Belushkin:2006qa} from
a simultaneous fit to the  data for all four FFs in both space-like and time-like regions including
the constraints from meson-nucleon scattering data, unitarity, and perturbative QCD.
The results of this analysis for the so-called ``superconvergence approach'' (SC)
are shown as blue bands  confined by the dashed lines in Fig.~\ref{fig:1NFF}.
An update of the analysis of Ref.~\cite{Belushkin:2006qa}  based on the
fit to the most recent MAMI data for electron-proton scattering and
simultaneously to the world data for the neutron form factors was made in
Ref.~\cite{Lorenz:2012tm} and shown in Fig.~\ref{fig:1NFF} by black long-dashed lines.
Another strategy was used in the latest dispersive analysis of Ref.~\cite{Lorenz:2014yda}.
First, the world experimental data on electron proton scattering were corrected in Ref.~\cite{Lorenz:2014yda}  for the TPE contributions,
which were calculated including the nucleon and $\Delta(1232)$-resonance intermediate states.  Then, the corrected data were fitted
using the proton FFs  evaluated in the dispersive approach. No updates of the neutron FFs were made.
The comparison of the results of the dispersive approach with those from the analysis in Refs.~\cite{Ye:smallrp,Ye:2017gyb} reveals that
the   electric and magnetic proton FFs from Ref.~\cite{Lorenz:2014yda}   are compatible with  the band from Ref.~\cite{Ye:smallrp}
at  small and intermediate $Q$, although  they visibly deviate from each other at  $Q$  larger than $4$~fm$^{-1}$ (cf.~dot-dashed curve with the red band).
A closer look at the small momentum range, which is particularly
sensitive to the proton charge radius,  shows a  very good agreement
between the results of these analyses,
see the zoomed plot for $\GE^p$ in  Fig.~\ref{fig:1NFF}.  This is not surprising given that the value for the proton charge radius predicted in
Ref.~\cite{Lorenz:2014yda}     is  consistent  with the latest CODATA-2018 update employed in Ref.~\cite{Ye:smallrp} as input, 
see also Ref.~\cite{Hammer:2019uab} for a mini-review on the status of the proton radius puzzle.

Last but not least, the lattice QCD simulations for the nucleon FFs are already
approaching the accuracy compatible with the experimental precision.  For example,
in Ref.~\cite{Alexandrou:2017ypw}, the electromagnetic FFs of the
nucleon are computed including both the connected and disconnected
contributions for the pion masses  basically at the physical point.
The resulting isoscalar and isovector nucleon FFs were found to
overshoot the experimental data by about  one standard deviation
which, as proposed in Ref.~\cite{Alexandrou:2017ypw}, could be due to
small residual excited state contamination.  Further simulations  should help in resolving this issue.

In this work, we will employ a set of different parametrizations
for the proton and neutron FFs  as input to calculate the deuteron FFs
and, in this way, to make a complementary nontrivial test of  our understanding of the nucleon FFs.
Indeed, since our current study is aimed at a high-accuracy
systematic investigation of the nuclear effects up to N$^4$LO in
chiral EFT, the comparison of  the calculated deuteron form factors
with data should provide useful insights into the consistency of the
single nucleon input with the elastic scattering data on the deuteron.
Since the  up-to-date dispersive results from Refs.~\cite{Lorenz:2012tm, Lorenz:2014yda} are given without uncertainties,
we will use the results of Ref.~\cite{Ye:2017gyb,Ye:smallrp} as our central input, while   the FFs from Ref.~\cite{Belushkin:2006qa}
will be employed as a consistency check.

Finally,  since the deuteron FFs involve only the isoscalar
combinations of the nucleon FFs, see Eq.~\eqref{eq:isoscalarFFs},
we plot these combinations  in Fig.~\ref{fig:1NFFscal}.
Notice further that the charge and quadrupole FFs of the deuteron  are
sensitive predominantly to the  isoscalar electric FF of the nucleon,
while the isoscalar magnetic FF  contributes only through the
numerically small spin-orbit correction.
For the isoscalar electric form factor, the dispersive results
of Refs.~\cite{Belushkin:2006qa,Lorenz:2012tm} are essentially consistent
with each other as well as with  those from the analysis~\cite{Ye:smallrp}
at least for $Q\lesssim 3.5$ fm$^{-1}$.

\subsection{Two-nucleon contributions to the charge density operator}\label{Sec:twoN}

The charge density operator is dominated by the LO single-nucleon
contribution, while the first two-nucleon (2N) terms appear
only at N$^3$LO~\cite{Kolling:2009iq,Kolling:2011mt}.
The dominant contributions to the 2N charge density operator stem
from one-loop diagrams of the one-pion exchange (OPE), two-pion exchange and
contact types, whose explicit expressions are parameter-free and can be found in
Refs.~\cite{Kolling:2009iq,Kolling:2011mt}. All these terms are of isovector type and, therefore,
do not contribute to the deuteron form factors. In addition to the
already-mentioned static (i.e.~order-${(1/m_N)}^0$) contributions, one also has to consider
tree-level one-pion exchange diagrams with a single insertion of
the kinetic energy or  $1/m_N$-corrections to the leading pion-nucleon
vertex.
In the  two-nucleon kinematics
\begin{eqnarray}\label{Eq:kinem}
  N(p_1) + N(p_2) + \gamma(k) \to N(p_1^\prime) + N(p_2^\prime)
\end{eqnarray}
with auxiliary three-momenta defined as
$\bm{q}_1 = \bm{p}_1^{\,\prime} - \bm{p}_1$
and $\bm{q}_2 = \bm{p}_2^{\,\prime} - \bm{p}_2$,
the isoscalar one-pion exchange charge density can be written as~\cite{Kolling:2011mt}
\begin{eqnarray}
  \label{eq:NNchargeDensityKolling}
  \rho_\text{2N}^{1\pi} &=&
  (1-2 \bar{\beta}_9)
  \frac{e g_A^2}{16 F_\pi^2 \mN}
  (\bm{\tau}_1 \cdot \bm{\tau}_2)
  \frac{
    (\bm{\sigma}_1 \cdot \bm{k})
    (\bm{\sigma}_2 \cdot \bm{q}_2)
    }
    {\bm{q}_2^2 + \mpi^2}
  \nn
  && + (2 \bar{\beta}_8 - 1)
  \frac{e g_A^2}{16 F_\pi^2 \mN}
  (\bm{\tau}_1 \cdot \bm{\tau}_2)
  \frac{
    (\bm{\sigma}_1 \cdot \bm{q}_2)
    (\bm{\sigma}_2 \cdot \bm{q}_2)
    (\bm{q}_2 \cdot \bm{k})
    }
    {{(\bm{q}_2^2 + \mpi^2)}^2}
   + (1 \leftrightarrow 2),
\end{eqnarray}
where the dimensionless quantities $\bar{\beta}_8$ and $\bar{\beta}_9$ parametrize
the unitary ambiguity of the long-range contributions to the nuclear forces and currents at N$^3$LO.
The explicit form of the corresponding unitary transformations is
given in Eq.~(4.23) of Ref.~\cite{Kolling:2011mt}. Further,
$\gA$ is the axial-vector coupling constant of the nucleon,
$\fpi$ is the pion decay constant
and $(1 \leftrightarrow 2)$ stands for a contribution resulting from
interchanging the nucleon labels. Notice that the OPE contribution has
also been taken into account in phenomenological studies, where it
represents a part
of the so-called meson-exchange currents,
see e.g.~\cite{Friar:1979by}.

It is important to emphasize that all terms of the OPE charge density in Eq.~\eqref{eq:NNchargeDensityKolling}
are proportional to unobservable unitary-transformation-parameters $\bar{\beta}_8$ and $\bar{\beta}_9$.
The same parameters also appear in the $1/\mN^2$- and $1/\mN$-contributions to the
two-~\cite{Epelbaum:2004fk} and three-nucleon forces at N$^3$LO~\cite{Bernard:2011zr},
see also Ref.~\cite{Friar:1999sj} for a related discussion. This unitary ambiguity reflects the
fact that nuclear forces and currents are not directly measurable and, in general,
scheme-dependent. In contrast, observable quantities such as e.g.~the form factors must, of course,
be independent of the choice of $\bar{\beta}_8$, $\bar{\beta}_9$ and
other off-shell parameters. This can only be achieved by using
\emph{off-shell consistent} expressions for the nuclear forces and
currents. In particular, to be consistent with the new semilocal
momentum-space regularized NN potentials of
Refs.~\cite{Reinert:2020mcu,Reinert:2017usi}  which we employ to
calculate the deuteron wave function (DWF) for our analysis, the
so-called minimal nonlocality choice with
\begin{equation}\label{Eq:minNON}
\bar{\beta}_8=1/4, \quad \quad \bar{\beta}_9=-1/4
  \end{equation}
has to be made.
Note that the employed calculational approach
relies on a numerically exact solution of the 2N Schr\"odinger
equation with a potential truncated at a given order. This way one
unavoidably includes certain higher-order contributions to the
scattering amplitude so that the
calculated observables are only expected to be approximately independent of
$\bar{\beta}_8$, $\bar{\beta}_9$. The residual dependence on these
parameters should be of a higher order, which provides a useful tool
to check consistency of the calculations. In Section~\ref{sec:betaindependence}, we
will demonstrate that the deuteron FFs calculated with different sets of
$\bar{\beta}_8$, $\bar{\beta}_9$ yield consistent results.

An important consequence of the unitary ambiguity associated with $\bar{\beta}_8$ and $\bar{\beta}_9$ is that
one can use unitary transformations to reshuffle the contributions to observables
between the charge density and DWF. One can even completely
eliminate the isoscalar 2N charge density operator at N$^3$LO. As will be
shown below, this also holds true for the short-range corrections
at N$^4$LO.%
\footnote{
	Notice, however, that the leading isovector contributions to the 2N charge
	density at N$^3$LO cannot be eliminated by means of unitary
	transformations~\cite{Kolling:2009iq,Hyuga:1977cj}.
	}
The expression in Eq.~(\ref{eq:NNchargeDensityKolling}) is thus to be understood as the
contribution induced by acting with the unitary operator specified
in Eq.~(4.23) of Ref.~\cite{Kolling:2011mt} on the isoscalar part of the leading
single-nucleon charge density operator $\rho_\text{1N}^\text{Main,  LO}  = e$,
where the electric nucleon FF at leading order (labeled by the superscript LO) was set to unity.
Since we do not rely on the chiral expansion of the nucleon FFs in our
analysis, it is more consistent and appropriate to define the isoscalar OPE
contribution as the one induced by $\rho_\text{1N}^\text{Main}$ rather than
$\rho_\text{1N}^\text{Main,  LO}$, which generalizes the
expression in Eq.~\eqref{eq:NNchargeDensityKolling} to
\begin{eqnarray}
  \label{eq:NNchargeDensityKollingWithGES}
  \rho_\text{2N}^{1\pi} &=&
  (1-2 \bar{\beta}_9) \GE^S(Q^2)
  \frac{e g_A^2}{16 F_\pi^2 \mN}
  (\bm{\tau}_1 \cdot \bm{\tau}_2)
  \frac{
    (\bm{\sigma}_1 \cdot \bm{k})
    (\bm{\sigma}_2 \cdot \bm{q}_2)
    }
    {\bm{q}_2^2 + \mpi^2}
  \\
  \nonumber
  && + (2 \bar{\beta}_8 - 1) \GE^S(Q^2)
  \frac{e g_A^2}{16 F_\pi^2 \mN}
  (\bm{\tau}_1 \cdot \bm{\tau}_2)
  \frac{
    (\bm{\sigma}_1 \cdot \bm{q}_2)
    (\bm{\sigma}_2 \cdot \bm{q}_2)
    (\bm{q}_2 \cdot \bm{k})
    }
    {{(\bm{q}_2^2 + \mpi^2)}^2}
   + (1 \leftrightarrow 2).
\end{eqnarray}
While this expression is equivalent to
Eq.~\eqref{eq:NNchargeDensityKolling} up to terms of a higher order,
using Eq.~(\ref{eq:NNchargeDensityKollingWithGES}) ensures
that our results for the deuteron FFs are independent
of the parameters $\bar{\beta}_8$ and $\bar{\beta}_9$ to a very high
degree, as will be explicitly demonstrated in Section~\ref{sec:betaindependence}.

Although the pionic  contributions to the isoscalar charge density at N$^4$LO have not been
worked out yet, the complete expression for the contact
operators at  N$^4$LO is derived and given in Appendix~\ref{sec:appIVCT}.
The expression for the antisymmetrized isoscalar contact operators at N$^4$LO reads:
\begin{eqnarray}
\rho_\text{Cont} &=&
2 e \left(A+B+\frac{C}{3}\right) \frac{\bm{\sigma}_1\cdot \bm{\sigma}_2 + 3}{4}
  \frac{1- \bm{\tau}_1\cdot \bm{\tau}_2 }{4} \bm{k}^2
  +
  2 e \, C
  \frac{1- \bm{\tau}_1\cdot \bm{\tau}_2 }{4}
    \left(
      (\bm{k} \cdot \bm{\sigma}_1)(\bm{k} \cdot \bm{\sigma}_2) - \frac{1}{3} \bm{k}^2 (\bm{\sigma}_1\cdot \bm{\sigma}_2)
    \right)
\nonumber
\\
	&& +
  2 e \, (A - 3 B - C) \frac{1 - \bm{\sigma}_1\cdot \bm{\sigma}_2}{4}
      \left(
      \frac{\bm{\tau}_1\cdot \bm{\tau}_2 + 3}{4}
       \right) \, \bm{k}^2,
\label{eq:contactchargedensityLO}
\end{eqnarray}
where the first (second) line in Eq.~\eqref{eq:contactchargedensityLO} contributes
to the isospin-0-to-isospin-0 (isospin-1-to-isospin-1) channel and $A$, $B$, and $C$ denote the corresponding LECs.
These LECs contribute to the deuteron FFs in two linear combinations $A+B+C/3$ and $C$.
The expression in Eq.~\eqref{eq:contactchargedensityLO}  agrees with
the isoscalar part of the result published in Ref.~\cite{Phillips:2016mov},
while the corresponding isovector terms are different, see Appendix~\ref{sec:appIVCT}.
Notice further that the contact operator relevant for the quadrupole moment of the deuteron
(the term $\sim C$ in Eq.~\eqref{eq:contactchargedensityLO}) was first derived in Ref.~\cite{Chen:1999tn}.

As already pointed out above, the short-range operators Eq.~\eqref{eq:contactchargedensityLO} can, in
principle, also be eliminated via a suitable unitary transformation at the cost of
changing the off-shell behavior of the NN potential. The corresponding
unitary transformation acting on two-nucleon states is given in  Ref.~\cite{Reinert:2017usi}  and
can be written as
\begin{eqnarray}
  U = e^{A {T}_1+B {T}_2+C {T}_3},
  \label{eq:unittransfABC}
\end{eqnarray}
where the anti-Hermitean generators read
\begin{eqnarray}
  \label{eq:unitarytransformGeneratorsNN}
  {T}_1 &=& \big( \bm{p}_1'^2 + \bm{p}_2'^2 - \bm{p}_1^2 - \bm{p}_2^2 \big),
  \nn
  {T}_2 &=& \big( \bm{p}_1'^2 + \bm{p}_2'^2 - \bm{p}_1^2 - \bm{p}_2^2 \big)
    (\bm{\sigma}_1 \cdot  \bm{\sigma}_2),
  \\
  {T}_3 &=&   \bm{\sigma}_1 \cdot (
    \bm{p}_1 - \bm{p}_2 + \bm{p}_1' - \bm{p}_2') \; \bm{\sigma}_2 \cdot (
    \bm{p}_1' - \bm{p}_2' - \bm{p}_1 + \bm{p}_2 )
+    \bm{\sigma}_1 \cdot (\bm{p}_1' - \bm{p}_2' - \bm{p}_1 + \bm{p}_2 )
    \; \bm{\sigma}_2 \cdot ( \bm{p}_1 - \bm{p}_2 + \bm{p}_1' - \bm{p}_2'
    ).
\nonumber
\end{eqnarray}
Here, $\bm{p}_i$ ($\bm{p}_i'$) denote the initial and final momenta of the nucleons.
However, in Refs.~\cite{Reinert:2017usi,Reinert:2020mcu}, the freedom to perform such
unitary transformations has already been exploited to eliminate the
redundant contact interactions contributing to the $^1$S$_0$ and
$^3$S$_1$ partial waves and the mixing angle $\epsilon_1$ at
N$^3$LO.%
\footnote{
	To eliminate the redundant terms in NN potential,
	the parameters $A$, $B$, $C$  have to be taken formally of the order
	$\mathcal{O}(m_N/\Lambda_b)$ rather than
	$\mathcal{O}({(m_N/\Lambda_b)}^0)$. This is the reason for the apparent mismatch in the chiral
	order of the off-shell contact terms in the NN potential (N$^3$LO) and the
	corresponding short-range charge density operators (N$^4$LO).
    }
Therefore, to be consistent with the choice of the off-shell behavior adopted
in the NN potentials of Ref.~\cite{Reinert:2020mcu},
the short-range contributions to the charge density in Eq.~(\ref{eq:contactchargedensityLO}) have to be
taken into account explicitly.
Here, we follow the same procedure as in the case of the OPE charge density and
employ the short-range charge density operator induced by
applying the unitary transformation in Eq.~\eqref{eq:unittransfABC}
to the charge density operator $\rho_\text{1N}^\text{Main}$ from Eq.~\eqref{eq:rho1Nmain}:
\begin{eqnarray}
  \label{eq:inducedchargedensity}
  \delta \hat \rho
  = \hat{U}^\dagger \hat \rho_\text{1N}^\text{Main} \, \hat{U} - \hat \rho_\text{1N}^\text{Main}
  \simeq \left[ \hat \rho_\text{1N}^\text{Main}, A \hat{T}_1+B \hat{T}_2+C \hat{T}_3 \right],
\end{eqnarray}
where square brackets denote a commutator
and $\hat X$ indicates that the quantity $X$ is to be
regarded as an operator rather than a matrix element with respect to momenta of the nucleons.
Evaluating the commutator in the given kinematics yields the
generalization of Eq.~(\ref{eq:contactchargedensityLO}) for the contact isoscalar charge density
\begin{eqnarray}
\rho_\text{Cont} &=&
2 e \GE^S(\bm{k}^2) \bigg[
	\left(A+B+\frac{C}{3}\right) \frac{\bm{\sigma}_1\cdot \bm{\sigma}_2 + 3}{4}
  \frac{1- \bm{\tau}_1\cdot \bm{\tau}_2 }{4} \bm{k}^2
  +
  C
  \frac{1- \bm{\tau}_1\cdot \bm{\tau}_2 }{4}
    \left(
      (\bm{k} \cdot \bm{\sigma}_1)(\bm{k} \cdot \bm{\sigma}_2) - \frac{1}{3} \bm{k}^2 (\bm{\sigma}_1\cdot \bm{\sigma}_2)
    \right)
\nonumber
\\
	&& +
   (A - 3 B - C) \frac{1 - \bm{\sigma}_1\cdot \bm{\sigma}_2}{4}
      \left(
      \frac{\bm{\tau}_1\cdot \bm{\tau}_2 + 3}{4}
       \right) \, \bm{k}^2
        \bigg],
\label{eq:contactchargedensity}
\end{eqnarray}
where the nucleon FF $\GE^S({\bm{k}}^2)$ coming from $ \rho_\text{1N}^\text{Main}$
accounts for a non-pointlike nature of the NN$\gamma$ vertex.
The linear combinations of the LECs $(A+B+C/3)$ and $C$ will be determined from the deuteron
data as discussed  in  Section~\ref{sec:Setup}. The combination $(A - 3 B - C)$
corresponds to the isospin-$1$-to-isospin-$1$ transition and should be
determined from other processes.
For the complete expression including isovector terms the reader is
referred to Appendix~\ref{sec:appIVCT}.

Finally, we emphasize that the above expressions do not provide the
complete contribution to the 2N charge density operator at N$^4$LO. It
is, however, conceivable that most (if not all) of the corrections of the one- and two-pion
exchange range, which still have to be worked out, are of isovector type and, therefore, do not
contribute to the deuteron FFs. We expect that isoscalar long-range contributions at N$^4$LO not considered in our study are, to some extent, effectively taken into account by the short-range operators for not too high values of the momentum transfer. For the
sake of brevity, we will refer to all results based on the
short-range part of the 2N charge density operator in
Eq.~(\ref{eq:contactchargedensity}) as being N$^4$LO.

\section{Regularization of the charge density operator}\label{sec:regul}

We now discuss regularization of the charge-density operators introduced in the previous section.
The single-nucleon charge density operator requires no regularization.
However, two-nucleon contributions (both OPE and contact) have to be regularized,
because of the divergent loop integrals appearing in the convolution
with deuteron wave function.
We specifically focus here on the consistency with the regularization of chiral NN potential~\cite{Reinert:2020mcu,Reinert:2017usi}.
The new generation of chiral NN potentials of Ref.~\cite{Reinert:2020mcu,Reinert:2017usi} employed in our analysis
make use of the local momentum-space regulator for pion exchange
contributions, which, by construction, maintains the long-range structure of the nuclear
force. The short-range part of the nuclear forces developed in
Refs.~\cite{Epelbaum:2014efa,Epelbaum:2014sza} is regularized with an angular-independent Gaussian
momentum-space regulator. The meaning of consistency of
the regularization procedure for nuclear forces and currents is discussed   in
Refs.~\cite{Krebs:2019uvm,Epelbaum:2019jbv}, where it is shown that the usage of dimensionally
regularized  loop contributions to the three-nucleon forces and
2N currents leads, in general, to incorrect results for
observables. In order to avoid this problem, loop contributions to the current operators  need to be rederived using a
regulator compatible with that employed in the  NN
potentials. The complications related to the loop operators are, however, irrelevant for our
analysis: thanks to the deuteron acting as an isospin filter, none of
the terms in  the 2N charge density stemming from  loop diagrams
at N$^3$LO  contribute to the deuteron FFs.
Still, it is important for our analysis to employ a proper regulator chosen in a way
compatible with the NN potentials of Refs.~\cite{Reinert:2020mcu,Reinert:2017usi}.
In particular, since the contribution of the single-nucleon charge density to
the deuteron FFs drops off rapidly with increasing values of the
momentum transfer, the calculated FFs at larger $Q$-values
become sensitive to the two-nucleon charge density operator which depends on the regulator.

We start with the OPE operators given by Eq.~\eqref{eq:NNchargeDensityKollingWithGES}.
These operators contain single and squared pion propagators.
The regularization of the contributions with the single pion propagator is defined in Ref.~\cite{Reinert:2017usi} and
can be effectively written as a substitution:
\begin{eqnarray}
  \frac{1}{\bm{p}^2+ \mpi^2}
  \to \frac{1}{\bm{p}^2+ \mpi^2} \exp \left( - \frac{\bm{p}^2+ \mpi^2}{\Lambda^2} \right),
  \label{eq:PionPropRegPresc}
\end{eqnarray}
where $\Lambda$ is a fixed cutoff chosen consistently with the employed NN potential
in the range of $400$--$550$~MeV.%
\footnote{
	In Ref.~\cite{Reinert:2017usi}, also the results for  $\Lambda = 350$~MeV are given.
	However, for such a soft cutoff one already observes a
	substantial amount of finite-regulator artifacts, and the description of NN data
	deteriorates noticeably. For this reason we do not use this cutoff
	value in our analysis.
	}

Apart from  the single pion propagator, the OPE charge density, Eq.~\eqref{eq:NNchargeDensityKollingWithGES},
also contains the pion propagator squared.
The prescription for regularizing the squared pion propagator
can be obtained from Eq.~\eqref{eq:PionPropRegPresc}
by taking a derivative with respect to $\mpi^2$, as done  in Ref.~\cite{Reinert:2017usi}, which yields
\begin{eqnarray}
  \frac{1}{{(\bm{p}^2+ \mpi^2)}^2}
  \to
  \left( \frac{1}{{(\bm{p}^2+ \mpi^2)}^2}  + \frac{1}{\Lambda^2 (\bm{p}^2+ \mpi^2)} \right)
  \exp \left( - \frac{\bm{p}^2+ \mpi^2}{\Lambda^2} \right).
  \label{eq:SquaredPionPropRegPresc}
\end{eqnarray}
Using the regularization procedure specified above, the
regularized expression for the
isoscalar part of the OPE charge density takes the form
\begin{eqnarray}
  \label{eq:NNchargeDensityKollingWithGESregularized}
  \rho_\text{2N}^{1\pi, \, \text{reg}} &=&
  (1-2 \bar{\beta}_9) \GE^S(Q^2)
  \frac{e g_A^2}{16 F_\pi^2 \mN}
  (\bm{\tau}_1 \cdot \bm{\tau}_2)
  \frac{
    (\bm{\sigma}_1 \cdot \bm{k})
    (\bm{\sigma}_2 \cdot \bm{q}_2)
    }
    {\bm{q}_2^2 + \mpi^2}
    \exp \left( - \frac{\bm{q}_2^2+ \mpi^2}{\Lambda^2} \right)
  \nonumber
  \\
  && + (2 \bar{\beta}_8 - 1) \GE^S(Q^2)
  \frac{e g_A^2}{16 F_\pi^2 \mN}
  (\bm{\tau}_1 \cdot \bm{\tau}_2)
    (\bm{\sigma}_1 \cdot \bm{q}_2)
    (\bm{\sigma}_2 \cdot \bm{q}_2)
    (\bm{q}_2 \cdot \bm{k})
  \nonumber
  \\
    && \times
    \left( \frac{1}{{(\bm{q}_2^2+ \mpi^2)}^2}  + \frac{1}{\Lambda^2 (\bm{q}_2^2+ \mpi^2)} \right)
    \exp \left( - \frac{\bm{q}_2^2+ \mpi^2}{\Lambda^2} \right)
   + (1 \leftrightarrow 2).
\end{eqnarray}

As a next step, we consider the regularization of the contact charge
density given by Eq.~\eqref{eq:contactchargedensity}.
To ensure consistency between regularizations of potential and charge
density and avoid ambiguity due to the dependence of the charge density operator
on the photon momentum, we exploit the fact that both the off-shell contact
NN potential and the short-range charge density operators can be generated by
the same unitary transformation acting on the kinetic energy
term and the single-nucleon charge density, respectively.
The contact part of the NN potential is regularized in Ref.~\cite{Reinert:2017usi} via a non-local
Gaussian cutoff
\begin{eqnarray}
  V_\text{Cont}^\text{reg} = V_\text{Cont} \;
  \exp \left( - \frac{ {(\bm{p}'_1-\bm{p}'_2)}^2 +
  {(\bm{p}_1-\bm{p}_2)}^2 }{4 \Lambda^2} \right).
\end{eqnarray}
The regularized off-shell
contact NN interactions can be obtained by applying the unitary transformation
given by Eq.~(\ref{eq:unittransfABC}) to the kinetic energy term
with the \emph{regularized} generators $T_i$
\begin{eqnarray}
  \label{eq:unitarytransformGeneratorsNNreg}
  {T}_i^\text{reg} = {T}_i \;
  \exp \left( - \frac{ {(\bm{p}'_1-\bm{p}'_2)}^2 + {(\bm{p}_1-\bm{p}_2)}^2 }{4\Lambda^2} \right)  ,
  \qquad
  i=1,2,3.
\end{eqnarray}
Then, by acting with this unitary transformation on the single-nucleon
charge density $\rho_\text{1N}^\text{Main}$ from Eq.~\eqref{eq:rho1Nmain},
we obtain the consistently regularized 2N short-range charge density operator:
\begin{eqnarray}
  \label{eq:contactchargedensityReg}
  \rho_\text{Cont}^\text{reg}
  = 2 e \GE^{S}({\bm{k}}^2) \left(
    (A + B \,  (\bm{\sigma}_1 \cdot \bm{\sigma}_2))
    F_1 \left( \frac{\bm{p}_1-\bm{p}_2}{2}, \frac{\bm{p}'_1-\bm{p}'_2}{2}, \bm{k} \right)
  + C F_2 \left( \frac{\bm{p}_1-\bm{p}_2}{2}, \frac{\bm{p}'_1-\bm{p}'_2}{2}, \bm{k} \right)
   \right),
\end{eqnarray}
where the functions $F_1$ and $F_2$ are defined as
\begin{eqnarray}
  F_i (\bm{p}, \bm{p}', \bm{k}) =
      E_i \left( \bm{p}-\frac{\bm{k}}{2}, \bm{p}' \right)
    + E_i \left( \bm{p}+\frac{\bm{k}}{2}, \bm{p}' \right)
    + E_i \left( \bm{p}'-\frac{\bm{k}}{2}, \bm{p} \right)
    + E_i \left( \bm{p}'+\frac{\bm{k}}{2}, \bm{p} \right),
 \end{eqnarray}
 with
 \begin{eqnarray}
  E_1 \left( \bm{p}, \bm{p}' \right) &=& \left( \bm{p}^2 - \bm{p}'^2 \right)
   \, \exp \left(-\frac{\bm{p}^2 + \bm{p}'^2}{ \Lambda ^2} \right), \nonumber \\
  E_2 \left( \bm{p}, \bm{p}' \right) &=& \left[
    (\bm{\sigma}_1 \cdot \bm{p}) (\bm{\sigma}_2 \cdot \bm{p}) -  (\bm{\sigma}_1 \cdot \bm{p}') (\bm{\sigma}_2 \cdot \bm{p}')
   \right] \,
   \exp \left(-\frac{\bm{p}^2 + \bm{p}'^2}{ \Lambda ^2} \right).
\end{eqnarray}

Note that, similarly to the procedure used for obtaining  the contact interactions,
the regularized OPE  contribution to the 2N charge density (as given in Eq.~\eqref{eq:NNchargeDensityKollingWithGESregularized})
can also be derived   by  regularizing the long-range unitary transformation (as given in Eq.~(4.23) of
Ref.~\cite{Kolling:2011mt})  and acting with it on the single-nucleon
charge density $\rho_\text{1N}^\text{Main}$ from Eq.~\eqref{eq:rho1Nmain}.

Equations (\ref{eq:NNchargeDensityKollingWithGESregularized}) and
(\ref{eq:contactchargedensityReg}) provide the final expressions for the 2N
charge density operator at N$^3$LO and N$^4$LO used in the calculation of the
deuteron FFs.

\section{Relativistic corrections}\label{sec:relcorr}

Although the deuteron FFs are Lorentz-invariant,
the individual ingredients (charge density operators and deuteron wave functions)
do depend on the reference frame.
At N$^2$LO and below, all frame-dependent corrections are irrelevant, but starting from
N$^3$LO, the relativistic corrections to each ingredient have to be systematically taken into account.
Frame dependence of the charge density operator is automatically accounted for by the kinematics,
because the operator is calculated explicitly including all relevant $1/\mN$ corrections.
In this section, we will focus on the relativistic corrections to the
deuteron wave functions stemming from the motion of the initial and final
deuterons.

The DWF is typically calculated for the deuteron at rest.
However, a calculation of the deuteron FFs always involves at least one moving deuteron.
Our calculation is carried out in the Breit frame, where the initial and final deuterons are moving in opposite directions.
To account for this motion, the rest-frame DWF needs to be boosted.
To the chiral order we are working (N$^4$LO),
the DWF boost corrections have to be considered only when calculating
the convolution integrals of the DWF
with the leading single-nucleon charge density  $\rho_\text{1N}^\text{Main}$  from Eq.~\eqref{eq:rho1Nmain}.

Since subleading corrections to the single-nucleon charge density as well as the first contributions to the two-nucleon charge density appear only at
N$^3$LO, see Sections~\ref{sec:charge} and~\ref{sec:Setup}, the corresponding DWF-boost corrections are beyond the scope of our study.%
\footnote{
It is reassuring that the relativistic corrections to
 the OPE charge density operator  considered in
Ref.~\cite{Arenhovel:1999nq} were found to have a tiny effect on the
deuteron charge radius and quadrupole moment.
}

Different  approaches have been considered in the literature to
include the DWF boost corrections and found to yield basically the same results.
In Ref.~\cite{Arnold:1979cg}, a
covariant  relativistic calculation of the deuteron form factor was
performed, and the final result was expanded in powers of $1/\mN$, see also Ref.~\cite{Gilman:2001yh} for a review.
Alternatively, boosted DWF was calculated  in
Refs.~\cite{Krajcik:1974nv,Friar:1977xh,Ritz:1996za,Wallace:2001nv,Schiavilla:2002fq} utilizing the $1/\mN$ expansion of the
generators of the Poincar\'e group. This is the approach we follow in
our analysis.
For a deuteron moving with the velocity $\bm{v}$, the boosted DWF operator has the form~\cite{Schiavilla:2002fq}
\begin{eqnarray}
  \label{eq:boostedDWF}
  \psi(\bm{p}, \bm{v})
  \simeq
  \left( 1 - \frac{\bm{v}^2}{4} \right)
  \left[ 1 - \frac{1}{2} (\bm{v}\cdot\bm{p})(\bm{v}\cdot\bm{\nabla}_{\!p})
           - \frac{i}{4 \mN} \bm{v}\cdot (\bm{\sigma}_1 - \bm{\sigma}_2) \times \bm{p} \right]
  \psi(\bm{p}, 0),
\end{eqnarray}
where $\bm{p}$ is the relative momentum of two nucleons, and  $\psi(\bm{p}, 0)$ is the rest-frame DWF  which is normalized as
\begin{eqnarray}
	\int \frac{d^3 p}{(2\pi)^3}\, |\psi\left(\bm{p}, 0\right )|^2     =
	\int \frac{d^3 p}{(2\pi)^3}\, |\psi\left(\bm{p}, \bm{v}\right )|^2  = 1\,.
\end{eqnarray}
Then, to the order we are working, the boost-corrected
 matrix element~\eqref{Eq:ME} evaluated with the leading density $\rho_\text{1N}^\text{Main}$ reads
\begin{eqnarray}
  \frac{1}{2P_0}\Braket{P', \lambda_d'| \rho_\text{1N}^\text{Main} | P, \lambda_d} &=& e\, \GE^S({\bm{k}}^2)\int \frac{d^3 p}{{(2\pi)}^3}\, \psi_{\lambda_d'}  ^\dagger\left(\bm{p} +\frac{\bm{k}_\text{boosted}}{4}, 0\right )\,
  \psi_{\lambda_d}\left(\bm{p} -\frac{\bm{k}_\text{boosted}}{4}, 0\right ),
\label{Eq:boost}
\end{eqnarray}
where
\begin{eqnarray}
  \label{eq:kboosted}
   \bm{k}_\text{boosted} = \bm{k}  \sqrt{1 - v_B^2} =   \frac{\bm{k}}{\sqrt{1+\eta}},
\end{eqnarray}
and we used the fact that the spin dependent term  in Eq.~\eqref{eq:boostedDWF}   vanishes for
spin-$1$-to-spin-$1$ transitions relevant for the deuteron FFs.

The  term on the rhs of Eq.~\eqref{Eq:boost}  is   related to the length contraction
of that part of the relative nucleon momentum in the deuteron which is parallel to $\bm k$.
As a consequence of this contraction, the matrix element  must be evaluated with the
deuteron wave function taken in its rest frame but with the Breit momentum $\bm k$ replaced by $\bm k_\text{boosted}$.

Finally, we remind the reader on the ambiguity of the relativistic
corrections to the NN potential associated with the employed form of
the Schr\"odinger equation. The corrections to the kinetic
energy of relative motion of the nucleons are most easily taken into account by replacing the
nonrelativistic expression $\bm{p}^2/\mN$  with $2 \sqrt{\bm{p}^2 +
  \mN^2} - 2 \mN$ instead of using the Taylor expansion, since
otherwise the spectrum of the 2N Hamiltonian is unbounded from below.
Instead of solving the corresponding relativistic Schr\"odinger
equation, it is more convenient to rewrite it in the equivalent
nonrelativistic form as explained in Ref.~\cite{Friar:1999sj}.
This choice was adopted  in the Nijmegen partial wave analysis~\cite{Stoks:1993tb}
and is made in the chiral NN potentials of Refs.~\cite{Epelbaum:2014efa,Epelbaum:2014sza,Reinert:2017usi,Reinert:2020mcu}.
While rewriting the Schr\"odinger equation does affect the $\mN^{-1}$
and $\mN^{-2}$ contributions to the NN potential, the
deuteron wave function remains unchanged so that we can directly
employ the DWF from Refs.~\cite{Reinert:2017usi,Reinert:2020mcu}.

\section{Anatomy of the calculation}\label{sec:analyticresults}

In this section, we summarize the analytic expressions for the deuteron
charge and quadrupole form factors, as well as for the charge radius and the
quadrupole moment.
We discuss the individual contributions to these quantities from different
types of the charge density
introduced in the previous sections.  We define the structure radius of the deuteron and argue, following Ref.~\cite{Filin:2019eoe},
that a high-accuracy calculation of this quantity along with  high-precision atomic data for the   1S-2S hydrogen-deuterium isotope shift
provide access to the neutron  charge radius.

\subsection{The charge form factor and structure radius of the deuteron}\label{Sec:chargFF}

The deuteron charge form factor $\GC$ can,  up to N$^4$LO,  be written as
\begin{eqnarray}
\label{eq:GCcontributions}
  \GC(Q^2) = \GC^\text{Main}(Q^2)
    + \GC^\text{DF}(Q^2)
    + \GC^\text{SO}(Q^2)
    + \GC^\text{Boost}(Q^2)
    + \GC^{1\pi}(Q^2)
    + \GC^\text{Cont}(Q^2),
\end{eqnarray}
where
$\GC^\text{Main}(Q^2)$,   $\GC^\text{DF}(Q^2)$, and $\GC^\text{SO}(Q^2)$
 arise from charge densities defined in Eq.~\eqref{eq:rho1Nmain},
$\GC^\text{Boost}(Q^2)$ is a relativistic correction due to initial and final deuteron motion,
$\GC^{1\pi}(Q^2)$ stems from the one-pion-exchange charge density in Eq.~\eqref{eq:NNchargeDensityKollingWithGESregularized},
and $\GC^\text{Cont}(Q^2)$ is generated by the contact charge density
in Eq.~\eqref{eq:contactchargedensityReg}.
The main contribution $\GC^\text{Main}(Q^2)$ can be factorized as
\begin{eqnarray}
  \label{eq:GCMainDecomposition}
  \GC^\text{Main}(Q^2) = (\GE^{p}(Q^2) + \GE^{n}(Q^2)) \GC^\text{matter}(Q^2),
\end{eqnarray}
where $\GE^{p}(Q^2)$ and $\GE^{n}(Q^2)$ are the electric
FFs of the proton and neutron, while
$\GC^\text{matter}(Q^2)$ is a functional of the deuteron wave function.

Charge conservation restricts the behavior of the charge form factor at $Q^2 = 0$.
In particular, $\GC(0) = \GC^\text{Main}(0) = \GC^\text{matter}(0) = \GE^{p}(0) = 1$,
while all other contributions to $\GC$ vanish at $Q^2=0$.

The deuteron charge radius can be expressed as a derivative of the charge form factor with respect to $Q^2$ at $Q^2=0$
\begin{eqnarray}
  r_d^2 = -6 {\left.\frac{\partial \GC(Q^2)}{\partial Q^2} \right|}_{Q^2=0}.
\end{eqnarray}
Taking derivatives of all terms in Eq.~\eqref{eq:GCcontributions},
we get the complete set of contributions to the deuteron charge radius up to N$^4$LO
\begin{eqnarray}
  r_d^2 = r_{m}^2 + r_{p}^2 + r_{n}^2  + r_{\rm DF}^2  + r_{\rm SO}^2  + r_{\rm Boost}^2  + r_{1\pi}^2  + r_{\rm Cont}^2,
  \label{eq:deutchargeradiuscontributions}
\end{eqnarray}
where the deuteron matter radius $r_{m}$, the proton charge radius $r_{p}$ and the neutron charge radius $r_{n}$
are defined as:
\begin{eqnarray}\label{Eq:r1}
  r_{m}^2 = -6 {\left.\frac{\partial \GC^\text{matter}(Q^2)}{\partial Q^2} \right|}_{Q^2=0},
  \qquad
    r_{p}^2 = -6 {\left.\frac{\partial \GE^p(Q^2)}{\partial Q^2} \right|}_{Q^2=0},
  \qquad
    r_{n}^2 = -6 {\left.\frac{\partial \GE^n(Q^2)}{\partial Q^2} \right|}_{Q^2=0},
\end{eqnarray}
and the remaining corrections to the deuteron charge radius are calculated as
\begin{eqnarray}\label{Eq:r2}
  r_i^2 = -6 {\left.\frac{\partial \GC^{i}(Q^2)}{\partial Q^2} \right|}_{Q^2=0},
  \qquad
  i = \{\text{DF}, \, \text{SO}, \, \text{Boost},\,  {1\pi}, \, \text{Cont} \}.
\end{eqnarray}

Since the $r_\text{DF}$-term and the charge radii of the individual nucleons
are not related to the two-body dynamics of the deuteron,
they can be conveniently subtracted from the deuteron charge radius.
The resulting quantity is referred to as  the \emph{deuteron structure radius}
and is defined as (see, e.g.~Ref~\cite{Jentschura:2011NOTinHep})
\begin{eqnarray}\label{Eq:r_str}
  r_{\rm str}^2 = r_d^2 - (r_{p}^2 + r_{n}^2  + r_{\rm DF}^2).
\end{eqnarray}

The deuteron-proton mean-square
charge radii difference $r_d^2 - r_{p}^2$ in Eq.~\eqref{Eq:r_str} can be extracted experimentally
with an extremely high precision from spectroscopic measurements of the 1S-2S hydrogen-deuterium isotope shift~\cite{Jentschura:2011NOTinHep}.
In particular, a series of  very precise measurements of the
1S-2S isotope shift, accompanied with an accurate theoretical QED analysis
(see Ref.~\cite{Pachucki:2018yxe} for the latest update up through $O(\alpha^2)$),
resulted in the  extraction of the deuteron-proton mean-square
charge radii difference~\cite{Jentschura:2011NOTinHep}
\begin{equation}
  r_d^2 - r_{p}^2 = 3.820 07(65) \text{fm}^2.
  \label{Eq:rd-rp}
\end{equation}
Due to its high accuracy, this difference provides a tight link between $r_d$ and $r_p$ and thus
is important  in connection with the light nuclear charge radius
puzzle.
For many years,  the values for $r_p$ extracted from  electron and
muon experiments showed more than a 5$\sigma$ discrepancy~\cite{Pohl:2013yb}.
The very recent atomic hydrogen measurements~\cite{Beyer:2017gug,Bezginov:2019mdi},
however, claim consistency with the analogous muonic hydrogen experiments.
The  recommended value for the proton root-mean-square charge
radius has been changed to $r_p=0.8414(19)$~fm in the latest  CODATA-2018 update~\cite{CODATA2018}, and the deuteron
charge radius was updated accordingly,  by virtue of the  difference
in Eq.~\eqref{Eq:rd-rp}.
The updated CODATA deuteron charge radius  is only  1.9$\sigma$ larger
than the spectroscopic measurement on the muonic deuterium~\cite{Pohl1:2016xoo}
but still 2.9$\sigma$ smaller than the $r_d$
value from electronic deuterium spectroscopy~\cite{Pohl:2016glp}.

As follows from Eq.~\eqref{Eq:r_str}, the deuteron-proton charge radii difference
from Eq.~\eqref{Eq:rd-rp} allows one to extract the difference  $r_{\rm str}^2 - r_{n}^2$   to a very high accuracy.
The neutron charge radius can be deduced  from  measurements of the coherent neutron-electron scattering length extracted from data on
 neutron scattering off $^{208}$Pb,  $^{209}$Bi  and other heavy atoms.
 The  value for the neutron charge radius quoted by the PDG is  $ r_{n}^2 = -0.1161(22) \text{fm}^2$,  where the estimated error
was increased by a scaling factor of $1.3$~\cite{Tanabashi:2018oca}.  This value  is based on averaging the results of
 four different experiments from years 1973 to 1997.
 In Ref.~\cite{Jentschura:2011NOTinHep},  the value of $ r_{n}^2 = -0.114(3) \text{fm}^2$, which is consistent with the PDG result,  was employed based on   the  measurement
on $^{208}$Pb from Ref.~\cite{Kopecky:1997rw}.  Using this neutron radius and Eq.~\eqref{Eq:rd-rp} for the deuteron-proton charge radii difference,
the value of $  r_\text{str} =1.97507(78) \text{ fm}$
for the structure radius was extracted~\cite{Jentschura:2011NOTinHep}.
On the other hand, as advocated in Ref.~\cite{Mitsyna:2009zz}, the uncertainty for the neutron radius given above might suffer from the underestimation of  systematic errors.  For example,
the central values   on $^{208}$Pb and  $^{209}$Bi quoted in the most recent investigation of Ref.~\cite{Kopecky:1997rw} differ from  each other by 0.0090  $\text{fm}^2$,  which is  much larger than even the increased uncertainty given
by the PDG.
Therefore,  a different logical chain was adopted  in Ref.~\cite{Filin:2019eoe}, namely,  (a)
by employing the nuclear forces and currents derived up through fifth order in chiral EFT, a very accurate determination of $ r_\text{str} $ is becoming possible based on the analysis of the deuteron charge form factor;
(b)  by  using the predicted value for the deuteron structure radius together
with  the atomic data for the  deuteron-proton charge radii difference, the charge radius of the neutron was  for the first time extracted from light nuclei.
In this investigation, we follow the same  logic to update the
analysis of Ref.~\cite{Filin:2019eoe}. In particular, we employ the
updated NN potentials which include isospin breaking corrections up
through N$^4$LO and provide a statistically perfect description of
neutron-proton and proton-proton scattering data up to the pion
production threshold~\cite{Reinert:2020mcu}
to extract the structure
radius from  a combined analysis of the charge and  quadrupole
deuteron FFs in the range of momentum transfer up to $Q=6$ fm$^{-1}$.
Then, we update the  value for the neutron charge radius, see
Sec.\ref{sec:str_rad}
for the results.

\subsection{The quadrupole form factor and quadrupole moment of the deuteron}
Deuteron quadrupole form factor can be decomposed in the same way as the charge form factor, namely:
\begin{eqnarray}
\label{eq:GQcontributions}
  \GQ(Q^2) = \GQ^\text{Main}(Q^2)
    + \GQ^\text{DF}(Q^2)
    + \GQ^\text{SO}(Q^2)
    + \GQ^\text{Boost}(Q^2)
    + \GQ^{1\pi}(Q^2)
    + \GQ^\text{Cont}(Q^2),
\end{eqnarray}
where the individual terms originate from different charge-density contributions
in  full analogy with Eq.~\eqref{eq:GCcontributions}.
The deuteron quadrupole moment is defined as the value of the
quadrupole form factor at $Q^2=0$, namely
\begin{eqnarray}\label{Eq_Q_d}
   Q_d = \frac{1}{\md^2}\GQ(0).
\end{eqnarray}
Taking the $Q^2=0$ limit in  Eq.~\eqref{eq:GQcontributions}
yields the individual contributions
to the deuteron quadrupole moment, which read
\begin{eqnarray}
  Q_d = Q_{0} + Q_{\rm SO}  + Q_{\rm Boost}  + Q_{1\pi}  + Q_{\rm Cont},
  \label{eq:deutqaudmomcontributions}
\end{eqnarray}
where we used the fact that $\GQ^\text{DF}(0) = 0$ and defined the individual terms as
\begin{eqnarray}\label{Eq:Qterms}
 Q_{0} &=& \frac{1}{\md^2}\GQ^\text{Main}(0),
   \qquad
  Q_i = \frac{1}{\md^2}\GQ^{i}(0),
  \qquad
  i = \{\text{SO}, \, \text{Boost}, \, {1\pi}, \, \text{Cont} \}.
\end{eqnarray}
The analytic expressions for various contributions to the deuteron
charge and quadrupole form factors as well as to the structure radius and the
quadrupole moment are collected in  Appendix~\ref{sec:analyticexpressions}.

\subsection{Calculational setup}\label{sec:Setup}

The deuteron FFs at different chiral orders are calculated as follows:
\begin{itemize}
  \item LO:\\ The main contribution to the single-nucleon charge
    density $\rho_\text{1N}^\text{Main} $ in Eq.~(\ref{eq:rho1Nmain})
    is convoluted with the LO deuteron wave function.
  \item NLO:\\ Same as LO but using the NLO deuteron wave function.
  \item N$^2$LO:\\ Same as LO but using the N$^2$LO deuteron wave
    function.
 \item N$^3$LO:\\
   The contributions $\rho_\text{1N}^\text{Main}$,
   $\rho_\text{1N}^\text{DF}$ and $\rho_\text{1N}^\text{SO}$ from Eq.~(\ref{eq:rho1Nmain})
 to the single-nucleon charge density and the OPE contribution in
 Eq.~(\ref{eq:NNchargeDensityKollingWithGESregularized}) are
 convoluted with  the N$^3$LO deuteron wave
    functions; the relativistic boost corrections to the single-nucleon
    contributions are calculated as
    explained in Section~\ref{sec:relcorr}.
  \item N$^4$LO:\\
    Same as N$^3$LO but using the N$^4$LO$^+$ deuteron wave
    function and including the 2N short-range charge density operators from
    Eq.~(\ref{eq:contactchargedensityReg}).
\end{itemize}
Unless specified otherwise, all results presented below are based on
the semilocal momentum-space NN potentials of
Ref.~\cite{Reinert:2017usi}, updated to incorporate a more complete
treatment of isospin-breaking corrections~\cite{Reinert:2020mcu}. In particular, the
updated potentials take into account the charge dependence of the
pion-nucleon coupling constants. The determination of the pion-nucleon
coupling constants from NN data in Ref.~\cite{Reinert:2020mcu} leads to the
average value of $g_{\pi N}$, which is about $1\%$ larger than the one
employed in Ref.~\cite{Reinert:2017usi}, and
the resulting change in the deuteron wave function leads to a visible effect on the
quadrupole FF of the deuteron at higher $Q$-values.
Clearly, in all cases, the same cutoff value chosen from
the set $\Lambda = \{400, \, 450, \, 500, \, 550\}$~MeV is used  in the
regularized  NN potential and in the 2N charge density.
For single-nucleon FFs, we employ the most up-to-date parametrization
by Ye {\rm et al.}~\cite{Ye:smallrp} for our central results. We propagate the
uncertainty in the determination of these FFs to estimate its impact on the deuteron FFs
in Section~\ref{subsec:Error1NFF}.
In the same section we also consider the impact of using different single-nucleon FFs parametrizations.

It remains to specify the values of the various parameters used in the
expressions for the 2N charge density operator in
Eqs.~(\ref{eq:NNchargeDensityKollingWithGESregularized}) and (\ref{eq:contactchargedensityReg}).  Following
Refs.~\cite{Reinert:2017usi,Reinert:2020mcu}, we employ
the value of $\gA=1.29$ for the
effective axial-vector coupling constant, which accounts for the
Goldberger-Treiman discrepancy, $\fpi=92.4$~MeV for the pion decay
constant, $\mN = 2 m_p m_n / (m_p+m_n) = 938.918$~MeV for the nucleon
mass and $\mpi = (2 {\mpi}_{\pm} + {\mpi}_0)/3 = 138.03$~MeV for the
pion mass. Notice that the expressions for the  2N charge density are
taken in the isospin limit as the corresponding isospin-breaking
corrections start to contribute at N$^5$LO, which is beyond the
accuracy of our analysis. Finally, the two linear combinations of LECs
entering the short-range part of the 2N charge density operator at
N$^4$LO are determined from the best combined fit to the
experimental data on the momentum-transfer dependence of the charge
and quadrupole deuteron FFs as described in Section~\ref{sec:results}.
This then allows us to make a parameter-free
prediction for the structure radius and the quadrupole moment of the deuteron.

\section{Results for charge and quadrupole deuteron form factors}\label{sec:results}

In this section, we present our results for the deuteron charge and quadrupole form factors.
We fix two LECs appearing in the N$^4$LO contact charge density by
fitting the calculated  FFs, $\GC^\text{th}(Q)$ and $\GQ^\text{th}(Q)$,
to the corresponding world experimental data for $Q < 6$ fm$^{-1}$.
Using the LECs extracted from the best fit,  we predict the structure radius
and the quadrupole moment of the deuteron.
Following Ref.~\cite{Filin:2019eoe}, we use the predicted structure radius
to extract the neutron charge radius from the precisely measured deuteron-proton
charge-radii difference.
We provide a detailed analysis of various uncertainties, discuss several important consistency checks,
and discuss the role of the individual contributions to the charge and quadrupole deuteron form factors.

\subsection{Fitting procedure}
The values of the LECs appearing in the N$^4$LO contact charge density
of Eq.~\eqref{eq:contactchargedensityReg} are determined from a $\chi^2$-fit
of our theoretical predictions for $\GC^\text{th}(Q)$ and $\GQ^\text{th}(Q)$
to the experimental data.
The analytic expressions for the individual contributions
to $\GC^\text{th}(Q)$ and $\GQ^\text{th}(Q)$ are given in
Appendix~\ref{sec:analyticexpressions},
and the experimental data set used in the fit is described in
Section~\ref{sec:exper}.  In the infinite cutoff limit,
$\GC^\text{th}(Q)$  depends only on one combination of the
LECs, namely  $A+B+C/3$, while
$\GQ^\text{th}(Q)$ depends only on the LEC $C$.  Once the regularization is applied,
both $\GC^\text{th}(Q)$ and $\GQ^\text{th}(Q)$ in general
depend on the two mentioned linear combinations of the LECs,
see Eqs.~\eqref{eq:gccontExplicit} and~\eqref{eq:gqcontExplicit} in
Appendix~\ref{sec:analyticexpressions}.
The function $\chi^2 ( A+B+C/3; C)$ to be minimized is defined as follows
\begin{eqnarray}\label{Eq:chisq}
	\chi^2 = \sum_i
	\frac{{\bigl(\GC^\text{th}(Q^2_i;  A+B+C/3; C) - \GC^\text{exp}(Q^2_i)\bigl)}^2}{\Delta \GC{(Q_i^2)}^2}
	+ \sum_i
	\frac{{\bigl(\GQ^\text{th}(Q^2_i;  A+B+C/3; C) - \GQ^\text{exp}(Q^2_i)\bigl)}^2}{\Delta \GQ{(Q_i^2)}^2},
	\label{eq:chisq}
\end{eqnarray}
where $\{Q_i\}$ are the set of momenta, for which experimental data are available,
and the summations are performed for $Q_i$ below $Q_\text{max} = 6$ fm$^{-1}$.
The intrinsic systematic uncertainty related to the choice of $Q_\text{max}$
will be discussed below.
Following Refs.~\cite{Carlsson:2015vda,Wesolowski:2018lzj},
the uncertainties $\Delta \GC (Q_i^2)$ and $\Delta \GQ
(Q_i^2)$  in $\chi^2$ include, apart from the experimental errors,
also theoretical uncertainties added in quadrature
\begin{eqnarray}\label{Eq:deltaG}
	\Delta G_\text{X}{(Q_i^2)}^2 &=&\Delta G_\text{X}^\text{exp}{(Q_i^2)}^2 + \Delta G_\text{X}^\text{th,trunc}{(Q_i^2)}^2 +\Delta G_\text{X}^\text{th,nuclFF}{(Q_i^2)}^2,
	\qquad
	(X= \text{C and Q}).
\end{eqnarray}
In this way, we take into account uncertainties from the truncation of the chiral expansion
and from the parametrization of the nucleon form factors.
As  the expansion parameter in  chiral EFT increases with the  momentum transfer,
the truncation errors also grow with $Q$, as discussed in Section~\ref{subsec:Truncation}.
Thus the inclusion of the truncation errors directly in the objective
function allows us to use  the deuteron data in a larger range of $Q$,
namely up to $Q_\text{max}=6$ fm$^{-1}$ and even higher.
The uncertainty related to the parametrization of the nucleon FFs is
yet another source of the theoretical uncertainty which we include
directly in the fit, see Section~\ref{subsec:Error1NFF} for details.
Other kinds of uncertainties such as the ones associated with the
choice of $Q_\text{max}$ and with the $\pi$N and 2N LECs used in the NN potential
are estimated separately and discussed below.

\begin{figure}
\includegraphics[width=\textwidth]{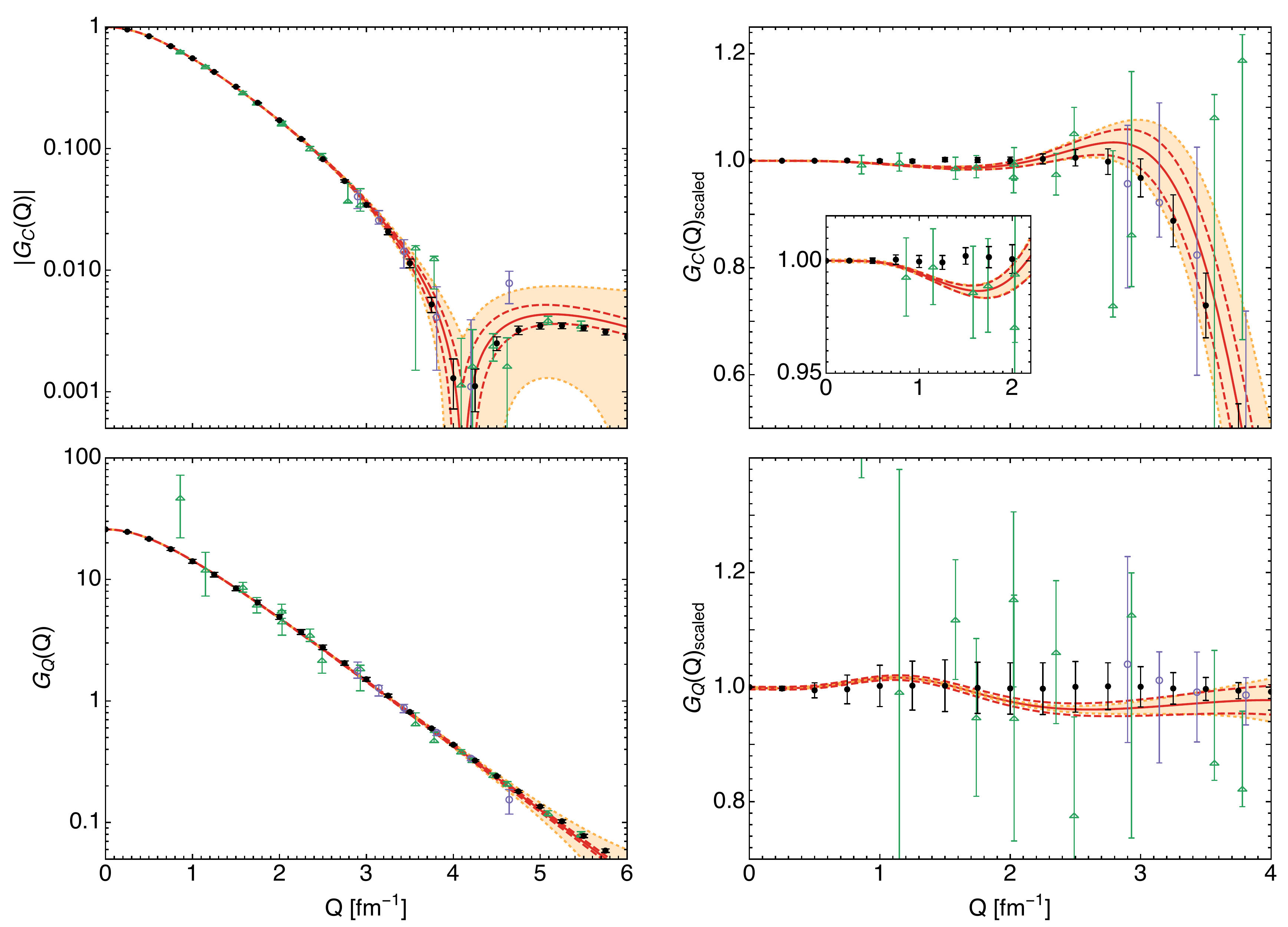}
\caption{\label{fig:GC_GQfinal}
 (Color online) Left panel: the deuteron charge and quadrupole FFs calculated at
  N$^4$LO for the cutoff choice of $\Lambda = 500$~MeV (solid red lines) along with
  the estimated truncation error ($68\%$ degree-of-belief) shown by the
  light-shaded band.  Bands between dashed (red) lines correspond to a 1$\sigma$ error
  in the determination of the two short-range contributions  at N$^4$LO.
  Right panel: the same form factors divided by the
  scaling functions as defined in Eq.~(\ref{eq:defGCrescaled}) and (\ref{eq:defGQrescaled}).
  Open violet circles and green triangles are experimental data from
  Refs.~\cite{Nikolenko:2003zq} and~\cite{Abbott:2000ak}, respectively.
  Black solid circles correspond to the parametrization of the deuteron
  FFs from Refs.~\cite{Marcucci:2015rca,Sick:priv}.
}
\end{figure}

Our central fit is performed for the cutoff  $\Lambda = 500$ MeV and $Q_\text{max} = 6$ fm$^{-1}$.
Assuming that the experimental data points are independent%
\footnote{
	Note that for the number of degrees of freedom we take just the number of data points minus the number of free parameters. Correlations between data points are neglected.
	},
the resulting $\chi^2$ and $\chi^2 / \text{d.o.f.} $ values are
\begin{eqnarray}
	\chi^2_\text{min} = 15.24,
	\qquad
	\chi^2_\text{min} / \text{d.o.f.} = 0.34.
\end{eqnarray}
The low value of $\chi^2_\text{min}/ \text{d.o.f.}$ may signal an overestimation of
the truncation errors, but it can also be caused by neglecting
correlations when estimating truncation errors at similar values of
the momentum transfer. The value of $\chi^2_\text{min}/ \text{d.o.f.}$, therefore, does not allow for a
straightforward statistical interpretation.
The obtained values of the two relevant linear
combinations of the LECs read
\begin{eqnarray}
	A+B + \frac{C}{3} &=& (-281  \pm 64) \text{ GeV}^{-5}  \simeq (-0.66 \pm 0.15) \frac{1}{\fpi^{2} \Lambda_b^{3} },
	\nonumber
	\\
	C &=& (-58  \pm 35) \text{ GeV}^{-5}  \simeq (-0.14 \pm 0.08) \frac{1}{\fpi^{2} \Lambda_b^{3} },
	\label{eq:fittedValuesOfLECs}
\end{eqnarray}
where the error corresponds to the 1$\sigma$ deviation of the $\chi^2$
and $\Lambda_b=650 \text{ MeV}$
refers to the breakdown scale of the chiral expansion,
see Sec.\ref{subsec:Truncation} for a discussion.
Notice that the both linear combinations of the LECs come
out of a natural size, see the second equalities in
Eq.~(\ref{eq:fittedValuesOfLECs}). This is an important consistency
check of our calculations, which is also fulfilled for the contact
interactions entering the employed NN potentials, see  Fig.~7 of
Ref.~\cite{Epelbaum:2019kcf}.
Finally, the correlation matrix for $A+B + C/3$ and $C$ reads
\begin{eqnarray}
	\rho =
	\begin{pmatrix}
	 1 & -0.4 \\
	 -0.4 & 1
	\end{pmatrix}.
\end{eqnarray}

\subsection{Results for the deuteron form factors}

The results for the  deuteron charge and quadrupole FFs from the best fit  to data up
to $Q=6$ fm$^{-1}$, evaluated for the cutoff  $\Lambda = 500$ MeV, are
visualized in Fig.~\ref{fig:GC_GQfinal}
together with the N$^4$LO truncation errors and statistical uncertainty
of the LEC's in $\rho_\text{Cont}^\text{reg}$ from Eq.~\eqref{eq:contactchargedensityReg}.
The plot contains two theoretical uncertainty bands: the light-shaded band   stands for
  the estimated truncation error corresponding to the $68\%$ degree-of-belief interval,  while
the band between long-dashed (red) lines corresponds to a 1$\sigma$ error in the determination of the two short-range contributions  at N$^4$LO.
In principle, these two uncertainty bands are not fully independent since the truncation error is also included in the estimate of the 1$\sigma$ error for the LECs in  the charge density operator as discussed
in  previous Section. In this way, however, the truncation error is estimated more conservatively.

Since the variation of  the FFs at small $Q$-values is difficult to
see on the logarithmic scale, we also plot  the rescaled FFs using a linear scale in the right panels of
Fig.~\ref{fig:GC_GQfinal}. Specifically, following
Ref.~\cite{Marcucci:2015rca}, we define the rescaled charge and quadrupole FFs via
\begin{equation}
  \label{eq:defGCrescaled}
\GC^{\rm scaled} (Q ) = \GC (Q )  {\left( \sum_{i=0}^{3} a_i \exp (- b_i
Q^2 ) \right)}^{-1}\,,
  \end{equation}
with $a_1=0.295$,  $a_2=0.637$, $a_3=0.010$,
  $b_0=3.149$~fm$^2$,   $b_1=1.183$~fm$^2$,  $b_2=0.346$~fm$^2$, $b_3=0.036$~fm$^2$ and $a_0 = 1 -
  a_1 - a_2 - a_3$ and
\begin{eqnarray}
	\label{eq:defGQrescaled}
	\GQ^{\rm scaled} (Q ) &=& \frac{\GQ (Q ) }{m_d^2 \, Q_d} {\left( \sum_{i=0}^{3} a_i \exp (- b_i Q^2 ) \right)}^{-1}\,,
\end{eqnarray}
with  $Q_d = 0.2859$~fm$^2$, $a_1=0.344$,  $a_2=0.275$, $a_3=0.035$,
  $b_0=1.483$~fm$^2$,   $b_1=0.475$~fm$^2$,  $b_2=0.222$~fm$^2$, $b_3=0.085$~fm$^2$ and $a_0 = 1 -
  a_1 - a_2 - a_3$.
In these plots,  along with the comparison of our theoretical results with the experimental data, we  also show the results of the parametrization of the deuteron FFs provided in Refs.~\cite{Marcucci:2015rca,Sick:priv}.
While the results for $\GC^\text{th}(Q)$ and $\GQ^\text{th}(Q)$ are generally quite consistent with this parametrization within errors,  a more close look in $\GC^{\rm scaled}(Q)$  reveals a discrepancy in the range of
intermediate $Q$'s from 1 fm$^{-1}$ to 2 fm$^{-1}$, where the
uncertainty from the chiral expansion is still very small.  Meanwhile, as will be discussed in Sec.~\ref{subsec:Error1NFF},
  this range of the transferred momentum is especially sensitive to
  the choice of  a parametrization of the nucleon FFs.
  In particular, the inclusion of  the uncertainty
  for the parametrization from Ref.~\cite{Ye:smallrp} results in the reduction of the discrepancy with Refs.~\cite{Marcucci:2015rca,Sick:priv}.
Nevertheless, the shape of $\GC^\text{th}(Q)$
in the range of $Q$'s from 1 fm$^{-1}$ to 3.5 fm$^{-1}$ appears to
change more rapidly as compared to the parametrization by Sick et al.

\subsection{Prediction for  structure radius and  quadrupole moment. Extraction of  neutron charge radius.}\label{sec:str_rad}

Using the fitted values of the LECs from Eq.~\eqref{eq:fittedValuesOfLECs}
and the theoretical expressions for $r_\text{str}$ and $Q_d$ collected in Appendix~\ref{sec:analyticexpressions},
we make
a parameter-free prediction for  the deuteron structure radius and the quadrupole moment, which read
\begin{equation}\label{rsrt_and_Q}
  r_{\rm str} = 1.9729 \substack{+0.0015\\ -0.0012}\ \text{fm},\quad \quad
  Q_d =  0.2854 \substack{+0.0038\\ -0.0017}\ \text{fm}^2,  
\end{equation}
where the uncertainties are obtained as a sum of
all individual uncertainties given in Table~\ref{Tab:uncert} taken in quadrature, see   Sec.~\ref{Sec:error} for discussion.
\begingroup
\squeezetable
\begin{table}[t]
\caption{\label{Tab:uncert} Deuteron structure radius squared and deuteron quadrupole moment predicted at
N$^4$LO in $\chi$EFT (2nd column) and the individual contributions to
the corresponding uncertainties from
the truncation of the chiral expansion (3rd column),  the statistical error in the
short-range charge density operator extracted from $\GC(Q^2)$
(4th column),
the
statistical uncertainty in $\pi$N LECs from the Roy-Steiner analysis (RSA) of Ref.~\cite{Hoferichter:2015tha,Hoferichter:2015hva}
propagated  through the variation of the deuteron wave functions (5th
column),
the statistical uncertainty in 2N LECs and $\pi$N coupling constants $f_i^2$ from the analysis of the 2N
data of Ref.~\cite{Reinert:2020mcu,Reinert:2017usi} (6th column) and
the choice of the maximal energy in the fit (7th column).
The total uncertainties evaluated as a sum of presented
uncertainties in quadrature are quoted in the 8th column.
}
\small
\begin{ruledtabular}
\begin{tabular*}{\textwidth}{@{\extracolsep{\fill}}crrrrrrr}
& {\rm central}& {{\rm truncation}} &   $\rho_\text{Cont}^\text{reg}$ &
                                                                    {$\pi$N LECs RSA} & {2N LECs and $f_i^2$}                                   & $Q$-range &   {total} \\[2pt]
\hline
 $r_{\rm str}^2$ [fm$^2$]        &         $3.8925$            & $\pm
                                                                 0.0030$
                                    & $\pm 0.0024$ & $\pm 0.0003$
                                                                                  &     $\pm 0.0025$       & $ \substack{+0.0035\\ -0.0005}$                                      & $ \substack{+0.0058\\ -0.0046}$               \\[2pt]
$Q_d$ [fm$^2$]             &      $0.2854$         & $\pm 0.0005$
                                    & $\pm 0.0007$ & $\pm 0.0003$
                                                                                  &     $\pm 0.0016$     &  $ \substack{+0.0035\\ -0.0005}$                                         & $ \substack{+0.0038\\ -0.0017}$               \\
\end{tabular*}
\end{ruledtabular}
\end{table}
\endgroup
As advocated in Ref.~\cite{Filin:2019eoe}, the knowledge  of the deuteron structure radius provides access to the neutron  charge radius, which  measures   the charge distribution inside the neutron.
Using Eqs.~\eqref{Eq:r_str},~\eqref{Eq:rd-rp}  and~\eqref{rsrt_and_Q},  we find
\begin{eqnarray}
	r_n^2 = -0.105 \substack{+0.005\\ -0.006} \text{ fm}^2,
\end{eqnarray}
which is consistent with our previous determination in Ref.~\cite{Filin:2019eoe}.
In  Section~\ref{sub:comparison_to_prl} we discuss some differences between the current result and the result of Ref.~\cite{Filin:2019eoe}.

\subsection{Comparison to PRL \textbf{124},  082501 (2020) (Ref.~\cite{Filin:2019eoe})}\label{sub:comparison_to_prl}

While this study is performed along the lines with Ref.~\cite{Filin:2019eoe},
there are several updates incorporated in the current analysis.
These updates can be summarized as follows:
(i)  the updated SMS potentials  of Ref.~\cite{Reinert:2020mcu} that
include isospin breaking corrections are employed to calculate the
deuteron wave functions;
(ii) we now simultaneously fit two linear combinations of the LECs and
use data for both the charge and quadrupole FFs;
(iii) our central result is based on the fit to data up to $Q_\text{max} = 6$ fm$^{-1}$;
(iv) statistical uncertainty of the  2N LECs in the NN potential is
propagated in a more reliable way.

The small difference in the predicted value for the deuteron structure radius and,
consequently, also for the neutron charge radius as compared to Ref.~\cite{Filin:2019eoe}
is largely caused by increasing the fitting range up to $Q_\text{max}=6$ fm$^{-1}$.
For such value of $Q_\text{max}$, both $r_\text{str}^2$ and $Q_d$ are basically saturated
with $Q_\text{max}$, that is they do not show any significant deviations in their magnitudes
when  $Q_\text{max}$ is increased further.
To estimate the  error related with the $Q_\text{max}$ dependence conservatively,
we vary $Q_\text{max}$ from 3 fm$^{-1}$ to 7 fm$^{-1}$.
The resulting uncertainties are shown in Table~\ref{Tab:uncert}.
The ``saturation'' of $r_\text{str}^2$ and $Q_d$ above $Q_\text{max} = 6$ fm$^{-1}$ also explains
the asymmetry of the $Q_\text{max}$ related uncertainties.

In addition, we want to make a remark about a finite-cutoff effect,
which was neglected in Ref.~\cite{Filin:2019eoe}.
In the infinite cutoff limit, $\GC^\text{th}(Q)$ at N$^4$LO
depends only on one linear combination of the LECs, namely $A+B+C/3$.
On the other hand, for  a finite cutoff, both combinations of the LECs $A+B+C/3$
and $C$ contribute to both   $\GC^\text{th}$ and $\GQ^\text{th}$,
which, therefore, can be  written  schematically as
\begin{eqnarray}\label{Eq:GC_LECs}
	\GC^\text{th} (Q^2) &=& G_\text{C,1}^\text{th} (Q^2) + \left( A+B+C/3 \right)  G_\text{C,2}^\text{th} (Q^2) + C \, G_\text{C,3}^\text{th} (Q^2),\\ \label{Eq:GQ_LECs}
	\GQ^\text{th} (Q^2) &=& G_\text{Q,1}^\text{th} (Q^2) + \left( A+B+C/3 \right)  G_\text{Q,2}^\text{th} (Q^2) + C \, G_\text{Q,3}^\text{th} (Q^2).
\end{eqnarray}
While the expressions for $G_\text{X,2}^\text{th} (Q^2)$ and $G_\text{X,3}^\text{th}
(Q^2)$  with $X$= C, Q are  very different a priori, as can be seen from  Appendix~\ref{sec:analyticexpressions}, in the actual calculations
it occurs numerically that the momentum-transfer dependence of $G_\text{C,2}^\text{th} (Q^2)$ and $G_\text{C,3}^\text{th} (Q^2)$ (and similarly of $G_\text{Q,2}^\text{th} (Q^2)$ and $G_\text{Q,3}^\text{th} (Q^2)$)
is basically identical. In practice, this means that  even for a
finite cutoff, the FFs in Eqs.~\eqref{Eq:GC_LECs} and~\eqref{Eq:GQ_LECs},
that depend on both linear combinations of the
LECs, largely decouple, so that one can study $\GC$ independently from $\GQ$.
For this reason, in Ref.~\cite{Filin:2019eoe},  only the charge FF was
considered,  in which the very last term in Eq.~\eqref{Eq:GC_LECs} was
not included as being redundant.  However, because this decoupling is
only approximate, in this study we make  a combined analysis of both
$\GC^\text{th} (Q^2) $ and $\GQ^\text{th} (Q^2) $.
By comparing the structure radius extracted  in this study with that of Ref.~\cite{Filin:2019eoe},
we conclude that they are completely consistent and that the effect of
considering both $\GC$ and $\GQ$ simultaneously is negligible.
On the other hand, since the LECs $ A+B+C/3$ and $C$  contribute also to  other reactions,
it is important to extract them individually. This goal can only be  achieved
if a combined analysis of $\GC^\text{th} (Q^2) $
and $\GQ^\text{th} (Q^2) $  is performed, which allows one to fix $A+B+C/3$ and $C$ separately.

\subsection{Error analysis}\label{Sec:error}

\subsubsection{Truncation error}\label{subsec:Truncation}

\begin{figure}
\includegraphics[width=\textwidth]{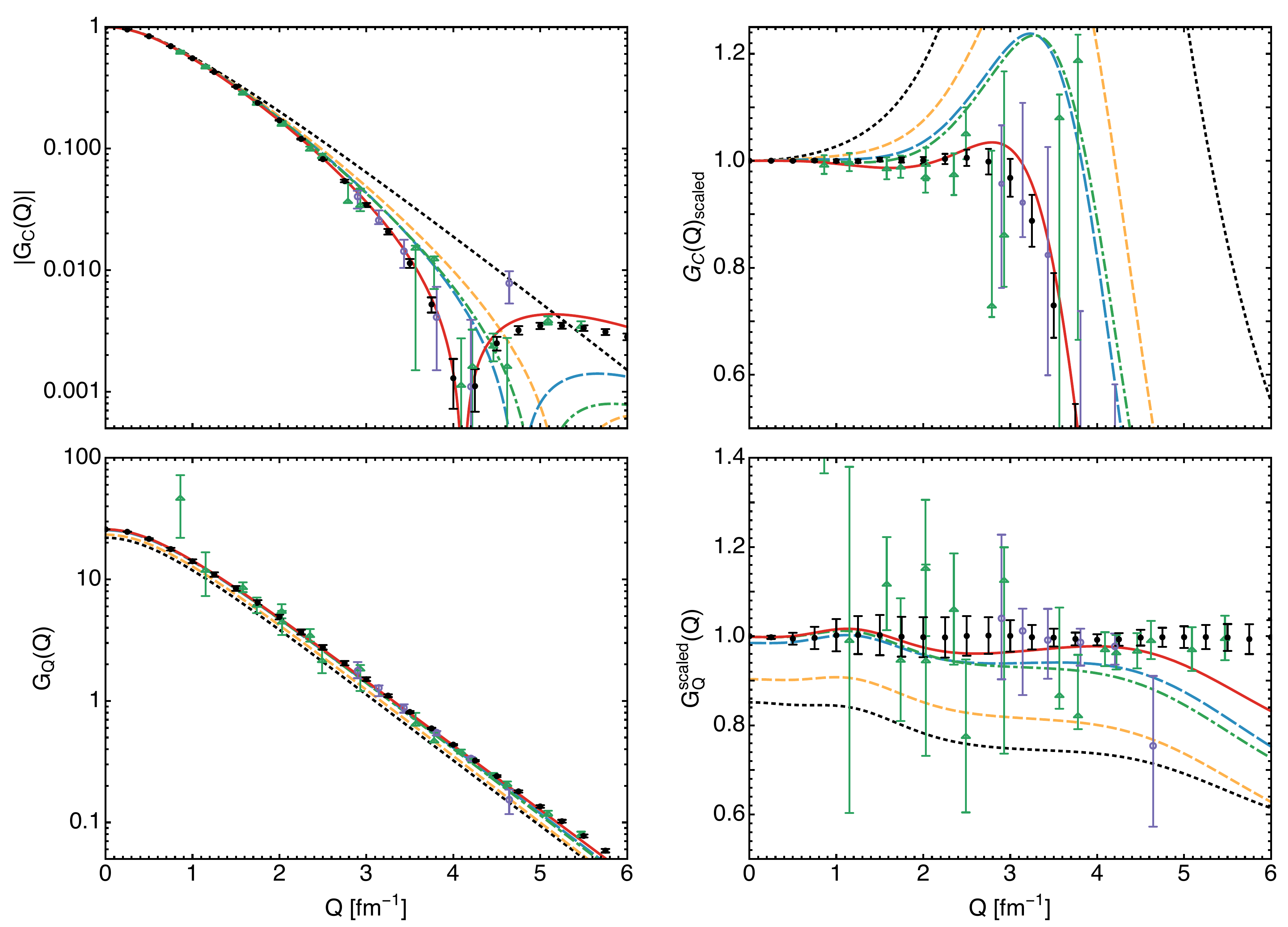}
\caption{\label{fig:GC_GQ_convergence}
 (Color online) Convergence of the chiral expansion for the charge (upper panel)
 and quadrupole (lower panel) deuteron FFs  for the cutoff $\Lambda = 500$~MeV.
   The curves correspond to different chiral orders, namely,  black dotted (LO),
   yellow dashed (NLO),  green dot-dashed (N$^2$LO), blue long-dashed (N$^3$LO)  and red solid (N$^4$LO).
      For remaining notation see Fig.~\ref{fig:GC_GQfinal}.
}
\end{figure}

We start from the discussion of the chiral expansion for the
deuteron form factors which is  important  for the truncation error estimate.
 The convergence pattern of the chiral expansion for the charge
 and quadrupole  deuteron form factors is  shown in Fig.~\ref{fig:GC_GQ_convergence}
 for the cutoff $\Lambda=500$ MeV.
Up to and including N$^3$LO, the calculation does not involve any free parameters,
while at  N$^4$LO,  two linear combinations of the LEC's are adjusted to achieve an
overall best description of the deuteron  FFs in the range of $Q$-values up to 6 fm$^{-1}$.
As a general pattern,  the chiral expansion of both form factors converges quite well.

For a given value of the cutoff $\Lambda$,  truncation errors can be
estimated from the convergence pattern of the chiral expansion using
the algorithm formulated in Ref.~\cite{Epelbaum:2014efa}. This simple
approach has, however, a disadvantage of not directly providing a
statistical interpretation of the estimated errors.
We therefore follow here the Bayesian approach developed in
Refs.~\cite{Furnstahl:2015rha,Melendez:2017phj,Wesolowski:2018lzj,Melendez:2019izc}, which allows one
to estimate truncation errors for a given degree-of-belief (DoB)
interval.  Throughout this analysis, we employ the
Bayesian model $\bar C_{0.5-10}^{650}$ specified in
Ref.~\cite{Epelbaum:2019zqc}   and assume the characteristic momentum
scale $p$ that determines the expansion parameter
\begin{equation}
  \label{ExpParam}
q = \max \left( \frac{p}{\Lambda_b}, \, \frac{M_\pi^{\rm
      eff}}{\Lambda_b} \right)
\end{equation}
to be given by $|\bm{k}|/2$~\cite{Phillips:2006im}.
In the impulse approximation valid up-to-and-including
N$^2$LO, it is easy to see that the deuteron wave function is being
probed at the momentum $| \bm{k} |/2$ rather than $| \bm{k} |$, see
Ref.~\cite{Phillips:2006im} for a discussion.
The quantity $M_\pi^{\rm  eff}$ in Eq.~(\ref{ExpParam}) serves to model
the expansion of few-nucleon observables around the chiral limit, while
$\Lambda_b$ denotes the breakdown scale of chiral EFT.

In Fig.~\ref{fig:GC_GQfinal}, we show the  charge and
quadrupole FFs calculated at N$^4$LO for the cutoff $\Lambda = 500$~MeV along
with the truncation error corresponding to the $68\%$ degree-of-belief (DOB)
interval estimated using Eq.~(21) from~\cite{Epelbaum:2019zqc} with $h=10$, $c_< = 0.5$ and
$c_> = 10$ and assuming  $\Lambda _b = 650$~MeV and $M_\pi^{\rm eff} = 200$~MeV~\cite{Epelbaum:2019wvf}.

The truncation errors for the structure radius and the quadrupole
moment given in Table~\ref{Tab:uncert} are estimated in exactly the same way.
To make this uncertainty estimate   conservatively
the truncation error in these quantities is, like in the deuteron FFs,  included twice: (i) by performing the
Bayesian analysis for  $r_{\rm str}^2$  and $Q_d$ explicitly and (ii) through the statistical
uncertainty in the short-range charge density extracted from the fit
to $\GC^\text{exp} (Q^2)$ and $\GQ^\text{exp} (Q^2)$
using  Eqs.~\eqref{Eq:chisq} and~\eqref{Eq:deltaG}.
We also provide  in Table~\ref{tab:convpattRQ} the results for
the deuteron
structure radius and the quadrupole moment at different orders of the
chiral expansion along with the corresponding truncation errors, which
show   a rather natural pattern of convergence for the considered
cutoff value of $\Lambda = 500$~MeV.

\begingroup
\squeezetable
\begin{table}[t]
\caption{\label{tab:convpattRQ}
Convergence pattern of the chiral expansion and the truncation errors
for the deuteron structure radius and the quadrupole moment.
All results are obtained for the cutoff $\Lambda = 500$~MeV and $Q_\text{max}=6$~fm$^{-1}$.
Truncation errors for  $r_\text{str}$ are recalculated from errors estimated for $r_\text{str}^2$
using the Bayesian approach as described in this Section.
}
\small
\begin{ruledtabular}
\begin{tabular*}{0.48\textwidth}{@{\extracolsep{\fill}}cccccc}
& LO
& NLO
& N$^2$LO
& N$^3$LO
& N$^4$LO
\\ \hline
$r_\text{str}^2$ [fm$^2$]
& $3.8 \pm 1.4$
& $3.86 \pm 0.13$
& $3.873 \pm 0.029$
& $3.877 \pm 0.008$
& $3.8925 \pm 0.0030$
\\
$r_\text{str}$ [fm]
& $1.9 \pm 0.4$
& $1.96 \pm 0.03$
& $1.968 \pm 0.007$
& $1.9689 \pm 0.0019$
& $1.9729 \pm 0.0008$
\\
$Q_d$ [fm$^2$]
& $0.24 \pm 0.10$
& $0.26 \pm 0.01$
& $0.282 \pm 0.006$
& $0.2854 \pm 0.0017$
& $0.2854 \pm 0.0005$
\end{tabular*}
\end{ruledtabular}
\end{table}
\endgroup

\subsubsection{Uncertainty from parametrizations of the nucleon form factors}\label{subsec:Error1NFF}

\begin{figure}
\includegraphics[width=\textwidth]{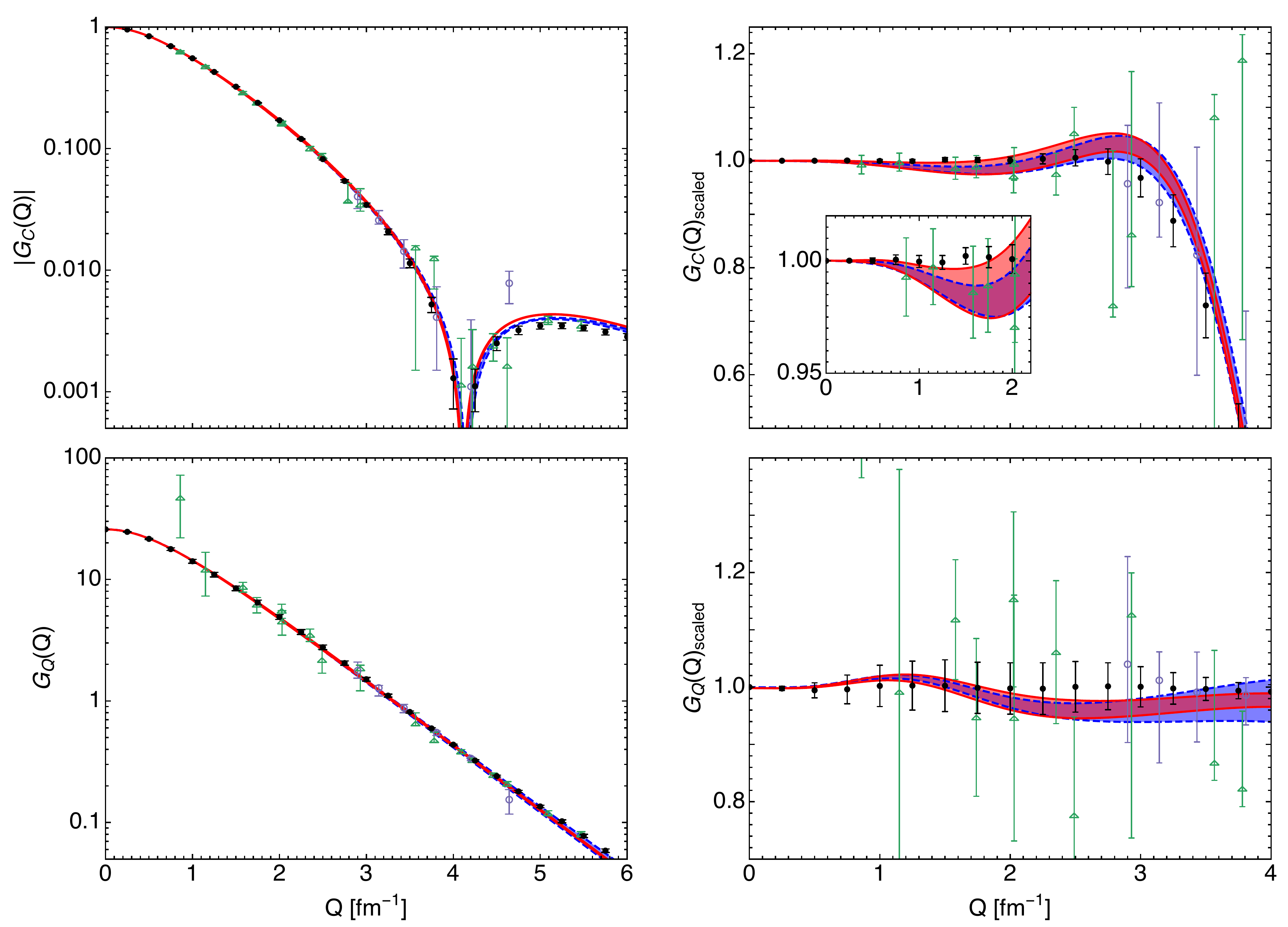}
\caption{\label{fig:Gdeut_vs_GN} (Color online) Effect of the uncertainty from various parametrizations of the nucleon form
  factors (see Fig.~\ref{fig:1NFFscal}) on the deuteron charge (upper panel) and quadrupole (lower panel) FFs.
 Red bands (between two solid lines) correspond to the  nucleon FFs
 extracted from the analysis  of Ref.~\cite{Ye:smallrp};  blue bands
 (between the dashed line)
  based on the nucleon FFs from Ref.~\cite{Belushkin:2006qa}.
 For remaining notation see Fig.~\ref{fig:GC_GQfinal}.
}
\end{figure}

In Fig.~\ref{fig:Gdeut_vs_GN}, we demonstrate the effect of the uncertainties from the nucleon FFs on the deuteron charge and quadrupole FFs.
Our central results, as given by red bands (between solid lines) in Fig.~\ref{fig:Gdeut_vs_GN}, rely on the  nucleon FFs extracted from a  recent global analysis of electron scattering data on
H, $^2$H and $^3$He targets carried out in Refs.~\cite{Ye:smallrp,Ye:2017gyb} using the proton charge radius from CODATA-2018 as input, see Sec.~\ref{Sec:NuclFF} for details.
The uncertainty from the nucleon FFs, as given in Ref.~\cite{Ye:smallrp},     is included in the statistical uncertainty of our calculation, see Eq.~\eqref{Eq:deltaG}.

 To investigate the sensitivity of the results to  parametrizations of the nucleon FFs, we refitted $\GC (Q^2)$ and  $\GQ (Q^2)$ using the nucleon FFs
from the dispersive analysis of Ref.~\cite{Belushkin:2006qa} (the SC approach),
where constraints from unitarity and analyticity were included.  The results are shown  as blue bands between dashed lines in Fig.~\ref{fig:Gdeut_vs_GN}. On the one hand, the results
obtained using  the parametrizations of Ref.~\cite{Ye:smallrp} and Ref.~\cite{Belushkin:2006qa} are generally consistent with each other as one may already expect from the comparison of the isoscalar nucleon FFs
in Fig.~\ref{fig:1NFFscal}.   On the other hand, the range where the
calculated deuteron FFs appear to be especially sensitive to the details of the nucleon FFs corresponds to the intermediate momentum transfers of $Q \simeq 1-2.5$~fm$^{-1}$.
In this range,  the errors related to the truncation of the chiral expansion
are still very small, which can be used to test the consistency of the employed up-to-date nucleon FFs with the deuteron FFs.
 In the regime of intermediate momenta, $\GC (Q^2)$
based on the one-nucleon input from Ref.~\cite{Belushkin:2006qa} is   systematically lower  than that for the nucleon FFs from Ref.~\cite{Ye:smallrp}.
This can be seen from  Fig.~\ref{fig:Gdeut_vs_GN} especially if one compares the theoretical results with the parametrization from Refs.~\cite{Marcucci:2015rca,Sick:priv}:  red bands based on the nucleon FFs
from Ref.~\cite{Ye:smallrp} are essentially consistent with this parametrization while the blue bands between dashed lines lie systematically lower.
This might be related to the fact that the analysis of Ref.~\cite{Belushkin:2006qa} was done before the new high-precision data from Mainz~\cite{Bernauer:2010wm,Bernauer:2013tpr} have become available.
Meanwhile, the updated versions of the dispersive approach~\cite{Lorenz:2012tm,Lorenz:2014yda}  including the MAMI data produce  larger values for the proton electric and magnetic FFs
at small and intermediate momenta  and, as shown in Fig.~\ref{fig:1NFF},
are in a  good agreement with the analysis of Ref.~\cite{Ye:smallrp}.
Since the results of Refs.~\cite{Lorenz:2012tm,Lorenz:2014yda}  are given without errors and no updates for a combined dispersive analysis of the  proton and neutron FFs was provided in Ref.~\cite{Lorenz:2014yda}, we refrain from using these results in the current investigation.

It is important to emphasize that  our results for
the structure radius and, therefore, also for the neutron charge radius
are only very weakly sensitive to the details of the nucleon FFs used in the fits.
This can be understood as follows.
The quality of the fits  to the world data for the  deuteron charge  form factor
(at least  for $Q_\text{max} \sim 4 $ fm$^{-1}$ and higher) increases significantly
if the momentum-transfer range around $Q\sim 4$ fm$^{-1}$,
where $\GC$ becomes small and changes its sign, is well reproduced.
Therefore, the contact interaction in the charge density at N$^4$LO
is adjusted  predominantly to reproduce this area.  Meanwhile, the comparison of
Figs.~\ref{fig:GC_GQfinal} and~\ref{fig:Gdeut_vs_GN} reveals that
by far the largest source of the uncertainty at  $Q\sim 4$ fm$^{-1}$
stems from the truncation of the chiral expansion while the nucleon FFs
in this $Q$-range have only a minor impact on the statistical uncertainty.
Therefore, the structure radius is insensitive to the choice of the parametrization of the nucleon FFs.

\subsubsection{Statistical uncertainty of the LECs determined from $\pi$N  and NN data}

The chiral SMS NN potential involves two groups of LECs:
(i) the $\pi$N LECs from the Roy-Steiner analysis
of Ref.~\cite{Hoferichter:2015tha,Hoferichter:2015hva},
and (ii) the 2N LECs and $\pi$N coupling constants,
which are adjusted to achieve the best fit of the neutron-proton
and proton-proton scattering data in Ref.~\cite{Reinert:2020mcu}.
We consider uncertainties coming from each group.

To account for the statistical uncertainty of the $\pi$N LECs from
the Roy-Steiner analysis, we generated a sample of 50 N$^4$LO$^+$ NN potentials
with normally distributed $\pi$N LECs. Then,  the propagation of this uncertainty
is performed through the variation in the deuteron wave functions.
By re-fitting the deuteron FF data we, therefore,
extracted the impact of this uncertainty on
$r_{\rm str}^2$  and $Q_d$, as shown in Table~\ref{Tab:uncert}.
The resulting uncertainty from these $\pi$N LECs appears to be very small.

The errors from the statistical uncertainty in the 2N LECs and $\pi$N coupling constants extracted
in Ref.~\cite{Reinert:2020mcu} were also propagated to $r_{\rm str}^2$ and $Q_d$ and the corresponding results are given in Table~\ref{Tab:uncert}.
These errors correspond to the maximum deviations from the central values of $r_m$ and $Q_0$,
which are compatible with the variation of the $\chi^2$ in the range $[\chi^2_\text{min}, \chi^2_\text{min} + 1]$
for the description of the neutron-proton and proton-proton data as done in Ref.~\cite{Reinert:2020mcu}.
This approach is similar to what was used to estimate the uncertainties
of the asymptotic deuteron wave function normalization $A_S$ and the $^1S_0$ NN scattering length in Ref.~\cite{Reinert:2017usi}.
Note also that in the present work, the method of error propagation from 2N LECs is different from what was done in Ref.~\cite{Filin:2019eoe},
where a covariance matrix was used. We found that the covariance-matrix approach overestimates the corresponding uncertainties for $r_{\rm str}^2$.
For the deuteron quadrupole moment, however, both approaches give very similar error estimates.

\subsubsection{$Q$-range dependence}\label{sec:qmaxdependence}

As long as the truncation error is included in the uncertainty employed in the fitting procedure,
as done in Eq.~\eqref{Eq:deltaG},  all data available can, in principle, be included in the fits.
This procedure allows us to utilize  the deuteron data in the
range of $Q$  up to $Q_\text{max}=6$ fm$^{-1}$ and even higher.
The effective weight of the data points at  higher transferred momenta is
reduced as compared to data points with  similar experimental errors
at lower $Q$ because the  truncation error increases with growing
values of $Q$.
To estimate (conservatively) the  error for the extracted deuteron
quantities related with the truncation of the $Q$-range in the fits,  we
consider  the variation of $Q_\text{max}$ from $3$~fm$^{-1}$ to $7$~fm$^{-1}$
and include  this error in the uncertainty budget,
as shown in Table~\ref{Tab:uncert}.
The results for both $r_{\rm str}^2$ and $Q_d$ appear to be quite stable to this variation.
\subsubsection{Consistency checks}\label{sec:betaindependence}

\begin{figure}
\includegraphics[width=\textwidth]{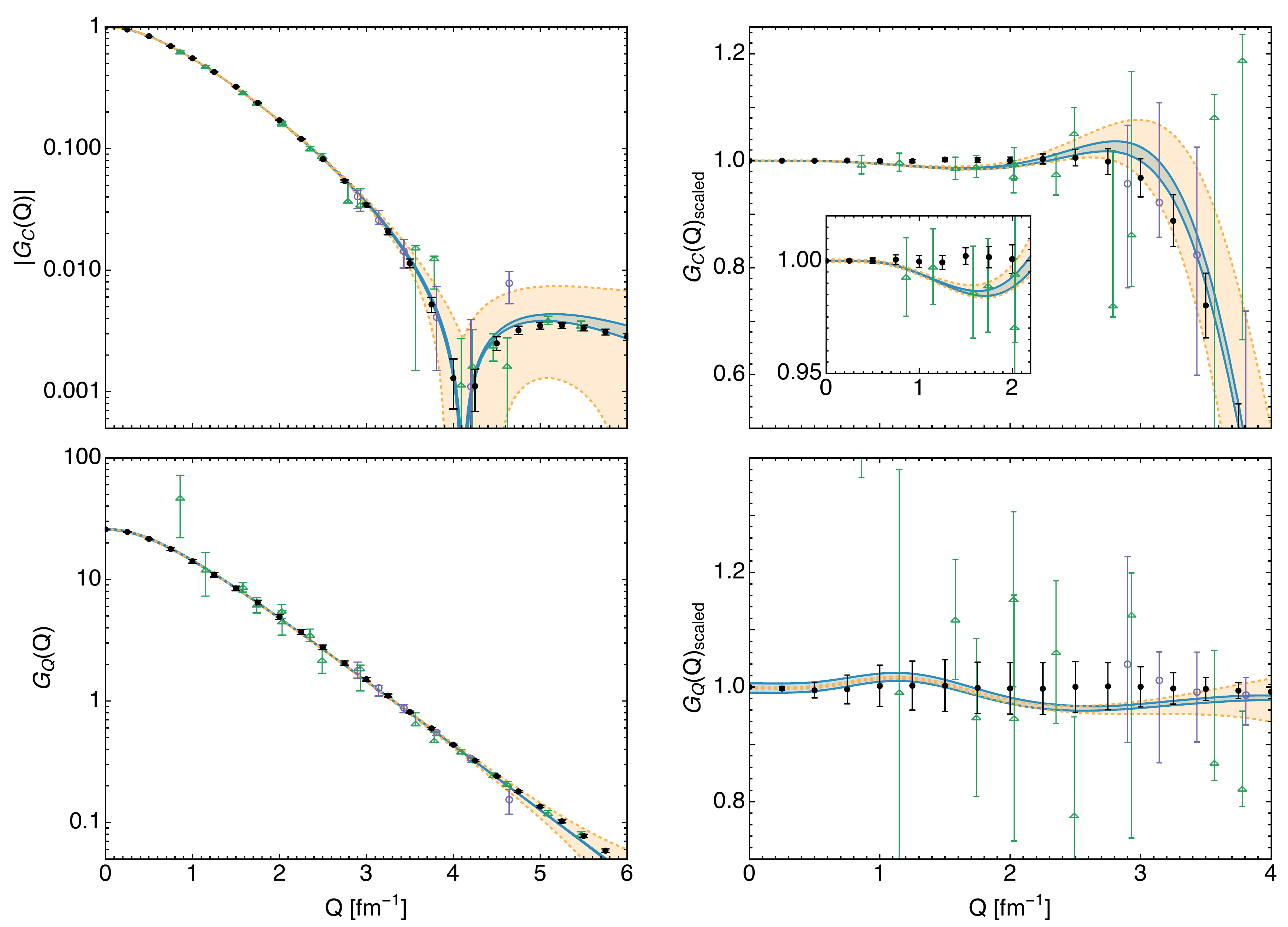}
\caption{\label{fig:GCQ_CutoffDependence}
 (Color online)
 Residual cutoff dependence versus the truncation error for the deuteron charge
 and quadrupole FFs at  N$^4$LO.\@
 Light-shaded blue bands between two solid lines correspond
 to the cutoff variation in the range of $\Lambda = 400 \ldots 550$~MeV.
 For remaining notation see Fig.~\ref{fig:GC_GQfinal}.
}
\end{figure}

We are now in the position to perform several consistency checks of our calculations.

As already pointed out in Sec.~\ref{Sec:twoN}, the two-body charge
density from OPE is proportional to unobservable
unitary-transformation parameters $\bar{\beta}_8$ and
$\bar{\beta}_9$. The observables must be independent of these
parameters at least approximately, i.e.~up to higher order effects.
The results presented above are based on   the minimal nonlocality
choice, Eq.~\eqref{Eq:minNON}, which is consistent  with the
employed chiral NN potentials of Ref.~\cite{Reinert:2020mcu} and also
with their predecessors from Ref.~\cite{Reinert:2017usi}.
To check the sensitivity of the deuteron FFs to $\bar{\beta}_8$ and $\bar{\beta}_9$,
we developed an approximately  phase-equivalent
version of the 2N potential using a different choice of the
unobservable phases, namely $\bar{\beta}_8=  \bar{\beta}_9=1/2$, by
re-doing the fit of NN data using exactly the same protocol as in
Ref.~\cite{Reinert:2020mcu}.
For this particular choice of $\bar{\beta}_8$, $\bar{\beta}_9 $,  the OPE
contribution to the charge density vanishes exactly: $\rho^{1\pi}_\text{2N} = 0$.
Repeating the fits of the calculated deuteron FFs to  the world data, we find for the central values
\begin{eqnarray}\label{rsrt_and_Q_beta89}
  r_{\rm str}^2= 3.8926\ \text{fm}^2, \quad \quad
  Q_d =  0.2849 \ \text{fm}^2, 
\end{eqnarray}
which should be compared with the values in Table~\ref{Tab:uncert}.
 As expected, the dependence on $\bar{\beta}_8$ and $\bar{\beta}_9$ for $r_{\rm str}^2$
 turns out to be very small, that is much smaller than the truncation
 error of the chiral expansion at N$^4$LO. For the quadrupole moment,
 the  dependence on these parameters is also consistent
 with   the truncation error.   Note that to achieve the independence of
$\bar{\beta}_8$, $\bar{\beta}_9 $ to such a high degree,
we found it to be crucial for  the nucleon FFs to be included not only
in the one-body  but also in the two-nucleon OPE charge density.
This can be  understood as follows:  as discussed in Sec.~\ref{Sec:twoN},
the derivation of the two-nucleon charge density operators relies on taking
the commutator of the leading one-body charge density with the
generators of the unitary transformation.
Because the one-body density is proportional to the nucleon FF,
the same should also hold for the two-body densities.
If one  neglects the nucleon FFs in the  OPE charge density,
a sizable violation of the $\bar{\beta}_8$, $\bar{\beta}_9$
independence would immediately reveal itself in the deuteron quantities.
Specifically, in this case one gets  $r_{\rm str}^2= 3.8825\ \text{fm}^2$ and
$Q_d =  0.2804 \ \text{fm}^2$, and one sees that the difference with
the values given in Table~\ref{Tab:uncert}  exceeds the truncation
error significantly.   The effect on these quantities of neglecting the nucleon FFs in
the short-range two-body charge-density operator at N$^4$LO is of
basically the same size.  Also, we would  like to emphasize that to observe $\bar{\beta}_8$, $\bar{\beta}_9 $ independence
in a  large range of momentum transfers it is important to follow the procedure, as described above:  
first, construct the phase-equivalent NN potentials for some choice of  $\bar{\beta}_8$, $\bar{\beta}_9$ by fitting NN data 
and then calculate corresponding deuteron wave functions.   If one applies the unitary   transformation to the existing wave function, 
then the unitary equivalence will hold only at small momentum transfers, as shown in Ref.~\cite{Adam:1993zz}.

Since the chiral expansion for the deuteron FFs is expected to converge
more rapidly  for not too soft values of the cutoffs,
our central results are obtained  for the cutoff $\Lambda=500$ MeV,
for which we also carried out a detailed error analysis as described in
the previous sections.
In Fig.~\ref{fig:GCQ_CutoffDependence}, as a consistency check,  we
confront the cutoff dependence of  $ \GC (Q)$ and $\GQ (Q)$
from the variation of the cutoff  from 400 to 550 MeV with the
truncation error.  We conclude that for  $\GC (Q)$, the cutoff
dependence lies well within the truncation error, while for the
quadrupole FF they are essentially compatible with each other
except for the region of small $Q$ where the cutoff dependence is a little larger.
We remind the reader, that the truncation error corresponds to the  $68\%$  DOB interval.
\begin{figure}
\includegraphics[width=\textwidth]{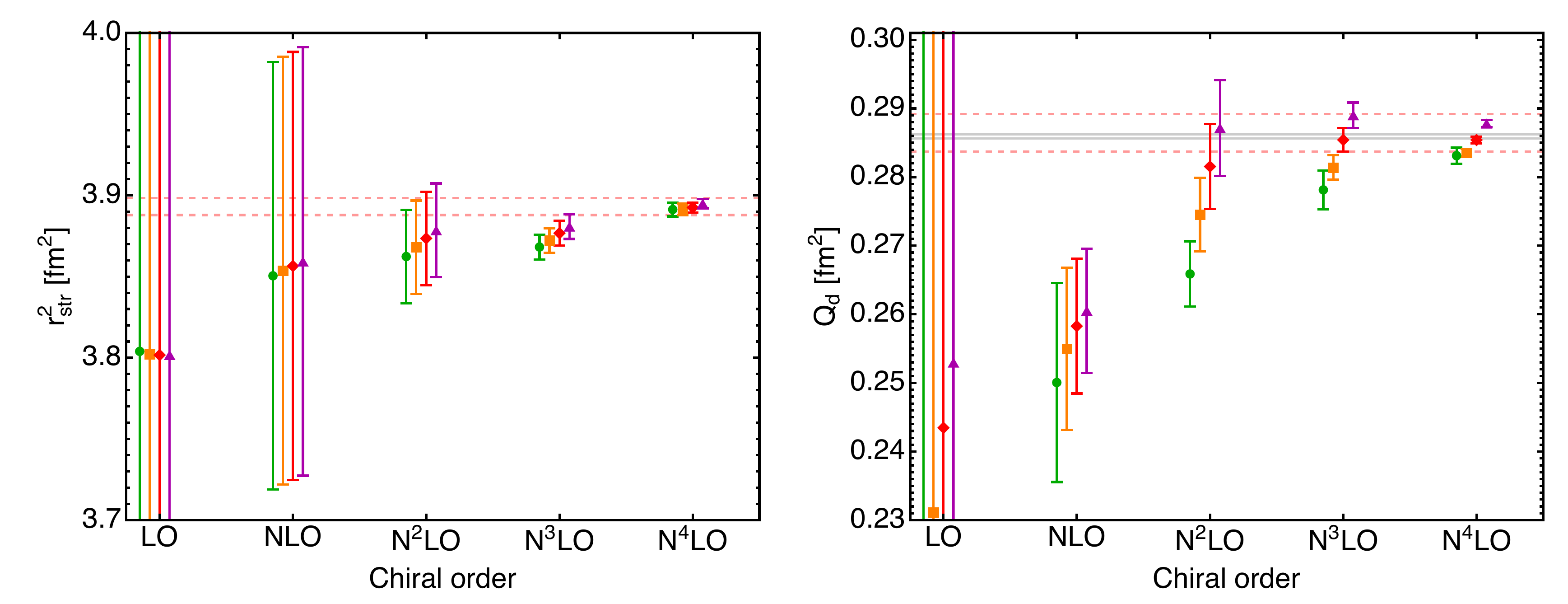}\hspace{1cm}
\caption{\label{fig:converg_rstrsq_Q}
  (Color online) Convergence of the chiral expansion for the deuteron structure radius squared (left panel) and the quadrupole moment (right panel).  Error bars correspond to the truncation errors in a given chiral order.
  Green circles stand for the results for $\Lambda=400$ MeV, orange squares    for $\Lambda=450$ MeV,  red diamonds  for   $\Lambda=500$ MeV, and purple triangles for $\Lambda=550$ MeV.
  The band bounded by solid horizontal gray lines on the right panel correspond to the quadrupole moment extracted in Refs.~\cite{Ericson:1982ei,Bishop:1979zz}.
  The bands bounded by the dashed horizontal red lines on both panels
  correspond to the \emph{total} uncertainties of our central results as given in Table~\ref{Tab:uncert}.
}
\end{figure}
 In Fig.~\ref{fig:converg_rstrsq_Q}, we show the convergence pattern
 of the chiral expansion for $ r_{\rm str}^2$ and $Q_d$
 along with the truncation errors and the cutoff dependence.
 While the results for $r_{\rm str}^2$ converge quite rapidly for any
 cutoff value, the quadrupole moment, in line with the discussion
 above,  shows a  lower rate of convergence. We further emphasize that
 the uncertainty of our result for the quadrupole moment at the
 highest considered order is dominated by the statistical errors in
 the NN LECs and by the uncertainty associated with the choice of the
 $Q$-range in the fit. Both of these error sources are considerably
 larger than the truncation uncertainty.

\subsection{Role of the individual contributions at N$^4$LO}\label{Sec:contribs}

The role of the individual charge density contributions to  the deuteron charge and quadrupole form factors   calculated   at N$^4$LO can be seen in  Tables~\ref{Tab:GC_contrib} and~\ref{Tab:GQ_contrib}.
Unlike the results in Fig.~\ref{fig:GC_GQ_convergence}, which
illustrate the convergence of the chiral expansion, all contributions
in Tables~\ref{Tab:GC_contrib} and~\ref{Tab:GQ_contrib}  were
evaluated using the deuteron wave function at N$^4$LO$^{+}$.
The results for the one-pion-exchange charge density were obtained
using the minimal nonlocality choice for the parameters
$\bar{\beta}_8$ and $\bar{\beta}_9$ from Eq.~\eqref{Eq:minNON}.
For the charge FF,  the most important correction beyond the main one stems from the CT  contribution  which dominates  in the whole domain of momenta $Q$ considered apart
from  the region of small $Q$ ($\lesssim$ 1 fm$^{-1}$), where the Darwin term is equally important.
Next in importance are the $1\pi$ and boost corrections which,
however, cancel each other to a large extent. The contribution from
the SO is basically negligible.
For the quadrupole FF at low $Q$ ($\lesssim$ 1 fm$^{-1}$), the dominant
corrections beyond the main term
originate from the $1\pi$, CT and SO contributions, in the order of
their importance, where the first two interfere constructively while the SO
is destructive. While the boost correction is negligible for all $Q$-values,
the Darwin term, being negligible at small momentum transfers, provides a sizable
contribution for $Q>1$~fm.

\begin{table}[t]
\caption{\label{Tab:GC_contrib} Impact of the individual contributions
  to the charge density operator on the charge form factor of the deuteron $\GC(Q^2)$ at N$^4$LO for the cutoff $\Lambda=500$ MeV.
  }
\small
\begin{ruledtabular}
\begin{tabular*}{\textwidth}{@{\extracolsep{\fill}}crrrrrrr}
{Q [fm$^{-1}$]} & {Main} & {SO} & {Darwin} & {Boost} & {$1\pi$} & {CT} & {Full} \\[3pt]
\hline
$0.0$       & $1.0000$          & $0.0000$        & $0.0000$            & $0.0000$           & $0.0000$         & $0.0000$        & $1.0000$          \\
$0.5$      & $0.8416$      & $0.0001$    & $-0.0012$       & $0.0001$       & $-0.0004$    & $-0.0005$   & $0.8397$      \\
$1.0$       & $0.5556$      & $0.0002$    & $-0.0031$       & $0.0006$       & $-0.0015$    & $-0.0017$   & $0.5501$      \\
$2.0$       & $0.1795$      & $0.0004$    & $-0.0040$        & $0.0018$       & $-0.0034$    & $-0.0047$   & $0.1696$      \\
$3.0$       & $0.0460$       & $0.0003$    & $-0.0023$       & $0.0019$       & $-0.0034$    & $-0.0062$   & $0.0363$      \\
$4.0$       & $0.0087$      & $0.0002$    & $-0.0008$       & $0.0012$       & $-0.0023$    & $-0.0056$   & $0.0014$      \\
$5.0$       & $0.0004$      & $0.0001$    & $-0.0001$       & $0.0005$       & $-0.0011$    & $-0.0041$   & $-0.0043$     \\
$6.0$       & $-0.0007$     & $0.0000$    & $0.0001$        & $0.0001$       & $-0.0004$    & $-0.0026$   & $-0.0034$     \\
\end{tabular*}
\end{ruledtabular}
\end{table}
 \begin{table}[t]
\caption{\label{Tab:GQ_contrib} Impact of the individual contributions
  to the charge density operator on the quadrupole form factor of the
  deuteron ($\GQ (Q^2)/m_d^2$) at N$^4$LO  for the cutoff $\Lambda=500$ MeV.
  The values are given in fm$^{2}$.
  }
\small
\begin{ruledtabular}
\begin{tabular*}{\textwidth}{@{\extracolsep{\fill}}crrrrrrr}
{Q [fm$^{-1}$]} & {Main} & {SO} & {Darwin} & {Boost} & {$1\pi$} & {CT} & {Full} \\[3pt]
\hline
$0.0$       & $0.2788$      & $-0.0018$   & $0.0000$         & $0.0000$           & $0.0063$     & $0.0022$    & $0.2854$      \\
$0.5$      & $0.2340$       & $-0.0017$   & $-0.0003$       & $-0.0001$      & $0.0059$     & $0.0021$    & $0.2399$      \\
$1.0$       & $0.1537$      & $-0.0013$   & $-0.0008$       & $-0.0003$      & $0.0050$      & $0.0019$    & $0.1582$      \\
$2.0$      & $0.0509$      & $-0.0006$   & $-0.0011$       & $-0.0001$      & $0.0027$     & $0.0013$    & $0.0531$      \\
$3.0$       & $0.0151$      & $-0.0002$   & $-0.0008$       & $0.0001$       & $0.0011$     & $0.0007$    & $0.0161$      \\
$4.0$       & $0.0043$      & $-0.0001$   & $-0.0004$       & $0.0001$       & $0.0004$     & $0.0004$    & $0.0048$      \\
$5.0$       & $0.0012$      & $0.0000$        & $-0.0002$       & $0.0001$       & $0.0001$     & $0.0002$    & $0.0014$      \\
$6.0$       & $0.0003$      & $0.0000$        & $-0.0001$       & $0.0001$       & $0.0000$         & $0.0001$    & $0.0004$      \\
\end{tabular*}
\end{ruledtabular}
\end{table}

\section{Summary and conclusions}\label{sec:summary}

In spite of the extensive progress in the understanding of the deuteron
structure that has been achieved since more than 50 years,
there is still strong motivation to reanalyze the deuteron form factors
in the framework of chiral EFT.
Being largely governed by the  leading-order single-nucleon charge density,
the  charge and quadrupole deuteron form factors (FFs) are
qualitatively described  in most of the calculations reported in the literature
(at least in some range  of the momentum transfer).
However, as long as higher-order corrections are concerned,   the
existing calculations show lack of  systematics, consistency and
controlled error  estimate.

In this paper, the deuteron charge and quadrupole form factors are calculated
using consistently regularized two-nucleon potentials and the charge density
 in chiral effective field theory.
This allowed to extract the important static properties of the
deuteron, namely  the structure radius and the quadrupole moment, with
unprecedented accuracy and to reliably estimate various sources of
uncertainty.
Our analysis provides a first step towards the understanding of
radii of medium-mass and heavy nuclei, which are currently known to be
significantly underpredicted.

The novel aspects of our study include:
\begin{itemize}
\item For the first time, the calculation of the deuteron FFs is
  pushed beyond N$^3$LO, which allows one to reduce the uncertainty from the
  truncation of chiral expansion and thus to extend the range of
 momenta considered.  To achieve this goal we  (i) employed the most recent
two-nucleon potentials up through N$^4$LO$^+$~\cite{Reinert:2020mcu}, which utilize a
complete treatment of isospin-breaking effects and provide a
statistically perfect description of NN data below pion production
threshold and (ii) implemented the charge density operator at N$^3$LO, supplemented with the 2N short-range operators at N$^4$LO.
\item  Regularization of the charge density operators is  carried out
  consistently  with the  two-nucleon potential using the same
unitarity transformations for the charge density operators and the
nuclear forces.
Specifically, the two-nucleon charge density operators are generated by
taking the commutator of the leading one-body charge density with the
 generators of the unitary transformation that incorporate the regulator as discussed in
Sec.~\ref{Sec:twoN}.
As a consistency check, we have demonstrated that
the residual cutoff dependence of the
deuteron charge FF and the extracted structure radius is much weaker
than the error estimated from the
truncation of the chiral expansion at   N$^4$LO.
 The cutoff dependence of the quadrupole moment (and in general of
 the quadrupole FF at low $Q$) at the highest considered order is
 somewhat larger than the estimated truncation error, but still of the
 same size as the total uncertainty of our result. Furthermore, the
 short-range charge density operators contributing to the charge and
 quadrupole FFs of the deuteron come out of a natural size.

\item  Instead of relying on the strict chiral expansion of the nucleon
    FFs known to converge slowly, we employed
    the most up-to-date phenomenological parametrizations of  experimental data from
    the global analysis of
    Refs.~\cite{Ye:2017gyb,Ye:smallrp}.
   The nucleon form factors from the dispersive approach of Ref.~\cite{Belushkin:2006qa}
     have also been used as a consistency check.
     We emphasize that making a reliable calculation of the
   deuteron FFs  requires the inclusion of the nucleon FFs both  in
   the one- and two-nucleon charge density operators, the feature that
   becomes obvious in the way we generate the two-body charge density
   by means of the unitary transformation.
We have verified this conclusion by explicitly checking the
insensitivity of our results for the FFs to the choice of
unobservable unitary phases $\bar{\beta}_8$ and
$\bar{\beta}_9$, which holds true to a very high degree of accuracy
when keeping the nucleon FFs in the OPE charge density.
The same conclusion applies when the nucleon FFs are neglected
 in the contact two-nucleon charge density at N$^4$LO.

  \item A comprehensive and systematic  analysis of various sources of
    uncertainties in the calculated deuteron FFs is performed.
    Specifically, we estimated the uncertainty from (i) propagating
    the statistical errors
  of the $\pi$N and NN low-energy constants (LECs)  entering the two-nucleon potentials
   (ii)  truncation of the chiral expansion evaluated using
   Bayesian methods  (iii)  statistical uncertainties in the N$^4$LO
   short-range charge density operators, (iv)
    employed parametrizations of the nucleon FFs and (v)
   fixing the maximum value of the momentum
    transfers $Q_\text{max}$ in the fits of the short-range charge
    operators.
\end{itemize}

Pushing the calculation to N$^4$LO and   using  the consistently
regularized charge density operators together with
the phenomenological nucleon form factors is found to
result in a very good description of the  deuteron form factors  at
least  up to $Q \simeq 6$ fm$^{-1}$.  Having adjusted the  two short-range
operators  to achieve the best fit of the world data on the charge and
quadrupole FFs of the deuteron,  we predict the deuteron structure
radius and quadrupole moment to have the values of
 \begin{equation}
  r_{\rm str} = 1.9729 \substack{+0.0015\\ -0.0012}\ \text{fm},\quad \quad
  Q_d =  0.2854 \substack{+0.0038\\ -0.0017}\ \text{fm}^2.  
\end{equation}
 Equipped with this  prediction for the  structure radius, we employ the
 high-accuracy data for the hydrogen-deuterium isotope shift in
 Eq.~(\ref{Eq:rd-rp}) to
 extract  the mean-square neutron charge radius, for which we obtain
\begin{eqnarray}
	r_n^2 = -0.105 \substack{+0.005\\ -0.006}  \ \text{fm}^2.
\end{eqnarray}
This result is consistent with our previous determination in Ref.~\cite{Filin:2019eoe} but
deviates by about $1.9\sigma$ from the current value $r_{n}^2 = -0.1161(22) \text{ fm}^2$ given by the Particle Data Group~\cite{Tanabashi:2018oca}
 and deviates by about $1.4\sigma$ from the very recent determination
 $r_n^2 = -0.122 \pm 0.004_{\rm (stat.)}  \pm 0.010_{\rm (syst.)}$~fm$^2$
 from the collective analysis of the nucleon form factors of Ref.~\cite{Atac:2020hdq}.

\begin{acknowledgments} 

 We are  grateful to U.-G.~Mei\ss ner for a careful reading of the manuscript and valuable comments and to Z.~Ye for providing us with the unpublished results for the nucleon form factors from Ref.~\cite{Ye:smallrp}.
 We also thank H.-W.~Hammer for providing us with the parametrization of the nucleon form factors from Ref.~\cite{Belushkin:2006qa} and   I.~Sick for  the parametrization of the deuteron form factors
from Ref.~\cite{Marcucci:2015rca}.  We are grateful to M.~Hoferichter and J.~Ruiz de Elvira   for the information on the central values and covariance matrix of the N$^3$LO $\pi$N LECs from the Roy-Steiner analysis.
This work was supported in part by DFG and NSFC through funds provided to the Sino-German CRC 110 ``Symmetries and
 the Emergence of Structure in QCD'' (NSFC Grant No. 11621131001, Grant No.~TRR110), the BMBF (Grant No. 05P18PCFP1)  and the Russian Science Foundation (Grant No.~18-12-00226).

\end{acknowledgments}

\appendix

\section{Analytic expressions for the contributions to $\GC$, $\GQ$,  $r_\text{str}^2$ and $Q_d$.}\label{sec:analyticexpressions}

In this appendix we list the analytic expressions for individual
contributions to the deuteron charge form factor $\GC$ (Eq.~\eqref{eq:GCcontributions}),
quadrupole form factor $\GQ$ (Eq.~\eqref{eq:GQcontributions}), structure radius squared $r_\text{str}^2$
(Eqs.~\eqref{Eq:r_str},~\eqref{eq:deutchargeradiuscontributions})
and quadrupole moment $Q_d$ (Eq.~\eqref{eq:deutqaudmomcontributions}).
Results are given in momentum space or in coordinate space depending
on which form is simpler for practical calculations.
All contributions are grouped according to the charge density operator which they are obtained from.
Specifically, we distinguish the following types of contributions:
main (LO) contributions, Darwin-Foldy-type contributions, spin-orbit corrections,
deuteron boost corrections, pion-exchange current contributions,
and contact contributions.
Results are expressed in terms of deuteron wave functions and single-nucleon form factors.
The deuteron WFs are normalized according to:
\begin{eqnarray}
  \int\limits_{0}^{\infty}  p^2 \left( {u(p)}^2 + {w(p)}^2 \right)  dp =
  \int\limits_{0}^{\infty} \left( {u(r)}^2 + {w(r)}^2 \right)     dr = 1,
\end{eqnarray}
and the deuteron $D$-state probability is:
\begin{eqnarray}
  P_D = \int\limits_{0}^{\infty}  p^2 {w(p)}^2 dp =
  \int\limits_{0}^{\infty} {w(r)}^2    dr.
\end{eqnarray}
We also introduce the following common combinations of the deuteron wave functions:
\begin{eqnarray}
  C(r) \equiv u^2 (r) + w^2(r),
  \qquad
  Q(r) \equiv u(r) w(r) - \frac{w^2(r)}{\sqrt{8}}.
\end{eqnarray}
Note that all momenta in this section are three-dimensional. For a
vector $ \bm{x}$, we use $x$ to denote  $ x \equiv |\bm{x}|$.

\subsection{Main (LO) contributions}

Main contributions stem from the LO charge density operator in Eq.~\eqref{eq:rho1Nmain};
\begin{eqnarray}
  \GC^\text{Main}(\bm{k}^2) = \GE^S(\bm{k}^2) \GC^\text{matter}(\bm{k}^2),
  \qquad
  \GQ^\text{Main}(\bm{k}^2)     =    \GE^S(\bm{k}^2)   \GQ^{(0)}(\bm{k}^2),
\end{eqnarray}
where we introduced the following auxiliary functions:
\begin{eqnarray}
    \GC^\text{matter}(\bm{k}^2) \equiv \int_0^\infty
    C(r)
  j_0 \left( \frac{k r}{2} \right)   dr,
 \qquad
 \qquad
    \GQ^{(0)}(\bm{k}^2) \equiv \frac{6 \sqrt{2} m_d^2}{\bm{k}^2}
     \int_0^\infty
     Q(r)
      j_2 \left( \frac{k r}{2} \right)   dr.
\end{eqnarray}
The LO contribution to the deuteron structure radius (so called deuteron matter radius) reads:
\begin{eqnarray}
  r_{m}^2 = r_0^2 = \frac14 \int\limits_{0}^{\infty} \left(
    p^2 \left(u'{(p)}^2+w'{(p)}^2\right)  + 6 w{(p)}^2
  \right)
  dp
  = \frac14 \int_0^\infty \left( {u(r)}^2 + {w(r)}^2 \right)  r^2   dr.
\end{eqnarray}
The LO contribution to the deuteron quadrupole moment reads:
\begin{eqnarray}
    Q_{0} = \int\limits_{0}^{\infty} \left(
    \frac{p^2 u'(p) w'(p)}{5 \sqrt{2}}-\frac{1}{20} p^2 w'{(p)}^2
     +\frac{3 p w(p) u'(p)}{5 \sqrt{2}}-\frac{3 {w(p)}^2}{10}
   \right)
  dp
     = \frac{\sqrt{2}}{10}
     \int_0^\infty
     Q(r)
        r^2    dr.
\end{eqnarray}

\subsection{Darwin-Foldy-type of contributions}

The Darwin-Foldy-type (DF) of contributions stem from the
charge density operator $\rho_\text{1N}^\text{DF}$ in Eq.~\eqref{eq:rho1Nmain}.
Since the DF charge density operator differs from the LO operator only by a pre-factor,
the resulting DF contributions to the form factors are trivially related to the LO ones.
Specifically, the DF contributions to the deuteron charge and quadrupole form factors read:
\begin{eqnarray}
  \GC^\text{DF}(\bm{k}^2)  = \GE^S(\bm{k}^2) \left( - \frac{\bm{k}^2}{8 \mN^2} \right)   \GC^\text{matter}(\bm{k}^2),
 \qquad
 \qquad
  \GQ^\text{DF}(\bm{k}^2)  =  \GE^S(\bm{k}^2) \left( - \frac{\bm{k}^2}{8 \mN^2} \right)
       \GQ^{(0)}(\bm{k}^2).
\end{eqnarray}
Deuteron \emph{structure} radius does, by definition, exclude the Darwin-Foldy contribution,
but the deuteron \emph{charge} radius receives a constant correction
$r_{DF}^2 = 3/(4 m_p^2) = 0.03317 \, \text{fm}^2$,
where $m_p$ is a proton mass.
Finally, the Darwin-Foldy term does not contribute to the deuteron
quadrupole moment since it is proportional to the photon momentum $k$,
while the quadrupole moment is defined at $k=0$.

\subsection{Spin-orbit contributions}

The spin-orbit contributions to the deuteron form factors stemming from the charge density operator
$\rho_\text{1N}^\text{SO}$ of Eq.~\eqref{eq:rho1Nmain} read:
\begin{eqnarray}
  \GC^\text{SO}(\bm{k}^2) = \left( \GE^S(\bm{k}^2)-2\GMNucl^S(\bm{k}^2) \right) \GC^\text{ang}(\bm{k}^2),
  \qquad
  \qquad
    \GQ^\text{SO}(\bm{k}^2) =
     \left( \GE^S(\bm{k}^2)-2\GMNucl^S(\bm{k}^2) \right)
     \GQ^\text{ang}(\bm{k}^2),
\end{eqnarray}
where
\begin{eqnarray}
  \GC^\text{ang}(\bm{k}^2) \equiv
     \frac{3}{2 \mN^2} \int_0^\infty
     \frac{{w(r)}^2}{r}
     \frac{\partial}{\partial r} \left( j_0 \left( \frac{k r}{2} \right)    \right)
     dr
   =
     -\frac{3 k}{4 \mN^2}
     \int_0^\infty
     \frac{{w(r)}^2}{r}
     j_1 \left( \frac{k r}{2} \right)
     dr,
  \\
    \GQ^\text{ang}(\bm{k}^2) \equiv
    (-1) \frac{6}{\sqrt{2} \bm{k}^2}
    \frac{3 m_d^2}{\mN^2}
    \int_0^\infty
      w(r) \left( \frac{\partial}{\partial r}\left( \frac{u(r)}{r} \right)
                - \frac{1}{\sqrt{2}} \frac{1}{r} \frac{\partial w(r)}{\partial r}  \right)
      j_2 \left( \frac{k r}{2} \right)
      dr.
\end{eqnarray}
The corresponding contributions to the deuteron structure radius and quadrupole moment read:
\begin{eqnarray}
  r_{\text{SO}}^2 = - \frac{3}{4 \mN^2} (2 \mu_n+2 \mu_p-1)   P_D,
  \qquad
  Q_{\text{SO}} &=& (1-2 \mu_n-2 \mu_p)   Q_{\text{angular}},
\end{eqnarray}
where $\mu_p$ and $\mu_n$ are the magnetic moments of the proton and the neutron, respectively, in units of nuclear magnetons, and
\begin{eqnarray}
     Q_{\text{angular}}  \equiv   (-1) \frac{6}{\sqrt{2}}
    \frac{3}{\mN^2} \int_0^\infty
      w(r) \left( \frac{\partial}{\partial r}\left( \frac{u(r)}{r} \right)
                - \frac{1}{\sqrt{2}} \frac{1}{r} \frac{\partial w(r)}{\partial r}  \right)
       \frac{r^2}{60}
      dr.
\end{eqnarray}

\subsection{Boost corrections}

Corrections to the deuteron form factors which appear due to the motion of initial and final deuterons
are discussed in the Section~\ref{sec:relcorr}. The final expressions
for the boost corrections to the charge and quadrupole form factors have the form:
\begin{eqnarray}
   \GC^\text{Boost}(\bm{k}^2) &=&  \GE^S(\bm{k}^2)  \GC^\text{ang}(\bm{k}^2)
  + \GE^S(\bm{k}^2) \left( \GC^\text{matter}(\bm{k}_\text{boosted}^2) - \GC^\text{matter}(\bm{k}^2) \right),
  \\
    \GQ^\text{Boost}(\bm{k}^2) &=&  \GE^S(\bm{k}^2)
     \GQ^\text{ang}(\bm{k}^2)
  + \GE^S(\bm{k}^2)
    \left( \GQ^{(0)\text{Boosted}}(\bm{k}^2)  -  \GQ^{(0)}(\bm{k}^2) \right),
\end{eqnarray}
where the boosted momentum $k_\text{boosted}$ is defined by Eq.~\eqref{eq:kboosted} and
the boosted version of $\GQ^{(0)}$ is
\begin{eqnarray}
    \GQ^{(0)\text{Boosted}}(\bm{k}^2) \equiv \frac{6 \sqrt{2} m_d^2}{\bm{k}^2}
     \int_0^\infty w(r) \left( {u(r)} - \frac{w(r)}{2\sqrt{2}} \right)
      j_2 \left( \frac{k_\text{boosted} r}{2} \right)   dr.
\end{eqnarray}
Boost corrections do not contribute to the deuteron structure radius and quadrupole moment.

\subsection{One-pion-exchange contributions}

One-pion-exchange (OPE) contributions to the deuteron form factors
originate from the charge density operator given by Eq.~\eqref{eq:NNchargeDensityKollingWithGESregularized}.
In momentum space, the expressions for the OPE contributions involve six-dimensional integration
and are somewhat cumbersome.  The Fourier transform to coordinate space
makes these expressions much shorter and the number of integrations reduces to one.
Below we give the OPE contributions in coordinate space.
For the sake of compactness, we introduce the functions $\bar{h}_1(x)$ and $\bar{h}_2(x)$
that correspond to the Fourier transforms of the regularized single and squared pion propagators,
respectively:
\begin{eqnarray}
  \bar{h}_1(r) \equiv \int \frac{d^3 l}{{(2 \pi)}^3} \frac{F_1 (\bm{l}^2, \Lambda) e^{i \bm{l} \cdot \bm{r}}}{\bm{l}^2+\mpi^2},
  \qquad
  \bar{h}_2(r) \equiv \int \frac{d^3 l}{{(2 \pi)}^3} \frac{F_2 (\bm{l}^2, \Lambda) e^{i \bm{l} \cdot \bm{r}}}{{(\bm{l}^2+\mpi^2)}^2},
\end{eqnarray}
where $F_{1} (\bm{l}^2, \Lambda)$ and $F_{2} (\bm{l}^2, \Lambda)$ are
the corresponding momentum-space regulators.
Without regularization (i.e.~when $F_{1}(\bm{l}^2, \Lambda)=F_{2}(\bm{l}^2, \Lambda)=1$),
the functions $\bar{h}_1(r)$ and $\bar{h}_2(r)$ take a simple form:
\begin{eqnarray}
  \label{eq:hnonreg}
  \bar{h}_1^\text{unreg}(r) = \frac{e^{- \mpi r}}{4 \pi r},
  \qquad
  \bar{h}_{2}^{\text{unreg}}(r) = \frac{e^{- \mpi r}}{8 \pi \mpi}.
\end{eqnarray}
For the regulator employed in the SMS NN potentials of
Ref.~\cite{Reinert:2017usi} with
\begin{eqnarray}
  F^{\text{SMS}}_{1} (\bm{l}^2, \Lambda) \equiv \exp \left( \frac{\bm{l}^2+\mpi^2}{  \Lambda^2} \right),
  \qquad
  F^{\text{SMS}}_{2} (\bm{l}^2, \Lambda) \equiv \exp \left( \frac{\bm{l}^2+\mpi^2}{  \Lambda^2} \right)  \left[ 1 + \frac{\bm{l}^2+\mpi^2}{  \Lambda^2} \right]
\end{eqnarray}
 we get the following closed form of the function $\bar{h}_1(r)$
\begin{eqnarray}
  \label{eq:hsms52}
  \bar{h}^\text{SMS}_{1}(r) &=&
  \frac{\exp (-\mpi r) \, \text{erfc}\left(\frac{\mpi}{\Lambda }-\frac{\Lambda  r}{2}\right)
       -\exp (\mpi r) \, \text{erfc}\left(\frac{\mpi}{\Lambda }+\frac{\Lambda  r}{2}\right)}
  {8 \pi r}.
\end{eqnarray}
The function $\bar{h}_2$ enters the final result only under a derivative operator.
To simplify the expressions even further we rewrite $\bar{h}'_2(r)$ in terms of $\bar{h}_1(r)$.
We employ the relation
\begin{eqnarray}
  \frac{\bm{l} }{ {(\bm{l}^2 + \mpi^2)}^2 } F_{2} (\bm{l}^2, \Lambda)
  = -\frac{1}{2} \bm{\nabla}_l \left( \frac{1}{\bm{l}^2 + \mpi^2} F_{1} (\bm{l}^2, \Lambda) \right),
  \label{eq:relationRegProp}
\end{eqnarray}
which is fulfilled by both the unregularized and SMS-regularized pion propagators.
Substituting the relation in Eq.~\eqref{eq:relationRegProp} in the
definition of $\bar{h}_2$, taking the derivative and integrating by parts
leads to the following relation in coordinate space:
\begin{eqnarray}
  \label{eq:relationh1h2}
  \bar{h}_2'(r) = \left( -\frac{r}{2} \right) \bar{h}_1(r).
\end{eqnarray}
Using the simplifications above, the OPE contribution to the deuteron charge form factor can be written as:
\begin{eqnarray}
  \GC^{1\pi}(\bm{k}^2) &=&
    \GE^S(\bm{k}^2)
  \frac{\gA^2}{16 \fpi^2 \mN}
  \int\limits_{0}^{\infty} dr
  \Bigg(
  (2 \bar{\beta}_8-1)
 k j_1\left(\frac{k r}{2}\right) \left(C(r) \left(r \bar{h}_{1}''(r)+4 \bar{h}_{1}'(r)\right)+4 \sqrt{2} Q(r) \left(r \bar{h}_{1}''(r)+\bar{h}_{1}'(r)\right)\right)
  \nonumber
  \\
  &&+ (1-2 \bar{\beta}_9)
  k j_1\left(\frac{k r}{2}\right) \left(C(r)+4 \sqrt{2} Q(r)\right) \bar{h}_{1}'(r)
  \Bigg),
\end{eqnarray}
where $j_n(x)$ are the spherical Bessel functions.
The OPE contribution to the deuteron quadrupole form factor reads:
\begin{eqnarray}
  \GQ^{1\pi}(\bm{k}^2) &=&
    \GE^S(\bm{k}^2)
  \frac{ \gA^2 m_d^2}{16 \fpi^2 \mN}
  \int\limits_{0}^{\infty} dr
  \Bigg\{
  (2 \bar{\beta}_8-1)
  \nonumber
  \\
  &&\times
  \Bigg(
    \frac{36}{k^2 r} j_2\left(\frac{k r}{2}\right) \left(-2 C(r) \left(\bar{h}_{1}'(r)-r \bar{h}_{1}''(r)\right)+\sqrt{2} Q(r) \left(4 \bar{h}_{1}'(r)-r \bar{h}_{1}''(r)\right)+9 w{(r)}^2 \bar{h}_{1}'(r)\right)
    \nonumber
    \\
    && -
    \frac{6}{k} j_1\left(\frac{k r}{2}\right) \left(2 C(r) \left(r \bar{h}_{1}''(r)+\bar{h}_{1}'(r)\right)+\sqrt{2} Q(r) \left(2 \bar{h}_{1}'(r)-r \bar{h}_{1}''(r)\right)\right)
    \Bigg)
  \nonumber
  \\
  &&+(1-2 \bar{\beta}_9)
  \left(
    \frac{324}{k^2 r} j_2\left(\frac{k r}{2} \right) w{(r)}^2 \bar{h}_{1}'(r)
  - \frac{24}{k} j_1\left(\frac{k r}{2}\right)
  \left(C(r)-\frac{Q(r)}{\sqrt{2}}\right) \bar{h}_{1}'(r) \right)
  \Bigg\}.
\end{eqnarray}
Finally, the OPE contributions to the deuteron structure radius and
quadrupole moment have the form:
\begin{eqnarray}
  r^2_{1\pi} &=&
  -\frac{\gA^2}{16 \fpi^2 \mN} \int\limits_{0}^{\infty} dr \, r
  \Bigg(
  (2 \bar{\beta}_8-1)
    \left(C(r) \left(r \bar{h}_{1}''(r)+4 \bar{h}_{1}'(r)\right)+4 \sqrt{2} Q(r) \left(r \bar{h}_{1}''(r)+\bar{h}_{1}'(r)\right)\right)
  \nonumber
  \\
  && + 2 (1-2 \bar{\beta}_9)
    \left(C(r)+4 \sqrt{2} Q(r)\right) \bar{h}_{1}'(r)
  \Bigg),
\\
  Q^{1\pi}
  &=&
  \frac{\gA^2}{16 \fpi^2 \mN} \frac{1}{5} \int\limits_{0}^{\infty} dr \, r
  \Bigg(
   (2 \bar{\beta}_8-1)
  \left(-4 C(r) \left(r \bar{h}_{1}''(r)+4 \bar{h}_{1}'(r)\right)
  +2 \sqrt{2} Q(r) \left(r \bar{h}_{1}''(r)+\bar{h}_{1}'(r)\right)+27 w{(r)}^2 \bar{h}_{1}'(r)\right)
  \nonumber
  \\
  &&-
   (1-2 \bar{\beta}_9)
  \bar{h}_{1}'(r) \left(20 C(r)-10 \sqrt{2} Q(r)-27 w{(r)}^2\right)
  \Bigg).
\end{eqnarray}
Our analytic expressions for OPE contributions agree with the ones of
Ref.~\cite{Friar:1979by} after the following notational changes are performed:
\begin{eqnarray}
  \bar{h}_1 \to \frac{\mpi}{4 \pi} h,
  \qquad
  \frac{\gA^2 \mpi^2}{16 \pi \fpi^2} \to f_0^2,
  \qquad
  \bar{\beta}_9 \to  \frac{\mu-1}{4}
  \qquad
  \bar{\beta}_8 \to  \frac{\nu}{2}.
\end{eqnarray}

\subsection{Contact charge density contributions}\label{subsec:contactGCGQ}

Contact N$^4$LO contributions to the deuteron form factors stem from
the corresponding short-range charge density operators in
Eq.~\eqref{eq:contactchargedensityReg}.
The contact contribution to the deuteron charge form factor is given by
\begin{eqnarray}
  \GC^\text{Cont}(\bm{k}^2) &=&
  \frac{1}{\pi^2}
  \GE^S(\bm{k}^2)
  \int\limits_{0}^{\infty} p^2 dp
  \int\limits_{0}^{\infty} p'^2 dp'
  F_{\Lambda} \left( p- \frac{k}{2}, p' \right)
  \nonumber
  \\
  &&\times
  \big[
  F_{\GC}^{uu}(p, p', k) u(p) u(p')
  + F_{\GC}^{uw}(p, p', k) w(p) u(p')
  \big]
  + (k \to -k),
  \label{eq:gccontExplicit}
\end{eqnarray}
where
\begin{eqnarray}
  F_{\Lambda} \left( p, p' \right) &=& \exp \left( - \frac{p^2+p'^2}{\Lambda^2} \right),
  \\
  F_{\GC}^{uu}(p, p', k) &=& \left( A+B+\frac{C}{3} \right)
    \frac{2  }{k p}
    \left(
         \Lambda ^4
        + \Lambda ^2  \left( {\left( p - \frac{k}{2}\right)}^2  -p'^2  \right)
    \right) ,
  \\
  F_{\GC}^{uw}(p, p', k) &=&
  \sqrt{2} C \left( \frac{\Lambda^6}{k p^3}
     +  \frac{\Lambda^4 \left( 4p-3k \right) }{3 k p^2}
    +  \frac{\Lambda^2 \left( k-4p \right) }{3 p}
      \right) ,
\end{eqnarray}
and $(k \to -k)$ means that the same contribution, but with opposite sign of $k$ should be added.
The contact contribution to the deuteron quadrupole form factor reads
\begin{eqnarray}
  \GQ^\text{Cont}(\bm{k}^2) =
  \frac{m_d^2}{\pi^2}
  \GE^S(\bm{k}^2)
  \int\limits_{0}^{\infty} p^2 dp
  \int\limits_{0}^{\infty} p'^2 dp'
  F_{\Lambda} \left( p- \frac{k}{2}, p' \right)
  \big[
  F_{\GQ}^{uu}(p, p', k) u(p) u(p')
  \nonumber
  \\
  + F_{\GQ}^{uw}(p, p', k) w(p) u(p')
  + F_{\GQ}^{ww}(p, p', k) w(p) w(p')
  \big]
  + (k \to -k),
  \label{eq:gqcontExplicit}
\end{eqnarray}
where
\begin{eqnarray}
  F_{\GQ}^{uu}(p, p', k) &=& 
  (-1) C \frac{\Lambda ^2 }{2 k^5 p}
  \left(
      k^2 {\left( p - \frac{k}{2} \right)}^2 + k (k-3p) \Lambda^2 + 3 \Lambda^4
  \right),
  \nonumber
  \\
  F_{\GQ}^{uw}(p, p', k) &=&
    (A+B) \frac{(-3)}{\sqrt{2} k^5 p^3} \Bigg(
       k^2  p^2 \left((k-2 p)^2-4 p'^2\right) \Lambda ^2
       - k  p \left(3 k^2-16 k p+12 (p^2-p'^2) \right) \Lambda ^4
  \nonumber
       \\
       &&+3  \left(k^2-12 k p+4 (p^2-p'^2)\right) \Lambda ^6
       +36 \Lambda ^8
     \Bigg)
  \nonumber
  \\
    &&+ \frac{C}{\sqrt{2} k^5 p^3}
    \left(
    k^2  p^2 \left({(k-2 p)}^2+4 p'^2\right) \Lambda ^2
    - k  p \left(3 k^2-4 k p+12 \left(p^2+p'^2\right)\right) \Lambda ^4
    + 3  \left(k^2+4 \left(p^2+p'^2\right)\right) \Lambda ^6
     \right) ,
    \nonumber
  \\
  F_{\GQ}^{ww}(p, p', k) &=& C \frac{8 p'^2 \left(k^2  p^2 \Lambda ^2 -3 k  p \Lambda ^4 + 3 \Lambda ^6\right)}{k^5 p^3}.
\end{eqnarray}
Next, the contact charge density contribution to the deuteron
structure radius has the form
\begin{eqnarray}
  r^{2}_{\text{Cont}} =
  \frac{1}{\pi^2}
  \int\limits_{0}^{\infty} p^2 dp
  \int\limits_{0}^{\infty} p'^2 dp'
  F_{\Lambda} \left(p, p' \right)
  \big[
  F_{r^2}^{uu}(p, p') u(p) u(p')
  + F_{r^2}^{uw}(p, p') w(p) u(p')
  \big],
  \label{eq:rsqcont}
\end{eqnarray}
where
\begin{eqnarray}
  F_{r^2}^{uu}(p, p') &\equiv& -2\left(A+B + \frac{C}{3} \right)
  \left( 3 - \frac{2 (p^2 + p'^2)}{\Lambda^2}  + \frac{{(p^2 - p'^2)}^2}{\Lambda^4} \right),
  \\ \label{Eq:Fuw}
  F_{r^2}^{uw}(p, p') &\equiv& \frac{8 \sqrt{2}}{3} \, C \, \left(
  \frac{  2 p^2 }{\Lambda ^2}
  + \frac{  p^2 \left(p'^2-p^2\right)}{\Lambda ^4}
   \right) .
\end{eqnarray}
Finally, the contact contribution to the quadrupole moment reads:
\begin{eqnarray}
  Q_{\text{Cont}} =
  \frac{1}{\pi^2}
  \int\limits_{0}^{\infty} p^2 dp
  \int\limits_{0}^{\infty} p'^2 dp'
  F_{\Lambda} \left(p, p' \right)
  \big[
  F_{Q}^{uu}(p, p') u(p) u(p')
  + F_{Q}^{uw}(p, p') w(p) u(p')
  + F_{Q}^{ww}(p, p') w(p) w(p')
  \big],
  \label{eq:qcont}
\end{eqnarray}
where
\begin{eqnarray}
  F_{Q}^{uu}(p, p') &\equiv&
  (-4C) \left( 1 - \frac{2 (p^2+p'^2)}{3 \Lambda^2} + \frac{2(p^4+p'^4)}{15 \Lambda^4}\right)
  ,
  \qquad
  F_{Q}^{ww}(p, p') \equiv \frac{16}{15} C \frac{p^2 p'^2}{\Lambda ^4},
  \\ \label{Eq:FQuw}
  F_{Q}^{uw}(p, p') &\equiv&
  \frac{4 \sqrt{2}}{15} p^2 \left( (A+B) \left( \frac{6}{\Lambda^2} + \frac{3(p'^2-p^2)}{\Lambda^4}\right) + C \left( - \frac{5}{\Lambda^2} + \frac{p^2+p'^2}{\Lambda^4}\right) \right).
\end{eqnarray}

\section{Complete expressions for the contact charge density at
  N$^4$LO including isovector terms}\label{sec:appIVCT}

In this appendix we present the N$^4$LO contact charge density operators
including isovector contributions.
The isovector components do not contribute to the deuteron observables
in the single-photon approximation,
but have to be taken into account when calculating the FFs and charge radii of heavier nuclei.
Charge-density operators presented here are derived using the same
procedure as used for derivation of
Eq.~\eqref{eq:contactchargedensity}, but keeping the isovector
terms.
After calculating and antisymmetrizing the commutators of the LO charge
density with the generators of the unitary transformation
Eq.~\eqref{eq:unitarytransformGeneratorsNN} we obtain the
following result for the N$^4$LO contact charge density:
\begin{eqnarray}
\rho_\text{Cont,AS}^{(A+B+C/3)} &=&
2 e \left(A+B+\frac{C}{3}\right) \frac{\bm{\sigma}_1\cdot \bm{\sigma}_2 + 3}{4}
\bigg[
  \GE^S(\bm{k}^2) \frac{1- \bm{\tau}_1\cdot \bm{\tau}_2 }{4} \bm{k}^2
\nonumber
\\
  && + \GE^V(\bm{k}^2)
    \left(
      \frac{{(\bm{\tau}_1 - \bm{\tau}_2)}_3}{2} \bm{k}\cdot (\bm{p}-\bm{p}')
      -
      \frac{i {(\bm{\tau}_1 \times \bm{\tau}_2)}_3 }{2} \bm{k}\cdot (\bm{p}+\bm{p}')
     \right)
  \bigg],
  \label{eq:rhoCT(A+B+C/3)}
\\
\rho_\text{Cont,AS}^{(A - 3 B - C)} &=&
  2 e \, (A - 3 B - C) \frac{1 - \bm{\sigma}_1\cdot \bm{\sigma}_2}{4}
  \bigg[
      \left(\GE^S(\bm{k}^2) \frac{\bm{\tau}_1\cdot \bm{\tau}_2 + 3}{4}
       +\GE^V(\bm{k}^2) \frac{{(\bm{\tau}_1 + \bm{\tau}_2)}_3}{2}
       \right) \, \bm{k}^2
  \nonumber
  \\
       &&+\GE^V(\bm{k}^2)
       \left(
          \frac{{(\bm{\tau}_1 - \bm{\tau}_2)}_3}{2} \bm{k}\cdot (\bm{p}-\bm{p}')
          +
          \frac{i {(\bm{\tau}_1 \times \bm{\tau}_2)}_3 }{2} \bm{k}\cdot (\bm{p}+\bm{p}')
        \right)
  \bigg],
  \label{eq:rhoCT(A-3B-C)}
\\
  \rho_\text{Cont,AS}^{(C)} &=&
  2 e \, C
  \bigg[
  \,\GE^S(\bm{k}^2) \frac{1- \bm{\tau}_1\cdot \bm{\tau}_2 }{4}
    \left(
      (\bm{k} \cdot \bm{\sigma}_1)(\bm{k} \cdot \bm{\sigma}_2) - \frac{1}{3} \bm{k}^2 (\bm{\sigma}_1\cdot \bm{\sigma}_2)
    \right)
  \nonumber
  \\
  && + \GE^V(\bm{k}^2) \frac{{(\bm{\tau}_1 - \bm{\tau}_2)}_3}{2}  
    \frac12 \left(
      (\bm{k} \cdot \bm{\sigma}_1) \bm{\sigma}_2 \cdot (\bm{p}-\bm{p}')
      + (\bm{k} \cdot \bm{\sigma}_2) \bm{\sigma}_1 \cdot (\bm{p}-\bm{p}')
      -\frac23 \bm{k}\cdot (\bm{p}-\bm{p}') (\bm{\sigma}_1\cdot \bm{\sigma}_2)
    \right)
 \label{eq:rhoCT(C)}   \\
  && - \GE^V(\bm{k}^2) \frac{i {(\bm{\tau}_1 \times \bm{\tau}_2)}_3 }{2}  
      \frac12 \left(
        (\bm{k} \cdot \bm{\sigma}_1) \bm{\sigma}_2 \cdot (\bm{p}+\bm{p}')
        + (\bm{k} \cdot \bm{\sigma}_2) \bm{\sigma}_1 \cdot (\bm{p}+\bm{p}')
        - \frac23 \bm{k}\cdot (\bm{p}+\bm{p}') (\bm{\sigma}_1\cdot \bm{\sigma}_2)
      \right)
     \bigg].
 \nonumber
\end{eqnarray}
Notice that all isoscalar operators are proportional to
$\GE^S({\bm{k}}^2)$, while all isovector ones are proportional to $\GE^V({\bm{k}}^2)$.

Finally we would like to make a remark about the ${}^1S_0 \to {}^1S_0$ contact operator
in the first line of Eq.~\eqref{eq:rhoCT(A-3B-C)}, which involves
the isospin operator ${(\bm{\tau}_1 + \bm{\tau}_2)}_3$.
This structure is remarkable in several ways.
First, from all presented isovector terms, this is the only one which
is allowed by the Pauli principle in S-to-S-wave transitions. Second,
this structure ensures that correct nucleon form factors appear in all isospin-1-to-isospin-1 channels,
namely:
\begin{eqnarray}
	\GE^S(\bm{k}^2) \frac{\bm{\tau}_1\cdot \bm{\tau}_2 + 3}{4}
       +\GE^V(\bm{k}^2) \frac{{(\bm{\tau}_1 + \bm{\tau}_2)}_3}{2}
       =
       \begin{cases}
       	2 \GE^p & \text{ for } pp \to pp\\
       	\GE^p + \GE^n & \text{ for } pn \to pn\\
       	2 \GE^n & \text{ for } nn \to nn
       \end{cases}
\end{eqnarray}
Our derivation of the contact charge density operator demonstrates
that the isovector structure in Eq.~\eqref{eq:rhoCT(A-3B-C)}
should be proportional to the same linear combinations of LECs as corresponding isoscalar part.
This is in contrast to Ref.~\cite{Phillips:2016mov},
where an extra LEC associated with the isovector terms was introduced.


\end{document}